\documentclass[trackchange,twocolappendix,twocolumn]{aastex701}
\usepackage[utf8]{inputenc}
\usepackage[italicdiff]{physics}
\usepackage{amsmath}
\usepackage{amssymb}
\usepackage{tensor}
\usepackage{tikz}
\usepackage{amsthm}
\usepackage{mathtools}
\usepackage{graphicx}
\usepackage{booktabs}
\usepackage{caption}
\usepackage{subdepth}
\usepackage{multirow}

\renewcommand{\fig}[1]{Fig.~\ref{fig:#1}}
\newcommand{\tab}[1]{Table~\ref{tab:#1}}
\newcommand{\eq}[1]{Eq.~(\ref{eq:#1})}
\renewcommand{\inf}{\infty}

\defcitealias{gottlieb_spinning_2025}{G25}
\defcitealias{burrows_black_2023}{B23}
\defcitealias{Burrows+2024b}{B24}
\defcitealias{Burrows+2025}{B25}
\defcitealias{Janka&Kreese24}{J24}

\begin{document}

\title{Strong Black Hole Natal Kicks in Magnetized Accretion-Powered Stellar Explosions}
\author[orcid=0000-0002-6341-4484,sname='Li']{Sean E.~Li}
\affiliation{Department of Physics and Columbia Astrophysics Laboratory, Columbia University, New York, NY 10027, USA}
\email[hide]{sean.li@columbia.edu} 
\author[orcid=0000-0003-3115-2456,sname='Gottlieb']{Ore Gottlieb}
\affil{Department of Physics and Kavli Institute for Astrophysics and Space Research, Massachusetts Institute of Technology,\\Cambridge, MA 02139, USA}
\email[hide]{ogot@mit.edu}
\author[orcid=0000-0002-4670-7509
,sname='Metzger']{Brian D.~Metzger}
\affiliation{Department of Physics and Columbia Astrophysics Laboratory, Columbia University, New York, NY 10027, USA}
\affiliation{Center for Computational Astrophysics, Flatiron Institute, 162 5th Avenue, New York, NY 10010, USA}
\email[hide]{bdm2129@columbia.edu}
\begin{abstract}
Asymmetric stellar explosions impart an impulse, or natal kick, to their compact remnants by linear momentum conservation. Black hole (BH) natal kicks are often assumed to be weaker than neutron star kicks because of greater mass accretion, and they are harder to constrain observationally because isolated BHs are electromagnetically faint. Using 3D general-relativistic magnetohydrodynamic simulations lasting up to $\sim30\,{\rm s}$, we show that BHs formed from the collapse of rapidly rotating massive stars (collapsars) threaded by strong large-scale magnetic fields can acquire large natal kicks $\sim 10^2$--$10^3~\mathrm{km~s^{-1}}$. Asymmetric electromagnetic outflows, magnetic-flux eruptions, and the gravitational pull of aspherical ejecta shape the kick magnitude, and their relative contributions depend on the progenitor structure, magnetic-flux history, and BH spin. More rapidly spinning BHs receive stronger, more nearly spin-aligned kicks, primarily through the gravitational pull of asymmetric jet-driven ejecta. Delayed transitions to the magnetically arrested state produce more asymmetric outflows and can generate even stronger recoils. Because the large-scale magnetic flux required in our models is also a key ingredient of relativistic gamma-ray burst jets and their associated energetic, jet-driven supernovae, natal kicks may be a natural consequence of magnetized collapsars. Such kicks could substantially alter BH retention in dense stellar environments, binary survival, spin-orbit misalignment, and the viability of magnetized collapsars in producing highly spinning BHs within the pair-instability mass gap.
\end{abstract}

\section{Introduction}\label{sec:intro}
Many massive stars end their nuclear burning evolution by imploding to form black holes (BHs) \citep{Oppenheimer&Synder39}. In some core-collapse supernova (CCSN) explosions, continued accretion after core bounce can push the proto-neutron-star (PNS) remnant of the iron core above its maximum stable mass \citep[e.g.,][and references therein]{Burrows+2025,Janka25}. In other cases, the PNS collapses more promptly to a BH before neutrinos can revive the stalled shock \citep[e.g.,][]{Chan:2017tdg}. Rapid rotation of the progenitor core at collapse may fundamentally alter this outcome. In the collapsar model \citep{woosley_gamma-ray_1993,macfadyen_collapsars_1999}, the infalling envelope of the rotating progenitor circularizes into an accretion disk around the nascent Kerr BH, and the resulting extraction of energy through jets and disk winds generates a powerful explosion (e.g., \citealt{Woosley&Bloom06}).
 
If sufficient large-scale magnetic flux accumulates on the BH horizon, the collapsar disk can enter a magnetically arrested state \citep[MAD;][]{narayan_magnetically_2003,Tchekhovskoy2011}, allowing the BH to launch bipolar relativistic jets via the Blandford-Znajek (BZ) mechanism \citep{blandford_electromagnetic_1977}. The requisite poloidal magnetic flux may originate from a magnetohydrodynamic (MHD) dynamo in the disk \citep[e.g.,][]{jacquemin-ide_magnetorotational_2024,Shibata2025,Chan2026}, or from a rotating magnetized PNS whose flux the BH inherits upon collapse \citep{Gottlieb2024}. If the magnetic energy density near the horizon exceeds the kinetic energy density of the accreting plasma, the jets can emerge from the BH ergosphere \citep{Komissarov2009,gottlieb_collapsar_2023,Issa2025}, propagate through the star to break out from the stellar surface \citep{Gottlieb2022b}, and power a gamma-ray burst \citep[GRB;][]{macfadyen_collapsars_1999}. In addition to canonical long-duration GRBs, magnetized collapsars may power a continuum of electromagnetic (EM) transients whose properties depend sensitively on the progenitor structure \citep{Gottlieb2022}.

The vigorous, highly aspherical explosions in collapsars provide a natural mechanism to impart net linear momentum to the ejecta and, by linear momentum conservation, a recoil kick to the BH. In principle, this means collapsar BHs could acquire natal kicks commensurate with or even exceeding those of promptly ``successful" CCSN remnants \citep[see, e.g.,][]{Burrows+2024b,Janka&Kreese24}. Kicks not only determine single BHs' retention in their host cluster environments \citep[e.g.,][]{Antonini+16}, affecting rates of dynamical interactions, but also shape the evolution of binary systems \citep[e.g.,][]{Vivanco+2026}, to which most massive stars belong \citep{Sana_2012} and which are favored progenitors of long GRBs \citep[e.g.,][]{fryer_explaining_2025}. In binaries, kicks can shorten the time to coalescence, induce spin--orbit misalignment and precession, or rip companions apart, depending on their magnitude and orientation relative to the orbital axis \citep[][and references therein]{Vigna-Gomez25}. 

Evidence of spin--orbit misalignment in LIGO-Virgo-KAGRA (LVK) gravitational-wave (GW) data suggests some BHs in isolated binaries may receive substantial natal kicks \citep[e.g.,][]{oshaughnessy_inferences_2017, wysocki_explaining_2018,callister_state_2021}, subject to assumptions about their natal spins and evolutionary pathways. More direct observational constraints suggest a broad range of kicks, ranging from systems consistent with complete implosion and negligible kick \citep[e.g.,][]{Shenar+2022,Vigna-Gomez+2024,Burdge_2024}, to X-ray binaries requiring kicks of tens to hundreds of $\rm{km\,s^{-1}}$ \citep[e.g.,][]{atri_potential_2019,DashwoodBrown2024}. Gaia kinematic data are consistent with a bimodal BH kick distribution \citep{nagarajan_mixed_2025}, with some BHs receiving comparable natal kicks to neutron stars \citep{Mapelli+2026}. Physically motivated kick prescriptions are needed to aid interpretation of GW observables and break degeneracies in population synthesis models.

At sufficiently high progenitor masses and compactness, collapsing stellar cores may form BHs promptly through direct horizon (DH) collapse, without the intermediate phase of a metastable PNS \citep[e.g.,][]{doi:10.1142/9789811282676_0003}.  Recently, \citet[hereafter G25]{gottlieb_spinning_2025} showed that DH collapsar progenitors above the pair-instability supernova (PISN) mass gap \citep{renzo24} could form BHs consistent with the components of the binary black hole (BBH) merger GW231123 \citep{GW231123}; see also \citet{Siegel+22}. The mass of the secondary, and possibly the primary, BH in GW231123 likely fell in the PISN gap, and the merger exhibited signs of spin--orbit misalignment \citep[for analysis of waveform systematics that reaffirms this interpretation, see][]{Bini2026}. Theoretical predictions for collapsar kicks would have implications for the viability of the collapsar channel for generating GW231123 and other LVK BBH mergers.

The kick depends on the stochastic asymmetries of the dynamics and how long they persist. Previous numerical studies \citep[e.g.,][]{scheck_multidimensional_2006, Nordhaus2010, wongwathanarat_three-dimensional_2013,stockinger2020,Janka&Kreese24,Burrows+2024b, sykes_long-time_2025,halevi_black_2025} have investigated recoil kicks in CCSNe, showing that anisotropic mass ejection, neutrino emission, fallback accretion, and the gravitational ``tugboat" attraction toward slower ejecta can accelerate neutron star (NS) and BH remnants to hundreds and even thousands of $\rm km~s^{-1}$ \citep[for review, see][and references therein]{popov_natal_2025}. However, for the most part these studies do not consider magnetic fields or rapid progenitor rotation; many are also limited to 2D or use approximate general relativity (GR). The main exception is \citet{halevi_black_2025}, who performed the first self-consistent simulation of the collapse of a rapidly rotating, weakly magnetized progenitor to a BH; however, they only tracked the post-BH collapsar evolution for $\lesssim 50~\mu\rm s$, which is insufficient time for the BH kick to saturate. \citet{Du+2018} analytically modeled the GW emission from a BH recoiling due to asymmetric jets, but the effect of jets on the BH kick remains to be explored numerically. Taken together, these limitations have left natal kicks from long-lived, magnetically driven collapsar explosions essentially unexplored.

\begin{figure*}
    \centering \includegraphics[width=\linewidth]{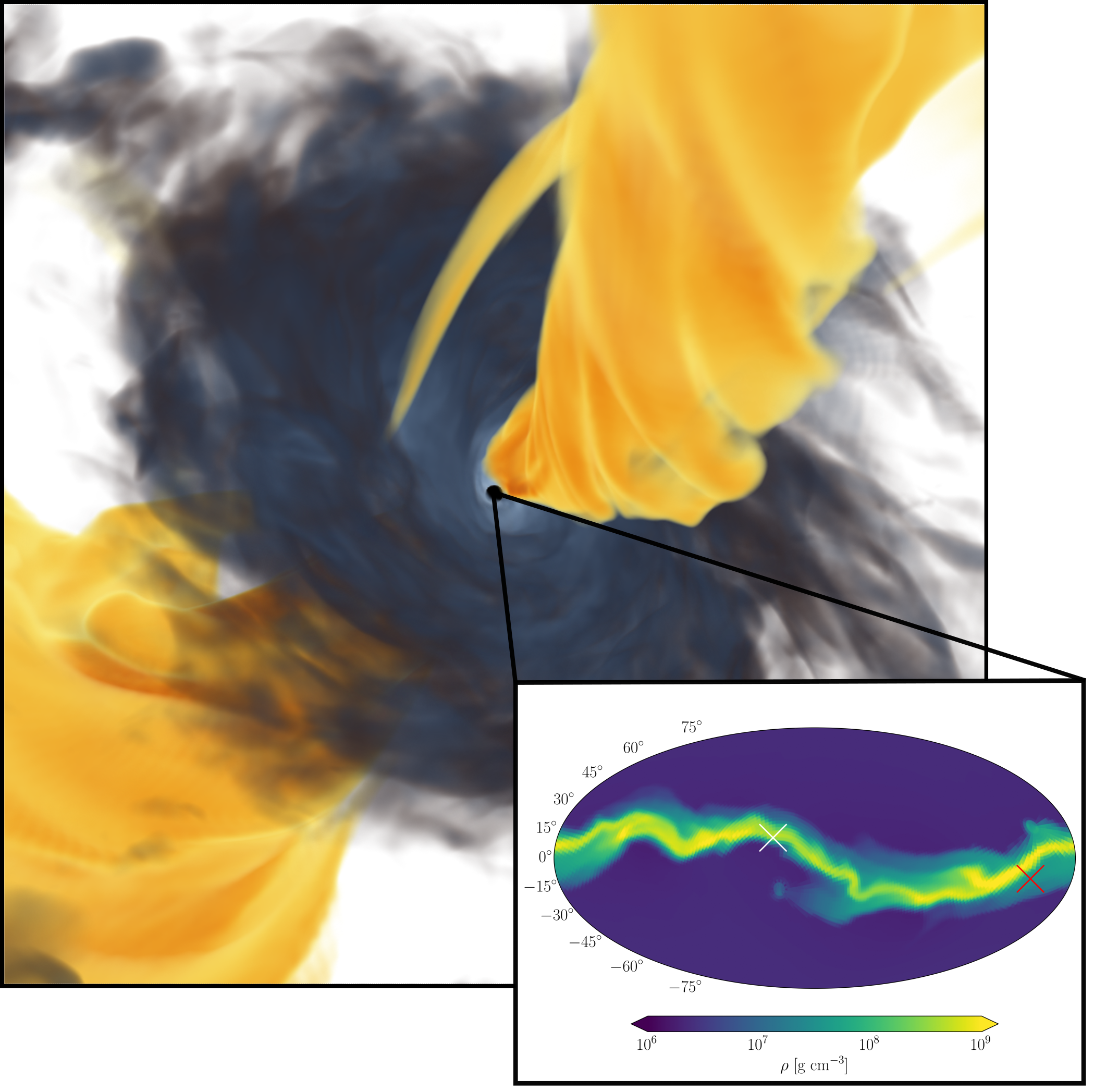}
    \caption{Volume rendering of the rest mass density on a logarithmic scale for our low-spin $a = 0.1$ fiducial collapsar simulation \texttt{a1} (but performed at slightly lower resolution). At this snapshot, the high-density accretion disk (shown in blue) has angular momentum nearly perpendicular to the BH spin axis. The jets are deflected as they propagate through falling gas, resulting in a helical low-density structure (shown in yellow). A narrow secondary funnel originates from a magnetic-flux eruption that punctures the disk tens of gravitational radii from the BH horizon, highlighting the extreme non-axisymmetric nature of the MHD morphology. The inset in the lower right shows a Mollweide projection of the rest-mass density on the BH horizon, viewed in a frame aligned with the disk angular momentum. The white (red) ``X" indicates (opposite) the direction of the instantaneous kinetic stress on the BH. The disk exhibits non-axisymmetric warping, primarily in the $m = 1$ and $m = 2$ modes (see Appendix \ref{app:warp} for analysis of the disk warp in our simulations).}
    \label{fig:a1_rho_volume}
\end{figure*}

In this paper, we investigate natal kicks to collapsar BHs numerically over long time scales for the first time.  Using global 3D general-relativistic magnetohydrodynamic (GRMHD) simulations, we evolve the collapsar system for several seconds to determine the kick velocity vector using global linear momentum balance. By separating the stresses on the horizon from gravitational interactions at a distance, and tracking the behavior of the disk and jets, we obtain a qualitative picture of the kick mechanisms. Our fiducial simulations, corresponding to low-mass collapsar progenitors with strong magnetic fields, exhibit a highly aspherical and non-axisymmetric ejecta morphology, as depicted in \fig{a1_rho_volume}. The BH-disk engine unbinds most of the progenitor star mass, enabling a sustained kick. We also apply our analysis to three of the more massive DH collapsar simulations of \citetalias{gottlieb_spinning_2025}. Our results indicate that BHs formed in strong accretion-powered explosions acquire strong kicks of hundreds of $\rm{km\,s^{-1}}$, with magnitudes and directions that vary with the BH spin and the available magnetic flux.

This paper is organized as follows.  In Sec.~\ref {sec:estimates}, we start with a general discussion of sources of ejecta asymmetries in collapsars and rough analytic estimates for the corresponding BH kick.  In Sec.~\ref {sec:methods}, we introduce the GRMHD simulations and describe our numerical methods for extracting BH kicks from the simulation data. We then overview our simulation results for the collapsar evolution, including sources of asymmetry, in Sec.~\ref {sec:accretion}. In Sec.~\ref{sec:results} we present the resulting BH kicks, and discuss the implications and limitations of our work in Sec.~\ref{sec:discussion}. We conclude with our main findings in Sec.~\ref{sec:conclusion}.

Unless otherwise noted, in this manuscript we adopt natural units $G = M_{\rm BH} = c = 1$, where $G$ is the gravitational constant, $M_{\rm BH}$ the BH mass, and $c$ the speed of light. As such, lengths are measured in gravitational radii $r_g \equiv GM_{\rm BH}/c^2$ and time in units of the light crossing time $r_g/c$. We restore factors of $M_{\rm BH}$ and $r_g$ for clarity as needed. We use Heaviside-Lorentz units for the magnetic field.
\section{Analytical Kick Estimates}\label{sec:estimates}
Several sources of asymmetry characterize the collapsar system, primarily associated with the disk and jets. These in turn imprint asymmetries on both the ejected and accreted material. Asymptotically, the recoil momentum of the remnant BH must balance the linear momentum carried by the asymmetric ejecta. Here, we provide simple estimates for the expected kick magnitudes, which shall help guide the interpretation of our quantitative results in Sec.~\ref{sec:results}. We consider direct jet recoil, indirect recoil mediated by the large-scale ejecta, and stochastic momentum transfer associated with the inner accretion flow.

If the powers of the two jets differ, bipolar relativistic jets can impart a kick to the BH via the reaction force. Even equally powerful jets can kick the BH if their axes are not antiparallel. In general, both effects likely play some role. For a fractional jet power asymmetry $\alpha_{\rm jet}$ along some axis (encompassing both aligned and misaligned jets) sustained over a coordinate time $T_{\rm jet}$, the kick imparted to a BH of mass $M_{\rm BH}$ is
\begin{align}
    v^{\rm BH}_{\rm jet} \simeq 21\,{\rm km\,s^{-1}}  \frac{\alpha_{\rm jet}}{0.1} \frac{P_{\rm jet}}{10^{50}\,\rm erg\,s^{-1}}
    \frac{4 M_\sun}{M_{\rm BH}} \frac{T_{\rm jet}}{50\,\rm s}.
    \label{eq:vkick_jet}
\end{align}
For typical (beaming-corrected) GRB jet power $P_{\rm jet} \sim 10^{50}\,{\rm erg\,s^{-1}}$ (e.g., \citealt{Wanderman&Piran10}), jets can directly impart a substantial recoil kick $\gtrsim 100$ km s$^{-1}$ if their asymmetry is very large, e.g.,~$\gtrsim 50\%$, and persistent over long times.

The potentially larger, indirect effect of the jets is via their impact on the surrounding star. As the jets propagate through the stellar envelope, much of their energy is converted into bulk kinetic energy of the sub-relativistic ejecta. Asymmetric jets carve out an asymmetric gas distribution, enhancing both the gravitational tugboat effect and the asymmetry of the accretion flow. The momentum crossing the horizon from freely-falling accreting matter cancels its gravitational pull on the way down. On large scales, however, asymmetric ejecta unbound by the jets may exert a persistent tugboat pull as it expands toward spatial infinity.

To estimate the recoil from this non-relativistic ejecta, i.e.~from the jet-driven supernova, we assume that a fraction $f_{\rm ej}$ of the stellar envelope of mass $M_\star$ (not including the BH mass) is ejected with a speed $v_{\rm ej} \ll c$, while the rest accretes onto the central BH of initial mass $M_{\rm BH,0} = f_{\rm BH,0}M_\star$. A time-averaged fractional asymmetry $\alpha_{\rm ej}$ in the outflow along some axis yields a kick 
\begin{align}
    v^{\rm BH}_{\rm ej} &\underset{\phantom{f_{\rm ej} = 0.5,\,f_{\rm BH,0} = 0.25}}{\simeq} \,\alpha_{\rm ej}\frac{f_{\rm ej}}{f_{\rm BH,0} + 1 - f_{\rm  ej} }v_{\rm ej} \label{eq:ej-kick} \\
    &\underset{f_{\rm ej} = 0.5,\,f_{\rm BH,0} = 0.25}{\approx} \,200\,{\rm km\,s^{-1}}\frac{\alpha_{\rm ej}}{0.01} \frac{v_{\rm ej}}{0.1 c},
    \label{eq:vkick_ej}
\end{align}
where the final remnant mass has been approximated as $M_{\rm rem} \simeq \left[f_{\rm BH,0}+(1-f_{\rm ej})\right]M_\star$. For example, hydrodynamic and light-curve modeling of the well-studied collapsar supernova SN1998bw found the progenitor to be a $M_\star \sim 14M_\sun$ C+O star with an ejecta mass $M_{\rm ej}\sim 10\,M_\sun$ and compact remnant mass $\sim 4M_\sun$ \citep{nakamura_1998bw_2001,Woosley&Bloom06}, i.e.~$f_{\rm ej} \sim 1$. For even a small asymmetry $\alpha_{\rm ej} = 1\%$ and $v_{\rm ej} = 10^4\,\rm km\,s^{-1}$ the resultant kick would be $\sim 350\,\rm km \,s^{-1}$. Achieving a comparable kick with $f_{\rm ej} = 0.5$ requires an asymmetry of $\alpha_{\rm ej} = 5.5\%$. For the same fractional asymmetry, this kick greatly exceeds that imparted by the jet directly (Eq.~\eqref{eq:vkick_jet}), a consequence of the fact that, for fixed energy, non-relativistic particles carry more momentum than radiation by a factor $c/v_{\rm ej} \gg 1$.  

Building a causal understanding of the BH kick requires also inspecting asymmetry near the BH horizon, i.e.~on the scale of the inner accretion disk. Collapsar disks are susceptible to non-axisymmetric perturbations, e.g., spiral waves, turbulent eddies, and overdense clumps \cite[e.g.,][]{henisey_variability_2012,Gottlieb2024b} that can transfer linear momentum to the BH. Periodic magnetic flux eruptions powered by reconnection can partially eject the disk, creating short-lived spiral depressions, as observed in previous GRMHD simulations \citep{chatterjee_flux_2022, ripperda_black_2022,Bopp2025}. The energy outflow rate during an eruption can be written $\dot E \sim \eta \dot M c^2$, where \citet{chatterjee_flux_2022} find an efficiency of $\eta \lesssim 10\%$. \citet{jacquemin_ide_2025} measure a recurrence time $\Delta T \simeq 1500\,r_g/c$ and a duty cycle $f_{\rm duty} = \Delta t/\Delta T \simeq 0.2$, where $\Delta t$ is the duration of the impulse, in agreement with their analytic expectations.

For disk ejecta moving at speed $v_{\rm disk}<c$, the momentum carried outward is $\Delta p = \dot E\Delta t / v_{\rm disk}$. If the impulse from each eruption is randomly oriented, the resultant kick speed after an accretion time $T_{\rm acc}$ follows a random walk, 
\begin{align}
    v^{\rm BH}_{\rm disk} &= \frac{\Delta p}{M_{\rm BH}}\sqrt{\text{\# of steps}} \\
    &\simeq \eta \frac{\langle \dot M \rangle \Delta t}{M_{\rm BH}}\frac{c^2}{v_{\rm disk}}\sqrt{\frac{T_{\rm acc}}{\Delta T}} \\
    \begin{split}
       &\approx 150\,\mathrm{km\,s^{-1}}\frac{\eta}{0.1} \frac{f_{\rm acc}}{0.1}\frac{0.5}{v_{\rm disk}/c} \\
       &\phantom{\approx\,}\times\qty(\frac{\Delta t}{300\,r_g/c})^{1/2}\qty(\frac{10^5\, r_g/c}{T_{\rm acc}})^{1/2}.
    \end{split}
\end{align}
where we have taken $\Delta T = 5\Delta t$. Thus, for a low-mass collapsar with $M_{\rm BH} = 4M_\odot$ accreting for $\sim 2\,\rm s$, flux eruptions with $v_{\rm disk} = 0.5c$ can power strong kicks provided the BH accretes $f_{\rm acc} = M_{\rm acc}/M_{\rm BH}\gtrsim 10\%$ of its initial mass through the disk, where $M_{\rm acc} = \langle\dot{M}\rangle T_{\rm acc}$. For a massive collapsar BH with $M_{\rm BH} = 40M_\odot$ accreting $10\%$ in the same time, the corresponding kick is larger by a factor of $\sim3$.

The powerful jets enabled by MAD-level magnetic flux suppress mass accretion and may thereby regulate the momentum transfer of flux eruptions. Moreover, if the eruptions are too weak to unbind material from the BH, they cannot contribute to the final kick. The disk may nevertheless mediate the BH kick in other ways; in particular, it is likely to be warped, i.e.~distorted vertically relative to the local orbital plane, by a combination of Lense-Thirring (LT) \citep{1918PhyZ...19..156L} and magnetic torques. Such non-axisymmetric disk warping may help communicate the global momentum budget to the BH by enhancing asymmetric accretion.

We emphasize that regardless of the horizon-scale physics mediating linear momentum transport, the BH can only acquire an asymptotic kick if outflows escape to infinity. A BH swallowing its progenitor whole may experience temporary net forces, but these must cancel in the end by momentum conservation. 
\section{Numerical Methods}\label{sec:methods}
\subsection{Numerical Setup}

To investigate collapsar kicks numerically, we use the 3D GRMHD finite-volume code \texttt{H-AMR} \citep{h-amr} to solve the equations of ideal MHD in the Kerr spacetime, following earlier collapsar simulation work \citep{Gottlieb2022,Gottlieb2022b,gottlieb_collapsar_2023,Gottlieb2024,gottlieb_spinning_2025}. Owing to GPU acceleration, adaptive mesh refinement (AMR), and local adaptive time stepping, \texttt{H-AMR} is well-suited for 3D collapsar simulations that require high resolution to resolve MHD turbulence and jets, and long runtimes to converge on BH remnant properties. Since the computational cost of numerical relativity is prohibitive for long runtimes in 3D, \texttt{H-AMR} does not evolve the spacetime metric. As a result, our simulations do not self-consistently capture the evolution of the BH mass or spin, the self-gravity of the gas, or the motion of the BH in the progenitor frame. However, we approximately account for the growth of the BH inertia when computing the kick, as we discuss in Sec.~\ref{subsec:kick}.

We initialize our simulations with a spinning BH at the center of a stripped (Wolf-Rayet-like) stellar envelope. We adopt spherical Kerr-Schild (KS) coordinates ($t,r,\theta,\phi$), in which the outer BH horizon is a coordinate sphere of radius $r_{+} = 1 + \sqrt{1 - a^2}$. The grid is uniform in $\ln r$, $\theta$, and $\phi$. For our fiducial models, the grid extends over $0.9r_{+} \lesssim r \lesssim 10^5\,r_g \simeq 1.6 r_\star$, where $r_\star$ is the progenitor radius, and the highest AMR level reaches a resolution of $N_r \times N_\theta \times N_\phi = 1536 \times 512 \times 640$ cells. For the grid setup of the direct horizon models, see Sec.~3 of \citetalias{gottlieb_spinning_2025}.

\begin{deluxetable*}{ccrrrrcrrrr}
\tablewidth{\textwidth}
    \tablecaption{Model parameters, including the Kerr spin parameter $a$, the logarithm of the polar magnetic field strength $B$ in Gauss evaluated at $100\,r_g$, simulation time $T_{\rm sim}$, initial BH mass $M_{\rm BH,0}$ and progenitor mass (excluding the BH mass) $M_{\star}$, the form of the initial vector potential $A_\phi$, the initial density profile $\rho(r)$, and maximum number of cells $N_r$, $N_\theta$, and $N_\phi$ in modified spherical KS coordinates ($t$, $\ln r$, $\theta$, $\phi$). The fiducial model names correspond to ten times the BH spin parameter. We denote the modified field model described in Sec.~\ref{subsec:cf} with ``\texttt{*}". The direct horizon models of \citetalias{gottlieb_spinning_2025} are labeled with ``\texttt{dh}" followed by their original names, which consist of ``\texttt{B}" and a suffix indicating the magnetic field strength (``\texttt{m}" for ``moderate", ``\texttt{s}" for ``strong", and ``\texttt{vs}" for ``very strong").
    \label{tab:models}}
    \tablehead{
    \colhead{Model} & \colhead{$a$} & \colhead{$\log_{10}(B_{100}[\rm G])$} & \colhead{$T_{\rm sim}\,[\rm s]$} & \colhead{$M_{\rm BH,0}\,[M_\sun]$} & \colhead{$M_{\star}\,[M_\sun]$} & \colhead{$A_\phi$} & \colhead{$\rho(r)$} & \colhead{$N_r$} & \colhead{$N_\theta$} & \colhead{$N_\phi$}}
    \startdata 
    \multicolumn{11}{c}{Fiducial} \\
    \midrule 
    \texttt{a1} & $0.1$ & 12.2 & 32.5 & 4 & 14 & \eq{a-phi} & \eq{rho-init} & 1536 & 512 & 640 \\ 
    \texttt{a5} & $0.5$ & 12.2 & 20.7 & 4 & 14 & \eq{a-phi} & \eq{rho-init} & 1536 & 512 & 640 \\
    \texttt{a8} & $0.8$ & 12.2 & 17.9 & 4 & 14 & \eq{a-phi} & \eq{rho-init} & 1536 & 512 & 640 \\
    \midrule
    \multicolumn{11}{c}{Modified Field} \\
    \midrule
    \texttt{a5*} & $0.5$ & 12.2 & 23.8 & 4 & 14 & Sec.~\ref{subsec:cf} & \eq{rho-init} & 1536 & 512 & 640 \\
    \midrule
    \multicolumn{11}{c}{Direct Horizon} \\
    \midrule
    \texttt{dhBm}  & $0.5$ & 11.0 & 6.5 & 40 & 100 & \eq{dh-field}, $\sim r^{-3}$ & Sec.~\ref{subsec:dh} & 4096 & 2304 & 4096 \\
    \texttt{dhBs} & $0.5$ & 11.7 & 2.8 & 40 & 100 & \eq{dh-field}, $\sim r^{-2}$ & Sec.~\ref{subsec:dh} & 4096 & 2304 & 4096  \\
    \texttt{dhBvs} & $0.5$ & 13.3 & 1.3 & 40 & 100 & \eq{dh-field}, $\sim r^{-1}$ & Sec.~\ref{subsec:dh} & 4096 & 2304 & 4096
    \enddata
    \tablecomments{In this work, we correct a typo in Table 1 of \citetalias{gottlieb_spinning_2025}: the magnetic field strength presented there is evaluated at $10\,r_g$, not $100\,r_g$.}
\end{deluxetable*}
\subsection{Fiducial Initial Conditions}

Our fiducial models assume a BH of initial gravitational mass $M_{\rm BH,0} = 4 M_\sun$ surrounded by a pre-collapse stellar envelope of mass $M_\star = 14 M_\sun$ and radius $r_\star = 3.75\times 10^{10}\,\rm cm$; these parameters roughly correspond to the stripped-envelope progenitors of collapsars with zero-age main-sequence stellar masses of $\sim$25--40$M_\sun$ (e.g., \citealt{1998bw,Woosley&Heger06}).  The initial BH mass moderately exceeds that expected right after the accretion-induced collapse of the PNS that forms first \citep[e.g.,][]{Margalit+22} and is not modeled here. We therefore focus on the post-BH-formation collapsar evolution as the subject of this work and do not model the preceding PNS collapse or any potential kick accumulated during that phase.

We consider three Kerr spin parameters $a \in \{0.1, 0.5, 0.8\}$, and adopt a fully relativistic polytropic equation of state with an adiabatic index $\gamma = 4/3$, consistent with the radiation-dominated conditions that characterize most regions of the collapsar system. We neglect cooling due to neutrino emission, leaving its effect on the inner accretion flow to future work. With these choices, our simulations are largely scale-free, so that ejecta masses and kick velocities can be rescaled across different BH and stellar masses to reasonable accuracy without breaking consistency with the initial density profile. 

Following \citet{Gottlieb2019}, we initialize the fluid-frame gas density profile of the stellar progenitor to
\begin{equation}
    \rho(r) =
        \rho_0 \qty(\frac{r}{r_g})^{-\alpha}\qty(1 - \frac{r}{r_\star})^{\beta}, \label{eq:rho-init}
\end{equation}
where $r_\star$ is the progenitor radius, and the normalization $\rho_0$ is chosen such that 
\begin{align}
    M_\star = \int D\sqrt{-g}\dd{r}\dd{\theta}\dd{\phi}.
\end{align}
Here $D \equiv \rho u^t$, where $u^t$ is the contravariant time component of the fluid 4-velocity, and $g \equiv \abs{g_{\mu\nu}}$ is the determinant of the spacetime metric. The power-law indices $\alpha$ and $\beta$ set the compactness of the progenitor. We choose $\alpha = 1.5$ and $\beta = 3$, motivated by collapsar progenitors \citep{Chan2026}. To keep the outer layers of the star bound in lieu of self-gravity, our progenitor is initially cold, with internal energy density
\begin{align}
    \epsilon(r) = 10^{-6}\rho(r).
\end{align}
To break spherical symmetry, we apply Gaussian perturbations to the internal energy density with an amplitude of $4\%$.

For all models except the DH collapsars, we adopt a specific angular momentum that corresponds to rigid body rotation at $r \lesssim 70\,r_g$, and becomes uniform outside \citep[see, e.g.,][]{Chan2026}, i.e.
\begin{align}
    l(r,\theta) = \Omega_0\left[\min\left(r, 70\,r_g\right)\right]^2\sin^2\theta.
\end{align}
We set the normalization $\Omega_0 = 5\,\rm rad\,s^{-1}$ to ensure prompt disk formation, consistent with the expectation that a disk is already present at the onset of BH formation in \texttt{MESA} stellar evolution models \citep{Gottlieb2024,Renzo2026}. The gas four-velocity is set to $u^\mu(t=0) = (u^t, 0, 0, u^\phi)$. Although in reality the star is already falling at BH birth, it quickly acquires $u^r < 0$, so our choice has little effect on the subsequent accretion morphology.

To model a scenario in which the BH inherits magnetic flux from the PNS, we initialize a central core of radius $r_{\rm c} \simeq 160\,r_g$ with a uniform magnetic field, and impose a dipole-like field for $r_{\rm c} < r < r_\star$. The corresponding 3-vector potential is $A_i = (0, 0, A_\phi)$, with
\begin{equation}
    A_\phi(r, \theta) = \mu \frac{\sin\theta}{r^2}
         \max\qty(\frac{r^2}{r^2 + r_{\rm c}^2} - \qty(\frac{r}{r_\star})^3, 0). \label{eq:a-phi}
\end{equation}
We choose the magnetic moment $\mu$ so that the BH attains a MAD-level amount of magnetic flux (see \tab{models} for the normalizations of the magnetic field), approximating flux inherited from the PNS. For numerical stability, we impose a density floor via a cap on the magnetization parameter $\sigma \equiv b^2 / \rho \leq \sigma_{\rm max}$, where $b^\mu$ is the fluid-frame magnetic four-vector, with $b^2 = b^\mu b_\mu$. In our fiducial models, $\sigma_{\rm max} = 150$.

\subsection{Modified Field Simulation}\label{subsec:cf}

Our fiducial models initialize the BH vicinity with sufficiently large-scale magnetic flux to produce powerful jets soon after the accretion disk forms.  However, in some collapsars the BH may not inherit an appreciable amount of magnetic flux from the PNS \citep[see discussion in][]{Chan2026}, in which case the flux required to power a GRB jet must instead be accumulated or amplified within the disk, potentially aided by a dynamo process.  To explore this possibility, we introduce a modified-field model, \texttt{a5*}, which is otherwise identical to the fiducial $a=0.5$ model but differs in the radial distribution of the initial poloidal magnetic flux.

Inside $r\simeq 200\,r_g$, we retain the fiducial field profile of Eq.~(\ref{eq:a-phi}), so that the material that accretes promptly supplies a comparable initial disk field.  At larger radii, we replace this profile with a rescaled, weaker large-scale poloidal field,
\begin{align}
    A_\phi(r,\theta) \propto \sin\theta \max\qty(\frac{r^2}{r^2 + r_{\rm c}^2} - \qty(\frac{r}{r_\star})^3, 0),
\end{align}
with the normalization chosen such that the flux advected to the BH from $\sim 10^2$--$10^4\,r_g$ is insufficient by itself to maintain a long-lived MAD state.  A later transition to the MAD regime must then rely on the accumulation and reorganization of the weaker field supplied from larger radii, potentially aided by disk dynamo action.

This configuration is not intended to represent a physically self-consistent stellar magnetic-field model.  Rather, it provides a controlled experiment in which the same progenitor model and symmetry-breaking perturbations pass through both sub-MAD and MAD-like phases, allowing us to assess how the BH kick depends on the origin and accumulation history of the large-scale magnetic flux.

\subsection{Direct Horizon Simulations}\label{subsec:dh}

We also consider three more massive collapsar models with $M_{\rm BH} = 40M_{\odot}$ and $M_{\star} \approx 100M_{\odot}$: ``moderately,'' ``strongly,'' and ``very strongly" magnetized DH collapsar models \textit{Bm}, \textit{Bs}, and \textit{Bvs} of \citetalias{gottlieb_spinning_2025}, which we denote here as \texttt{dhBm}, \texttt{dhBs}, and \texttt{dhBvs}. These models provide a qualitatively distinct regime in which the greater progenitor mass and compactness produce much larger accretion rates and jet powers, allowing us to test whether the kick mechanisms identified in the lower-mass collapsars persist in the DH regime.

Direct-horizon collapsars are also possible progenitors for highly spinning BHs within or near the upper PISN mass gap, such as the binary components responsible for GW190521 \citep{Siegel+22} and GW231123 \citepalias{gottlieb_spinning_2025}. In particular, \citetalias{gottlieb_spinning_2025} found that the spin and mass of the remnant BH in model \texttt{dhBm} were consistent with those inferred for the GW231123 secondary. (We do not consider the weakly magnetized model \textit{Bw} in this work, since the BH in that simulation is expected to retain a near-zero asymptotic kick because it accretes nearly the entire progenitor.) 

The stellar density profile of \texttt{dhBm}, \texttt{dhBs}, and \texttt{dhBvs} was derived by evolving a $\sim 250M_\sun$ He core in the 1D stellar evolution code \texttt{MESA} \citep{Paxton2011,Paxton2013,Paxton2015,Paxton2018,Paxton2019,Jermyn2023} up to core-collapse (at a mass of $\approx 149M_{\odot}$), then tracking its collapse to a BH with the general-relativistic hydrodynamics code \texttt{GR1D} \citep{OConnor2010}. The angular momentum profile is assumed to be a power law fit to 120 lower-mass progenitors evolved in \texttt{MESA} \citep{Chan2026}.  In the DH models, each BH is initialized with a moderate spin of $a = 0.5$, calculated by integrating the initial angular momentum up to the mass coordinate enclosing the BH. In addition to its strength, the structure of the magnetic field varies between the DH models; the vector potential is given by
\begin{equation}
    A_\phi = \mu\frac{\sin\theta}{r} r^n \max\qty(\frac{1}{r} - \frac{1}{r_\star}, 0), \label{eq:dh-field}
\end{equation}
with $n = -1$, $n = 0$, and $n = 1$ for \texttt{dhBm}, \texttt{dhBs}, and \texttt{dhBvs}. 

Notably, owing to their higher masses and more compact structures, the DH progenitors result in much higher mass accretion rates and jet powers than the less massive fiducial models for the same initial magnetic flux. Although models \texttt{dhBs} and \texttt{dhBvs} do not replicate any known BBH merger components, we include them in our analysis to further explore the relationship between the kick and magnetic flux available to the BH. See \citetalias{gottlieb_spinning_2025} for further details on the setup and motivation of the DH models.

\subsection{Black-Hole Kick Calculation}
\label{subsec:kick}

Since the BH is fixed at the coordinate origin, we do not measure its kick velocity directly. Instead, we infer the BH momentum as the negative of the total matter-plus-field momentum remaining on the grid and that which has crossed the outer boundary. We then use the horizon flux to define a diagnostic accretion contribution, and we identify the remainder as the gravitational momentum exchange between the plasma and the BH.

In an asymptotically flat spacetime in asymptotically Minkowskian coordinates, the total linear 4-momentum is defined by
\begin{align}
    p_\mu = \int (\tensor{T}{^0_\mu} + \tensor{t}{^0_\mu})\sqrt{-g}\dd[3]{x},
\end{align}
where $T^{\mu\nu}$ is the energy-momentum tensor and $t^{\mu\nu}$ the energy-momentum pseudo-tensor. This quantity is globally conserved and coordinate-invariant \citep[for a pedagogical discussion, see Sec.~20.3 of][]{MTW}. In the Kerr spacetime, the net spatial contribution from $t^{\mu\nu}$ vanishes, so
\begin{align}
    p_i = \int \tensor{T}{^0_i}\sqrt{-g}\dd[3]{x}.
\end{align}
In our work, we numerically calculate the total momentum in the spacetime at coordinate time $t$ via
\begin{align}
    p_i(t) = \int \tensor{T}{^t_i}\sqrt{-g}\dd{r}\dd{\theta}\dd{\phi} + p_i^{\rm esc}(t), \label{eq:p-total}
\end{align}
where $p_i^{\rm esc}$ is the momentum that escapes, if any, through the outer boundary of the grid at $r = r_{\rm max}$,
\begin{align}
    p^{\rm esc}_i(t) = \int_0^t \dd{t'} \oint_{r_{\rm max}} \tensor{T}{^r_i}\sqrt{-g}\dd{\theta}\dd{\phi}. \label{eq:p-out}
\end{align}
Here $\tensor{T}{^r_i}$ is the energy-momentum tensor in a mixed spherical-Cartesian coordinate basis, representing the flux of linear momentum in the positive radial direction. The index $i$ runs over spatial Cartesian KS coordinates, $i = x, y, z$.

Neglecting gravitational radiation, perturbations to the metric from matter and fields around the BH, and other higher-order effects, we approximate the linear momentum carried by the BHs with respect to their progenitors as opposite the total momentum carried by the gas at constant coordinate time,
\begin{align}
    p^{\rm BH}_i(t) = -p_i(t). \label{eq:p-bh}
\end{align}
We calculate the total BH kick velocity by normalizing the momentum given by \eq{p-bh} to the BH gravitational mass $M_{\rm BH}$,
\begin{align}
    v^{\rm BH}_i = p^{\rm BH}_i M^{-1}_{\rm BH}. \label{eq:v-kick}
\end{align}
We approximate the evolution of $M_{\rm BH}$ by integrating the mass accretion rate $\dot M$ (\eq{m-dot}) minus the rate of rest energy extraction by the BZ jets,
\begin{align}
    M_{\rm BH}(t) = M_{\rm BH,0} + \int_0^t \dot M \qty(1 - \eta_{\rm jet})\dd{t'},
\end{align}
where $\eta_{\rm jet}$ is the jet efficiency,
\begin{align}
    \eta_{\rm jet} = P_{\rm jet}\dot M^{-1}. \label{eq:eta}
\end{align}
We discuss the calculation of the total jet power $P_{\rm jet}$ in Appendix \ref{app:diag}.

Analogously with previous kick calculations \citep[e.g.,][]{scheck_multidimensional_2006,wongwathanarat_three-dimensional_2013,Janka&Kreese24}, we decompose the forces on the BH into
\begin{align}
    \dot p^{\rm BH}_i = \dot p_i^{\rm acc} + \dot p_i^{\rm grav} + (\dot M_{\rm BH} - \dot M)v_i^{\rm BH}, \label{eq:force}
\end{align}
where a dot denotes a total derivative with respect to coordinate time.

The first term in \eq{force}, $\dot p_i^{\rm acc}$, is the rate of net linear momentum capture through the BH horizon (``momentum accretion rate"), given by
\begin{align}
    \dot p_i^{\rm acc} = -\oint_{r_+} \tensor{T}{^r_i} \sqrt{-g}\dd{\theta}\dd{\phi}. \label{eq:f-mom}
\end{align}
To assess the relative contributions of asymmetric mass accretion, gas internal energy and pressure, and Maxwell (EM) stress to the BH kick, we write the ideal MHD energy-momentum tensor $T^{\mu\nu}$ as the sum of three terms,
\begin{align}
    \tensor{T}{^{\mu\nu}} \equiv T^{\mu\nu}_{\rm kinetic} + T^{\mu\nu}_{\rm thermal} + T^{\mu\nu}_{\rm EM}.
\end{align}
These are given by
\begin{align} 
    T^{\mu\nu}_{\rm kinetic} \equiv \rho u^\mu u^\nu, \label{eq:kinetic}
\end{align}
where $u^\mu$ is the fluid four-velocity,
\begin{align}
    T^{\mu\nu}_{\rm thermal} \equiv (\epsilon + p) u^\mu u^\nu + \tensor{g}{^{\mu\nu}} p, \label{eq:thermal}
\end{align}
where $\epsilon$ is the internal energy density and $p$ the gas pressure, and
\begin{align}
    T^{\mu\nu}_{\rm EM} \equiv b^2 u^\mu u^\nu + \frac{1}{2}\tensor{g}{^{\mu\nu}} b^2 - b^\mu b^\nu.
\end{align} 
We integrate these terms separately over the horizon, thus further breaking up \eq{f-mom} into 
\begin{align}
    \dot p_i^{\rm acc} = \sum_x\dot p_i^{(x)},
\end{align}
where $x \in \qty{\text{kinetic, thermal, EM}}$.
As a precaution against artificial contributions from the density floor in vacuum regions, we integrate the kinetic and thermal terms, \eq{kinetic} and \eq{thermal}, only over grid cells on the horizon for which $\sigma < (2/3)\sigma_{\rm max}$, following the approach discussed in \citet{lowell_rapid_2024}. Our results are not sensitive to the choice of $\sigma$ cutoff.

The second term in \eq{force}, $\dot p_i^{\rm grav}$, is the rate of momentum exchange between the gas and the BH via gravitational interaction. The gravitational pull of the slower ejecta in aspherical CCSN explosions is often referred to as the ``gravitational tugboat mechanism" and has been found to dominate NS natal kicks in numerical simulations (for detailed discussions of the tugboat effect, see Sec.~3.3 of \citet{wongwathanarat_three-dimensional_2013} and \citet{Janka2017}). We implicitly calculate the gravitational momentum transfer to the BH by subtracting \eq{f-mom} from the time derivative of \eq{p-total}, 
\begin{align}
    \dot p_i^{\rm grav} = -\dot p_i - \dot p_i^{\rm acc}. \label{eq:f-grav}
\end{align}
Since the Kerr spacetime has no Killing vector corresponding to spatial translations, \eq{f-mom} is coordinate-dependent, and hence so is \eq{f-grav}. We thus caution that the separate kicks we compute for momentum accretion and gravity should be interpreted only as rough indicators of their relative importance. Our calculation for the total kick, however, is substantially more robust. For more detail on the relativistic conservation equation on which our calculation relies, see Appendix \ref{app:cons}, and for a rough consistency check, see Appendix \ref{app:consistency}. Note that we found Newtonian calculations for the BH acceleration due to gas gravity to be unreliable; see Appendix \ref{app:gravity} for a discussion and tests.

The third term in \eq{force} corresponds to the linear momentum change of the BH in the progenitor frame due to changes in its rest energy via the BZ mechanism. In practice, it prevents the BH from accelerating in the progenitor frame if the BZ power is emitted symmetrically in the BH frame. (We neglect the Lorentz factor of the BH in this calculation as the BH kick must be non-relativistic, $v^{\rm BH} \ll c$, by energy considerations alone.)

The BH momentum evolves due to changes in both velocity and gravitational mass,
\begin{align}
    \dot p_i^{\rm BH} = M_{\rm BH}\dot v_i^{\rm BH} + \dot M_{\rm BH}v_i^{\rm BH}.
\end{align}
The partial velocity contributions from momentum accretion through the horizon then obey
\begin{align}
    \dot v_i^{(x)}(t) = \frac{\dot p^{(x)}_i(t)}{M_{\rm BH}(t)} - \frac{\dot M(t) v_i^{(x)}(t)}{M_{\rm BH}(t)}.
\end{align}
We numerically integrate this equation to obtain solutions for $v_i^{(x)}$ at each time step, and their sum gives $v_i^{\rm acc}$. We then take the contribution from gravity to be
\begin{align}
    v_i^{\rm grav}(t) = v^{\rm BH}_i(t) - v_i^{\rm acc}(t).
\end{align}

\begin{deluxetable*}{crrrrrrrrrr}
\tablewidth{\textwidth}
\tablecaption{Simulation diagnostics, including peak diagnostic explosion energy, $E_{\rm ej}$, given by \eq{e-diag}; stellar mass accreted, $M_{\rm acc}$, and unbound, $M_{\rm ej}$, at the end of the simulation; fraction $f_{\rm mass}$ of progenitor mass accreted or unbound by the end of the simulation, $T_{\rm sim}$; peak momentum asymmetry parameter, $\max(\alpha_{\rm ej})$; value of $\alpha_{\rm ej}$ at $T_{\rm sim}$; average radial speed $\bar v_{\rm ej}$ of the ejecta at $T_{\rm sim}$, and; time averages (denoted with angle brackets) of the mass accretion rate $\dot M$, dimensionless magnetic flux $\phi_{\rm BH}$, and jet power $P_{\rm jet}$. Time averages are taken over $10^4\,r_g/c < t < 10^5\,r_g/c$ for the fiducial models and \texttt{a5*} and $10^3\,r_g/c < t < 10^4\,r_g/c$ for the DH models. For $E_{\rm ej}$, $M_{\rm ej}$, and $f_{\rm mass}$, we report values using both the Bernoulli and geodesic criteria, with the latter in parentheses. \label{tab:scalars}}
\tablehead{
    \colhead{Model} & \colhead{$\log_{10} \rm max(E_{\rm ej})$} & \colhead{$M_{\rm acc}$} & \colhead{$M_{\rm ej}$} & \colhead{$f_{\rm mass}$} & \colhead{$\max(\alpha_{\rm ej})$} & \colhead{$\alpha_{\rm ej}$}& \colhead{$\bar v_{\rm ej}$} & \colhead{$\langle\dot M\rangle$} & \colhead{$\langle\phi_{\rm BH}\rangle$} & \colhead{$\log_{\rm 10}\langle P_{\rm jet}\rangle$} \\
     \colhead{} & \colhead{$(\rm erg)$} & \colhead{$(M_\sun)$} & \colhead{$(M_\sun)$} & \colhead{$(\%)$} & \colhead{$(\%)$} & \colhead{$(\%)$} & $(0.01c)$ & $(M_\sun\,\rm s^{-1})$ & \colhead{} & \colhead{$(\rm erg\,s^{-1})$}
}
\startdata
\multicolumn{11}{c}{Fiducial} \\
\midrule 
\texttt{a1} & 52.0 (51.9) & 0.24 & 12.80 (8.24) & 93 (61) & 16.4 & 1.7 & 2.5 & 0.031 & 48.05 & 50.13 \\
\texttt{a5} & 52.5 (52.4) & 0.08 & 13.64 (11.00) & 98 (79) & 3.4 & 2.0 & 3.5 & 0.015 & 55.03 & 51.71 \\
\texttt{a8} & 52.7 (52.6) & 0.04 & 13.76 (11.23) & 99 (81) & 2.5 & 2.1 & 3.9 & 0.012 & 54.21 & 52.19 \\
\midrule 
\multicolumn{11}{c}{Modified Field} \\
\midrule
\texttt{a5*} & 52.6 (52.4) & 0.31 & 12.45 (8.79) & 91 (65) & 30.0 & 3.6 & 3.7 & 0.026 & 7.89 & 50.07 \\
\midrule
\multicolumn{11}{c}{Direct Horizon} \\
\midrule 
\texttt{dhBm} & 53.4 (53.2) & 59.03 & 36.94 (26.17) & 98 (87) & 6.3 & 4.9 & 7.0 & 23.454 & 2.63 & 52.18 \\
\texttt{dhBs} & 54.2 (54.0) & 28.94 & 68.07 (44.50) & 99 (75) & 4.0 & 2.1 & 10.4 & 10.874 & 24.68 & 53.40 \\
\texttt{dhBvs} & 54.5 (54.4) & 20.31 & 65.32 (32.57) & 89 (57) & 9.0 & 9.0 & 16.1 & 13.447 & 40.22 & 54.38
\enddata
\end{deluxetable*}

The proximity of the kick velocity given by \eq{v-kick} at $t = T_{\rm sim}$ to the ``true" asymptotic kick depends on the remaining amount of bound matter, i.e.~matter that has not yet been either accreted or ejected. If the BH is on track to accrete most of the remaining envelope for $t > T_{\rm sim}$, its final kick will likely be suppressed; if instead jets and disk winds are set to continue ejecting significant envelope mass, the kick may rise. To quantify how far each model is from its asymptotic state, we track the fraction of the stellar mass that has either been accreted or ejected,
\begin{align}
    f_{\rm mass} \equiv (M_{\rm acc} + M_{\rm ej})M_\star^{-1},
\end{align}
where the cumulative accreted mass
\begin{align}
    M_{\rm acc}(t) = \int_0^{t} \dot M \dd{t'},
\end{align}
and $M_{\rm ej}$ is the mass of the ejecta. We bracket $M_{\rm ej}$ using the geodesic criterion,
\begin{align}
    u_t < -1,
\end{align}
and the relativistic Bernoulli criterion,
\begin{align}
    \mathcal{B} \equiv -\qty(1 + \frac{\epsilon}{\rho} + \frac{p}{\rho} + \frac{b^2}{\rho})u_t - 1 > 0,
\end{align}
for identifying unbound fluid elements. The former is known to underestimate, and the latter to overestimate, the ejecta mass and velocity in GRMHD simulations \citep[e.g.,][]{Ka2015,Vi+2020}.

To account for uncertainty in the future evolution after $t = T_{\rm sim}$, we provide rough upper and lower bounds for the asymptotic kick. As a heuristic upper bound we simply assume the kick grows linearly with $f_{\rm mass}$, i.e.
\begin{align}
    v_{i}^{\rm BH,\,\inf,\,+} = \frac{v_i^{\rm BH}(T_{\rm sim})}{f_{\rm mass}(T_{\rm sim})}. \label{eq:v-high}
\end{align}

As a lower bound, we assume the BH proceeds to accrete all of the inertia and linear momentum carried by the bound matter at the end of the simulation (strictly speaking, the kick may rise in this case, but in practice we find the bound matter carries negligible net linear momentum compared to the ejecta). We thus provide two lower bounds corresponding to the two criteria for unbound matter,
\begin{align}
    v_i^{\rm BH,\,\inf} = \frac{-p_i^{\rm ej}(T_{\rm sim})}{M_{\rm BH}(T_{\rm sim}) + M_{\rm bound}(T_{\rm sim})}, \label{eq:v-low}
\end{align}
where $p_i^{\rm ej}$ is the momentum of the ejecta,
\begin{align}
    p_i^{\rm ej} = \int \tensor{T}{^t_i}\sqrt{-g}\dd{r}\dd{\theta}\dd{\phi}.\label{eq:p-ej}
\end{align}
The integral is restricted to either $\mathcal{B} > 0$ or $u_t < -1$.
See Appendix \ref{app:diag} for definitions of various diagnostic quantities we refer to in the text.
\begin{figure*}
    \centering 
    \includegraphics[width=\linewidth]{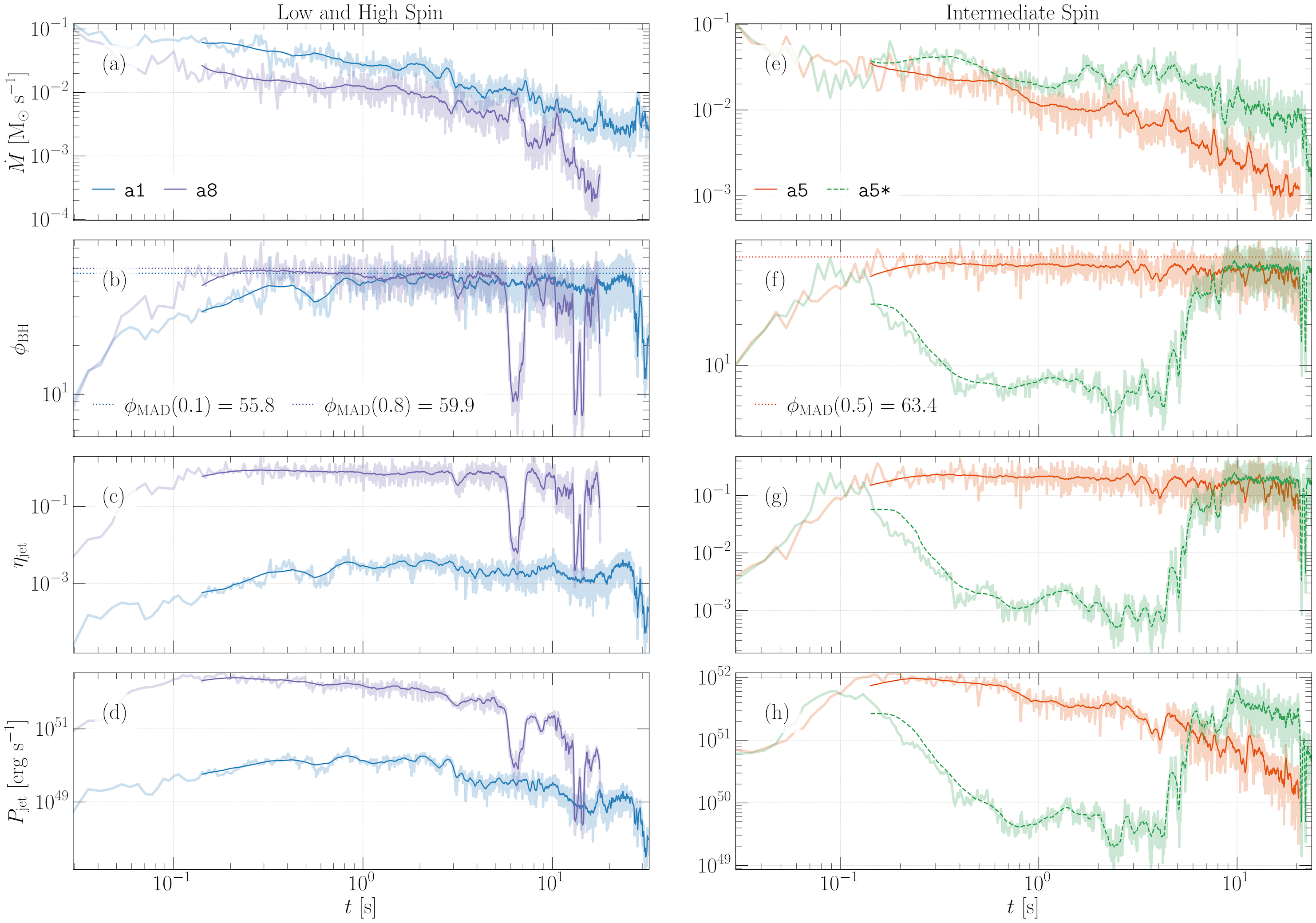}
    \caption{Time evolution of mass accretion rate $\dot M$, dimensionless flux parameter $\phi_{\rm BH}$, jet efficiency $\eta_{\rm jet}$, and total jet power $P_{\rm jet}$ for the low-mass collapsar models.  In the left panels we compare the low-spin \texttt{a1} ($a = 0.1$) and high-spin \texttt{a8} ($a = 0.8$) models, while in the right panels we compare the intermediate-spin ($a = 0.5$) models with fiducial (\texttt{a5}) and modified (\texttt{a5*}) progenitor magnetic field structure. The bold lines correspond to simple moving averages computed using a window of $3000\,r_g/c$. The dotted lines on the plots of $\phi_{\rm BH}$ correspond to the saturation value as a function of $a$ predicted by \eq{phi-sat}, i.e.~Eq.~(9) of \citet{narayan_jets_2022}.}
    \label{fig:diag}
\end{figure*}
\section{Collapsar Evolution and Asymmetry}\label{sec:accretion}
\begin{figure*}
    \centering
    \includegraphics[width=\linewidth]{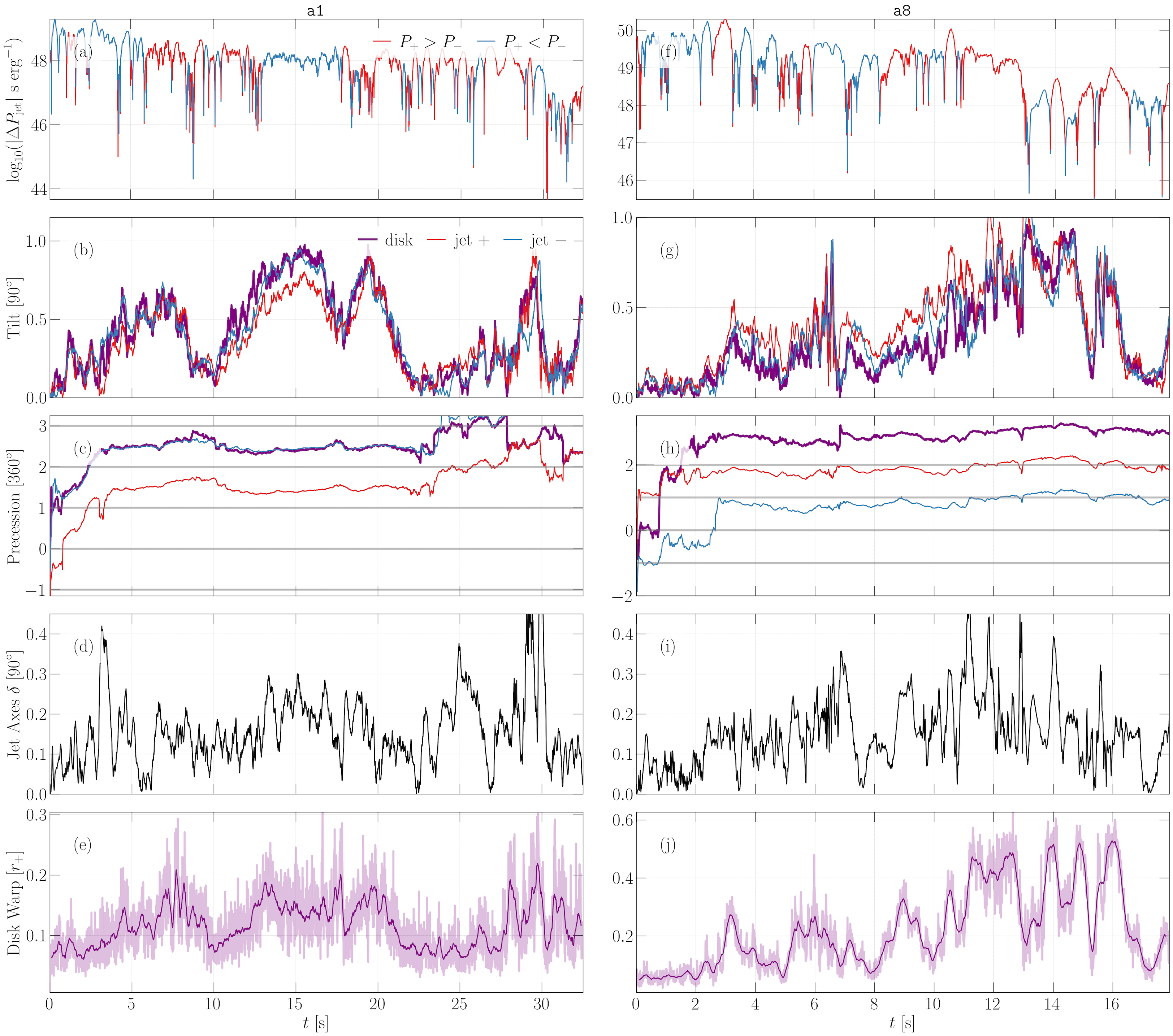}
    \caption{Fiducial models \texttt{a1} (lowest spin, $a = 0.1$; left) and \texttt{a8} (highest spin, $a = 0.8$; right): time evolution of the absolute value of the difference in jet power, $\abs{\Delta P_{\rm jet}} = \abs{P_+ - P_-}$ (red for $P_+ > P_-$ and blue otherwise; [a] and [f]), jet and disk nutation angles (``Tilt''; [b] and [g]), jet and disk cumulative azimuthal phases (``Precession''; [c] and [h]), separation angle between the jet axes (``Jet Axes $\delta$''; [d] and [i]), and RMS vertical displacement of the disk from its midplane on the horizon (``Disk Warp''; [e] and [j]). Gray horizontal lines in (c) and (h) demarcate integer multiples of $2\pi$. Decreasing (increasing) $\phi$ corresponds to (counter)clockwise precession about the $+z$-axis. The jet power difference and the bold line in the disk warp plots are simple moving averages with a window of $3000\,r_g/c$.}
    \label{fig:angles}
\end{figure*}
\begin{figure*} 
    \centering \includegraphics[width=\linewidth]{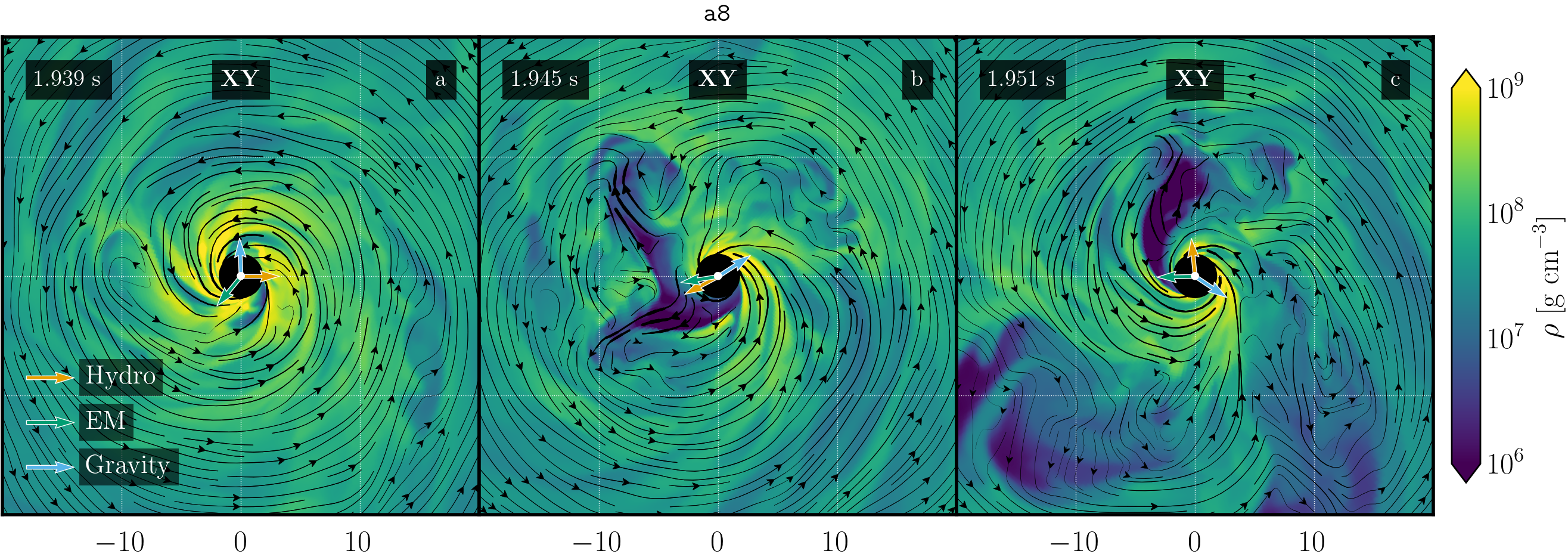}
    \caption{Series of closely spaced snapshots of the rest mass density $\rho$ in the rapidly spinning ($a = 0.8$) fiducial model \texttt{a8} through the equatorial ($xy$) plane in the innermost $20\, r_g$ at times $t = 1.939$ (a), $1.945$ (b), and $1.951\,\rm s$ (c). The boundary of the black ellipse corresponds to the Kerr outer horizon. A magnetic flux eruption partially ejects the disk (b), creating a non-axisymmetric low-density pocket extending over several $r_g$. We indicate the directions of the instantaneous force on the BH due to hydrodynamic (kinetic and thermal; ``Hydro'') stress,  Maxwell (``EM'') stress, and gravity projected onto each plane with arrows (orange, green, and blue, respectively). The hydrodynamic stress points towards lower density, while the gravitational force tends to point oppositely, toward higher density. The in-plane hydrodynamic and Maxwell stress vectors are mostly aligned during the initial eruption (b) but are misaligned before (a) and after (c). Black lines represent fluid velocity streamlines, and their thickness indicates the relative velocity magnitudes. The outflowing matter expelled by the eruption moves faster than the matter accreting at the horizon; see (b) and (c).}
    \label{fig:a8_eruption}
\end{figure*}
\begin{figure*}
    \centering
    \includegraphics[width=\linewidth]{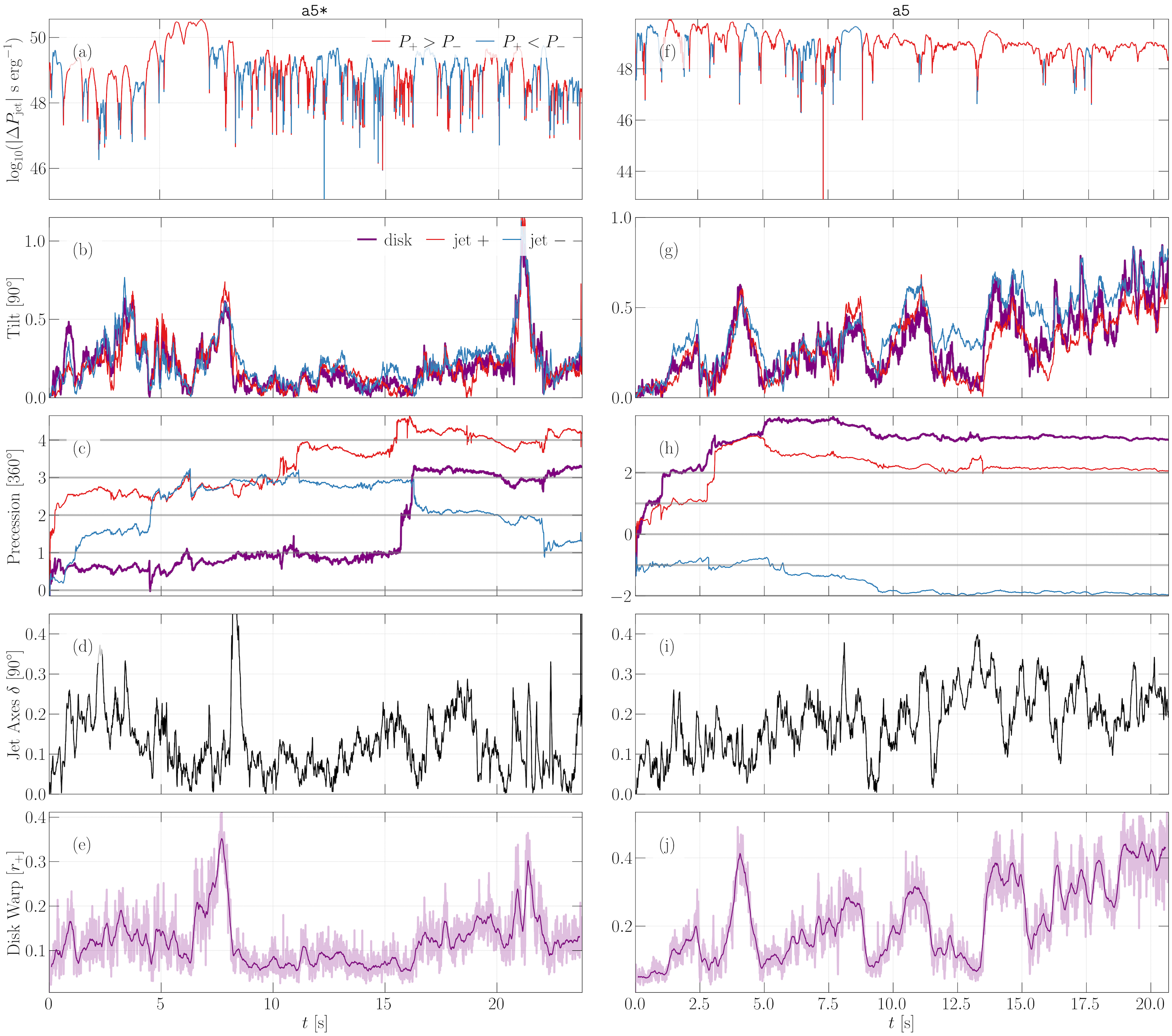}
    \caption{Same as \fig{angles}, but for the two intermediate-spin ($a=0.5$) models \texttt{a5} and \texttt{a5*}. Model \texttt{a5} uses the fiducial progenitor magnetic-field configuration, whereas \texttt{a5*} uses the modified radial magnetic-flux distribution described in Sec.~\ref{subsec:cf}.}
    \label{fig:angles_a5}
\end{figure*}

In this section, we discuss the overall behavior of our collapsar simulations and examine the asymmetries that power the BH kick. Table \ref{tab:scalars} summarizes several diagnostic quantities that characterize the simulation results.
Figure \ref{fig:diag} shows the time evolution of the mass accretion rate (\fig{diag}(a)),
\begin{align}
    \dot M = -\oint_{r_+} \rho u^r \sqrt{-g}\dd{\theta}\dd{\phi}, \label{eq:m-dot}
\end{align}
dimensionless magnetic flux threading the horizon (\fig{diag}(b)),
\begin{align}
    \phi_{\rm BH} = \frac{1}{2\sqrt{\dot M}}\oint_{r_+} \abs{B^r} \sqrt{-g}\dd{\theta}\dd{\phi},
\end{align}
where $B^r$ is the radial component of the coordinate-frame magnetic field,
\begin{align}
    B^r = u^t b^r - u^r b^t,
\end{align}
jet efficiency $\eta_{\rm jet}$ (\fig{diag}(c)), and jet power $P_{\rm jet}$ (\fig{diag}(d)) for the low-mass collapsar simulations \texttt{a1}, \texttt{a5}, \texttt{a8} (fiducial) and \texttt{a5*} (modified magnetic field). The left column compares the low and high spin models \texttt{a1} and \texttt{a8}, and the right column the two intermediate-spin models \texttt{a5} and \texttt{a5*}.

\subsection{\texttt{a1} and \texttt{a8}: Effects of BH Spin}

To highlight the role of BH spin on the collapsar dynamics, we first compare the lowest and highest BH spin simulations, $\texttt{a1}$ and $\texttt{a8}$. Following collapse, the infalling plasma initially undergoes near-free-fall motion toward the BH, with the gas beginning to circularize into a relativistic accretion disk within a few milliseconds. The mass-accretion rate falls from $\sim 0.1\,M_\odot\,{\rm s}^{-1}$ to
$\sim 0.03\,M_\odot\,{\rm s}^{-1}$ during the initial circularization phase, subsequently declining by another one and two orders of magnitude in models \texttt{a1} and \texttt{a8}, respectively; see \fig{diag}(a). 
The incoming gas carries a strong poloidal magnetic flux, which becomes anchored onto the BH horizon.  In both models, the disk becomes magnetically arrested within a few tenths of a second, with $\phi_{\rm BH}$ oscillating near its spin-dependent saturation value \citep{narayan_jets_2022}, 
\begin{align}
    \phi_{\rm MAD}(a) \simeq -20.2a^3 - 14.9a^2 + 34a + 52.6; \label{eq:phi-sat}
\end{align}
see \fig{diag}(b). 

The BH ergosphere twists the poloidal fields, launching bipolar BZ jets that entrain matter in the polar regions and drive shocks that propagate toward the stellar surface. For the low-spin model, after MAD onset the jets attain a time-averaged power $\langle P_{\rm jet}\rangle \sim 10^{50}\,\rm{erg\,s^{-1}}$ corresponding to a low efficiency $\eta_{\rm jet} \lesssim 1\%$. The high-spin BH launches much more robust jets at $\approx 80\%$ efficiency, with an average power of $\langle P_{\rm jet}\rangle \sim 10^{52}\,\rm erg\,s^{-1}$, roughly consistent with the expected quadratic dependence of the BZ power on the BH spin.  The lower-spin model produces a jet power closer to the characteristic beaming-corrected luminosities of GRB jets (e.g., \citealt{Wanderman&Piran10}), broadly consistent with previous arguments that collapsar BHs are likely slow-spinning \citep{gottlieb_collapsar_2023}.

At late times---beginning around $t\approx 6$ s in \texttt{a8} and $t\approx 14$ s in \texttt{a1}---the jets undergo brief disruptions associated with a sharp loss of poloidal magnetic flux from the horizon and a transition to sub-MAD accretion, admitting brief enhancement of the mass accretion rate. Apart from these transient disruptions, the secular decline in jet power largely follows the declining mass-accretion rate as progressively lower-density material falls in from larger radii. 

Initially tightly collimated and diametrically opposed, the jets widen, wobble, and bend as they propagate through the collapsing star and encounter non-axisymmetric plasma ram pressure, as illustrated in \fig{a1_rho_volume}. Thus, even when launched along the BH spin axis, they produce shocks that expand in various directions, resulting in a disordered envelope morphology. The breaking of reflection symmetry across the disk midplane allows the stochastic momentum transfer to the BH to develop substantial components both parallel and perpendicular to the initial spin axis.

These asymmetries are illustrated in the left column of \fig{angles}, which shows for model \texttt{a1} the time evolution of the difference in jet across the disk plane, the jet and disk tilt angles, and their cumulative azimuthal phases, as well as a measure of the vertical warping of the disk. To track jet power asymmetry when the disk is inclined, we measure the Poynting flux through hemispheres separated by the plane of the disk, given by its net angular momentum vector $\mathbf{J}_{\rm disk}$. We define the ``$+$" jet as the jet that propagates into the hemisphere that $\mathbf{J}_{\rm disk}$ points into; the ``$-$" jet propagates into the opposite hemisphere. See Appendix \ref{app:diag} for details of the calculation.

At early times $t \lesssim 6\,\rm{s}$ the ``$-$" jet, generally dominates by up to $10^{49}\,\rm{erg\,s^{-1}}$, before being overtaken by the ``$+$" jet. The disk and jets undergo strong nutation and precession, driven in part by interactions between the disk and bound material in the shocked cocoons that collimate the jets \citep{Gottlieb2022}. The relative weakness of the jets in the slowly spinning model makes them more susceptible to deflection by the gas ram pressure and disk winds than the rapidly spinning model, at least at early times. The two jet axes generally remain close to diametrically opposed, with separation angles typically $\lesssim 20^\circ$; see \fig{angles}(d) and \fig{angles}(i).  They are also approximately aligned with the inner-disk angular momentum because the orientation of the magnetic flux threading the horizon is set by the accretion flow \citep{Chan2026}. The greatest misalignment between the disk and BH spin axis occurs at late times ($t\approx15\,\rm s$ and $20\,\rm s$) when the disk angular momentum lies nearly in the $z = 0$ plane ($\theta_{\rm disk} \approx 90^\circ$) prior to re-aligning with the BH spin. 

Although the jets undergo substantial tilting, they complete only a few precession cycles over the simulation (see the cumulative azimuthal phases in \fig{angles}[c]). Their directional asymmetries therefore persist long enough to be imprinted on the large-scale ejecta and contribute coherently to the BH kick. These relatively steady collapsar jets contrast with the ``jittering'' jets driven solely by stochastic angular momentum accretion in more spherical stellar collapse \citep{Oded2011,Antoni&Quataert23}.

The contrast between \texttt{a1} and \texttt{a8} illustrates a key role of spin: stronger jets can promote early-time disk alignment, but this does not eliminate non-axisymmetric evolution at later times. In contrast to \texttt{a1}, the inner disk in model \texttt{a8} remains generally more aligned with the BH spin during the first $\sim10$ s; see \fig{angles}(g), except for the brief sub-MAD phase that occurs at $t \approx 6\,\rm s$. While the Bardeen-Petterson effect \citep{BP_effect} is known to cause inner disk alignment via Lense--Thirring torques, previous simulations indicate that EM torque by jets \citep[``magneto-spin alignment";][]{magneto-spin,Polko2017} is the dominant mechanism for aligning radiatively inefficient, geometrically thick disks, and can enforce alignment to larger radii. In tilted MAD simulations, \citet{Chatterjee_magneto-spin} found that the degree of disk alignment increases with jet power, which depends on the BH spin and the magnetic flux threading the horizon. For \texttt{a8}, the powerful jets may help maintain the inner disk in relatively close alignment despite stochastic torques, but after $\sim 10$s the disk becomes unstable to extreme wobbling (tilting up to $90^\circ$) as the jet power deteriorates and the jets are no longer launched along the BH spin axis. The other effect of the jet power may be to simply decrease the amount of turbulent plasma impacting the disk from above, relieving the need for stabilization in the first place.

Aside from the impact of the jet on large scales, the accretion flow can be strongly non-axisymmetric on small scales close to the horizon.  Magnetically arrested disks exhibit recurrent magnetic flux eruptions, in which accumulated magnetic flux is episodically expelled from the BH magnetosphere through reconnection, driving magnetized outflows into the surrounding disk \citep{ripperda_black_2022}. Such an eruption is illustrated in \fig{a8_eruption} for model \texttt{a8}, showing that it partially ejects the inner parts of the disk over a limited azimuthal range and leads to a non-axisymmetric momentum flux on the horizon (Sec.~\ref{subsec:kick}). 

The disks are also narrowly constricted on the horizon and exhibit non-axisymmetric vertical displacement (``warping'') of the midplane. We analyze the disk warp in a rotated coordinate system ($\tilde\theta$, $\tilde\phi$) whose north pole is aligned with the disk angular momentum, $\vb{J}_{\rm disk}$. We define the ``warp parameter" to be the root mean square (RMS) of the displacement of the local disk midplane above the mean disk plane, i.e.~the plane intersecting the origin whose normal points along $\vb{J}_{\rm disk}$, and we take the local height of the disk at each angle $\tilde\phi$ around its axis to be a density-weighted average in $\tilde\theta$. In Appendix \ref{app:warp} we show power spectra of the azimuthal modes of disk warp for simulations \texttt{a1} and \texttt{a8}. The disk around the higher-spin BH achieves greater maximum warping than its counterpart in \texttt{a1}, as shown in \fig{angles}(e),(j), especially when the disk is highly tilted; this is expected if the warp is driven primarily by Lense-Thirring precession, for which the frequency is proportional to $a$. The warping may contribute to the asymmetric momentum flux into the BH, but we leave detailed analysis of its origins and connection with global asymmetries to future study.

Thus, increasing BH spin strengthens the jets and promotes early-time axial symmetry, while late-time disk wobbling, jet deflection, and magnetic-flux eruptions preserve significant non-axisymmetry. This competition between jet power and asymmetry is central to the spin dependence of the BH kick discussed in Sec.~\ref{subsec:properties}.

\subsection{\texttt{a5} and \texttt{a5*}: Effects of Progenitor Magnetic Flux}

Figure \ref{fig:angles_a5} shows the same quantities as in \fig{angles}, but for the two intermediate-spin ($a = 0.5$) simulations.  Model \texttt{a5}, which adopts the standard progenitor magnetic-field configuration, evolves qualitatively similarly to \texttt{a1} and \texttt{a8}. The disk becomes magnetically arrested by $t\approx 0.2\,\rm s$ and launches a pair of jets with a peak power $P_{\rm jet}\approx 10^{52}\,\rm erg\,s^{-1}$, intermediate between those of \texttt{a1} and \texttt{a8}.

In contrast, \texttt{a5*}, which has the same BH spin and progenitor structure but a modified radial distribution of progenitor magnetic flux, undergoes a qualitatively different magnetic evolution. Poloidal flux initially advected to the BH from the innermost, highly magnetized region ($\lesssim 200\,r_g$) drives the disk into a brief MAD state by $t\approx0.1\,\rm{s}$; see \fig{diag}(f). Unlike \texttt{a5}, however, this initial MAD state is short-lived: by $t\approx1,\mathrm{s}$, the dimensionless magnetic flux falls to $\phi_{\rm BH}\lesssim10$, well below the MAD regime. The resulting decline in jet power occurs despite a comparatively steady mass-accretion rate, indicating that the reduction in jet power is primarily associated with the reduced coherent poloidal flux supplied from radii $\gtrsim200\,r_g$.

Beginning around $t\approx4\,\rm{s}$, the horizon flux begins to recover, eventually returning the disk to a MAD state and reaching the expected saturation level by $t\approx9\,\rm{s}$. The late-time recovery may reflect direct accumulation of the weaker coherent poloidal flux advected inward from larger radii, dynamo-driven amplification and reorganization of the disk magnetic field, or a combination of these processes; in the present work we do not distinguish between these possibilities. Near $t\approx 21$ and $22\,\rm{s}$, the jets undergo two brief disruption episodes similar to those seen in \texttt{a8}, both coinciding with extreme tilting of the inner disk.

The changing magnetic state also alters the jet and disk geometry (Fig.~\ref{fig:angles_a5}). During the early sub-MAD phase, $t\lesssim5\,\rm{s}$, the weaker jets exhibit larger and more variable tilt, together with moderate precession. This behavior is consistent with weaker magneto-spin alignment: as the jet power declines with the loss of horizon-threading magnetic flux, the jets become less effective at keeping the disk in alignment with the BH spin. As the horizon flux recovers and the system transitions back to the MAD state for $5\lesssim t\lesssim10\,\rm{s}$, the disk tilt becomes more strongly suppressed, except during the jet disruption episode at $t\approx22\,\rm{s}$. The jet power asymmetry peaks at $P_{\rm jet}\gtrsim10^{50}\,\rm{erg\,s^{-1}}$ during the transition to MAD as the flux on the BH increases while the asymmetry is still prominent. We also observe greater jet-axis misalignment during the sub-MAD phase than during the MAD phase. Thus, the radial distribution of magnetic flux affects not only the jet energetics but also the duration and geometry of the non-axisymmetric outflow.

The comparison between \texttt{a5} and \texttt{a5*} demonstrates that the radial supply and accumulation history of coherent poloidal magnetic flux can qualitatively alter both the energetics and geometry of the outflows, even at fixed BH spin and progenitor structure. In \texttt{a5*}, the extended sub-MAD phase is associated with weaker but more asymmetric outflows, leading to a substantially larger net ejecta momentum asymmetry than in the fiducial \texttt{a5} model.

\subsection{Direct-Horizon Models}

The DH models exhibit analogous behavior but in a quantitatively different regime of much higher accretion rates and jet powers. Model \texttt{dhBm} remains sub-MAD and develops asymmetric disk winds together with a one-sided jet, whereas the more strongly magnetized models transition to sustained MAD accretion and launch progressively more powerful jets. As in the fiducial models, these differences in magnetic-flux evolution and jet geometry generate markedly different large-scale outflow asymmetries.

The differences in jet power, magnetic-flux evolution, and outflow geometry described above naturally lead to different explosion asymmetries and, ultimately, different BH natal kicks. We therefore postpone discussion of the global ejecta properties, explosion morphology, and resulting recoil until the next section, where we relate them directly to the kick.

\begin{figure*} 
    \centering \includegraphics[width=\linewidth]{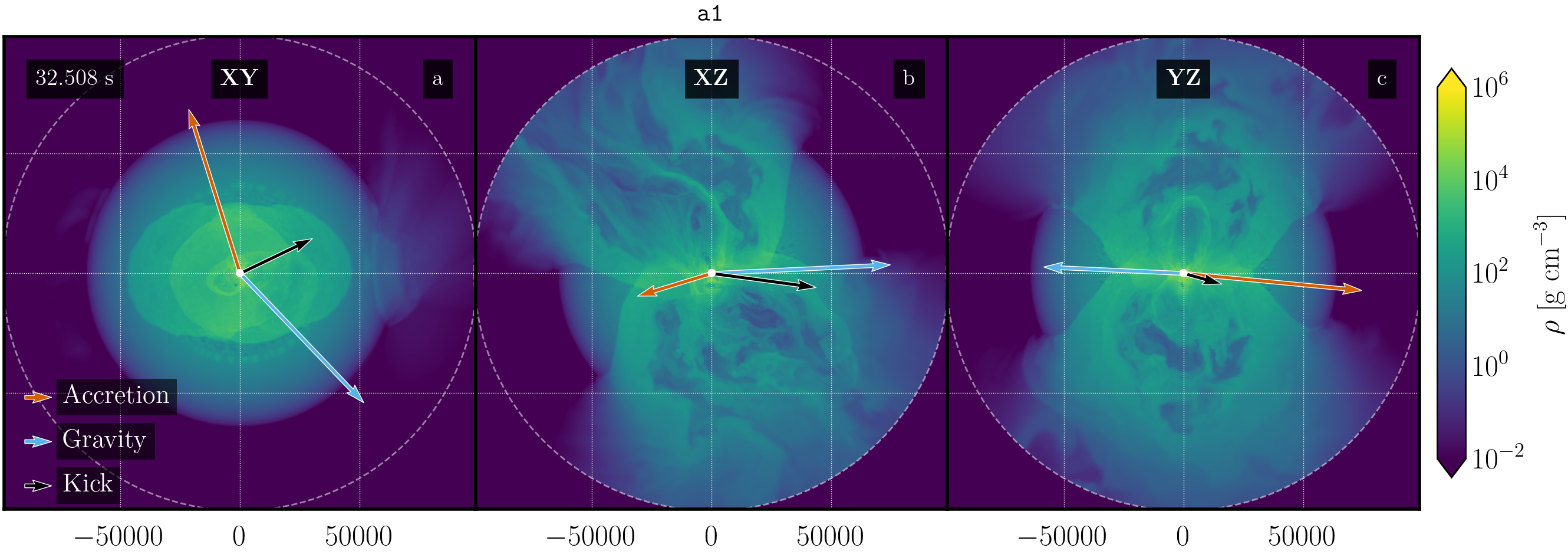} 
    \includegraphics[width=\linewidth]{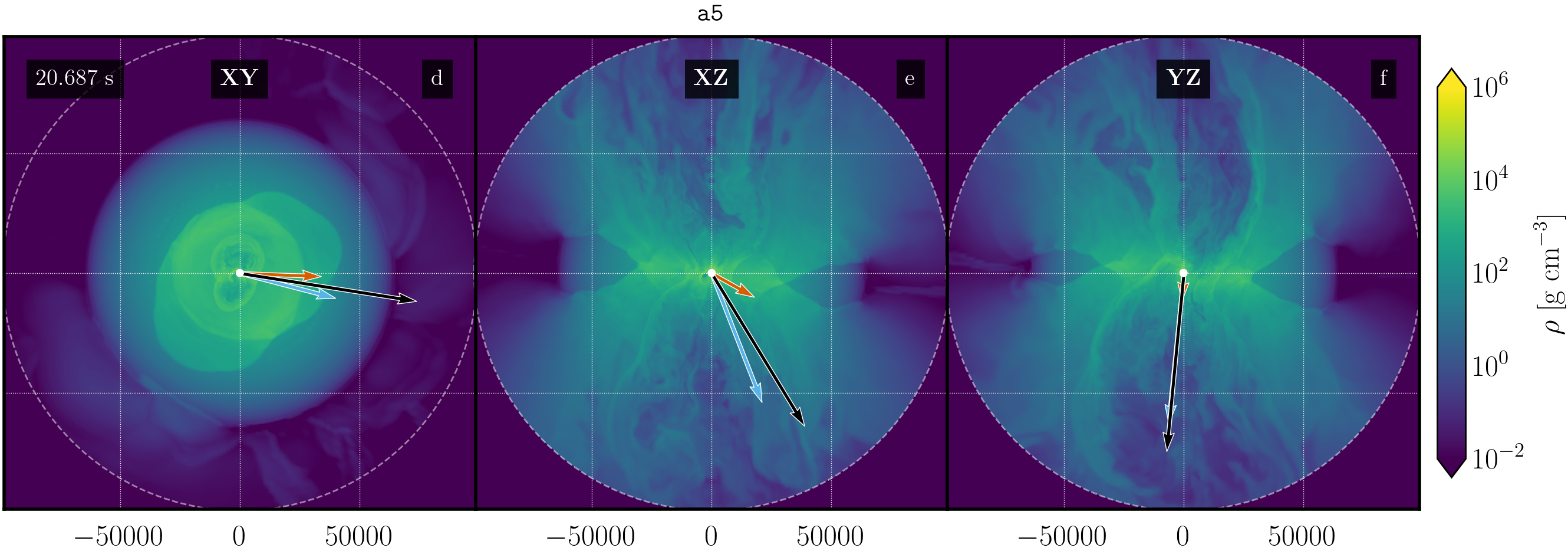} \includegraphics[width=\linewidth]{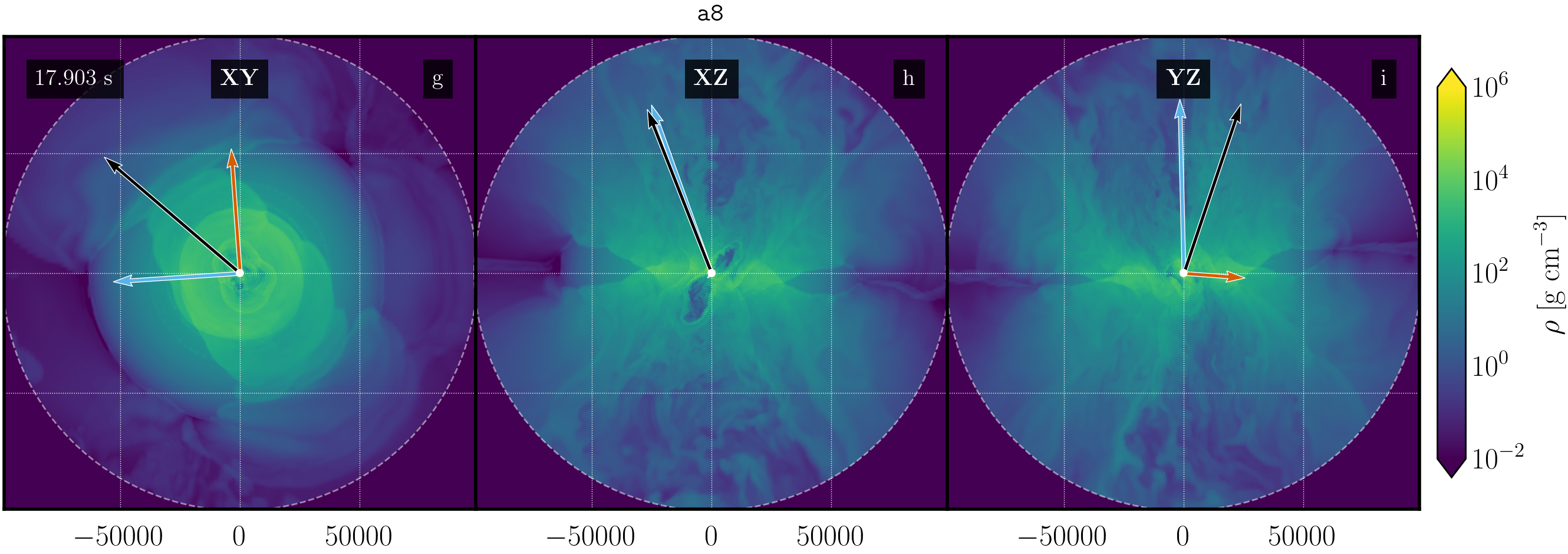}
    \caption{Equatorial ($xy$) and vertical ($xz$ and $yz$) slices of the rest-mass density $\rho$ in Cartesian KS coordinates at the end of the fiducial simulations \texttt{a1}, \texttt{a5}, and \texttt{a8}. Lengths are shown in $r_g$, and the gray dashed circles mark the outer boundary of the grid. Arrows show the final kick-velocity contributions from momentum accretion, $v_i^{\rm acc}$ (red), and gravity, $v_i^{\rm grav}$ (blue), together with the total kick $v_i^{\rm BH}$ (black), projected onto each plane. Arrow lengths show the relative magnitudes of the projected vectors; the overall scaling is arbitrary. For \texttt{a1}, the accretion and gravity kicks point roughly oppositely. The equatorial map (a) shows shocks from tilted jets interacting with an oblate-spheroidal accretion shock, while panels (b) and (c) show that the gravity kick is predominantly equatorial despite the ejecta appearing concentrated toward the poles. For \texttt{a5} ([d]--[f]) and \texttt{a8}
([g]--[i]), the total kick is more nearly aligned or anti-aligned with the BH spin axis, and the gravitational contribution more clearly dominates but is not opposite to the accretion contribution.}
    \label{fig:fiducial_explosion}
\end{figure*}

\begin{figure*} 
    \centering \includegraphics[width=\linewidth]{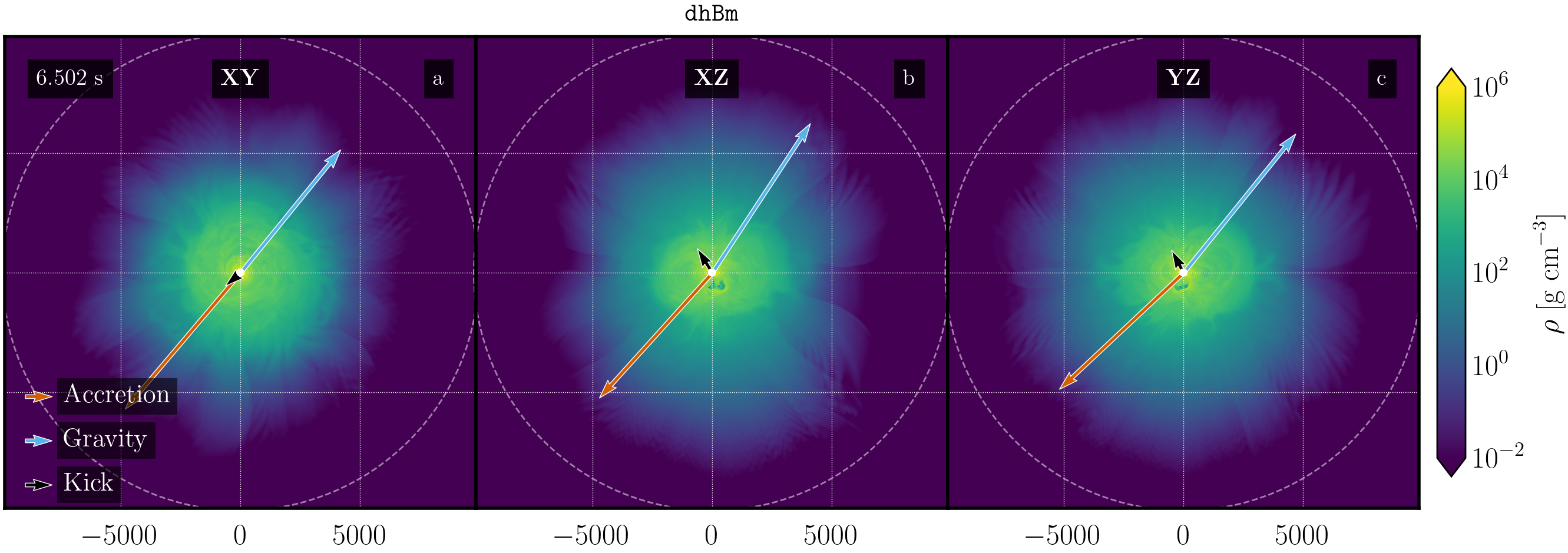} \includegraphics[width=\linewidth]{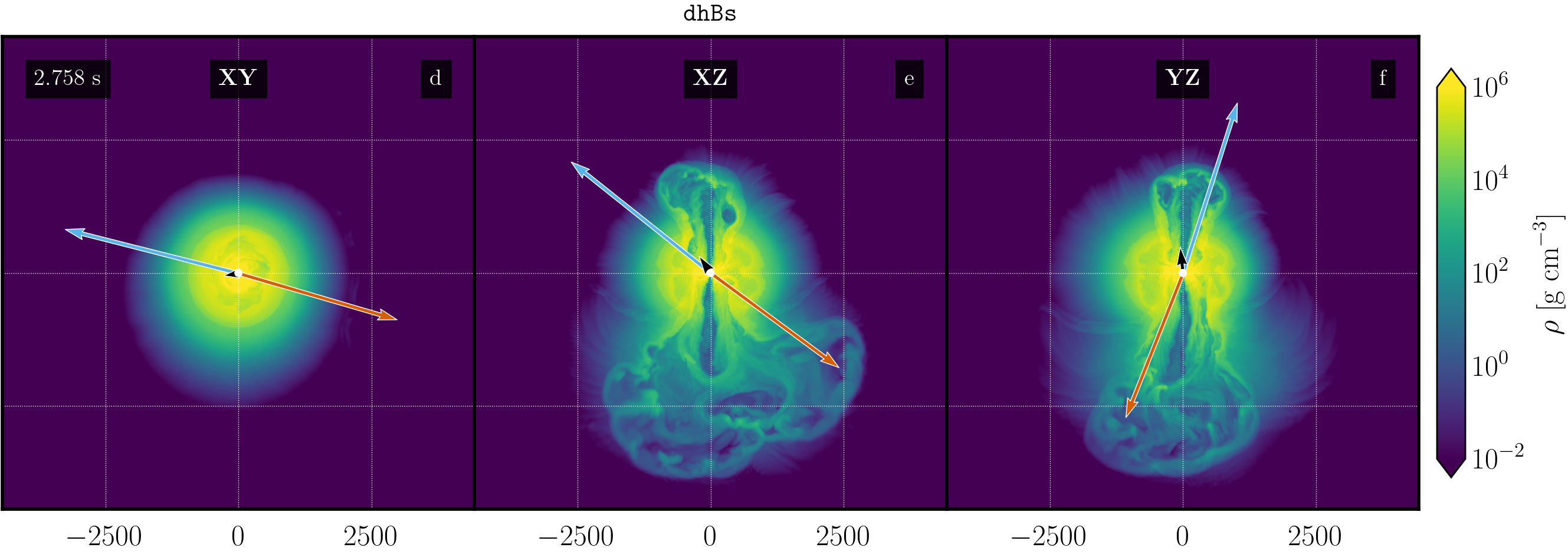} \includegraphics[width=\linewidth]{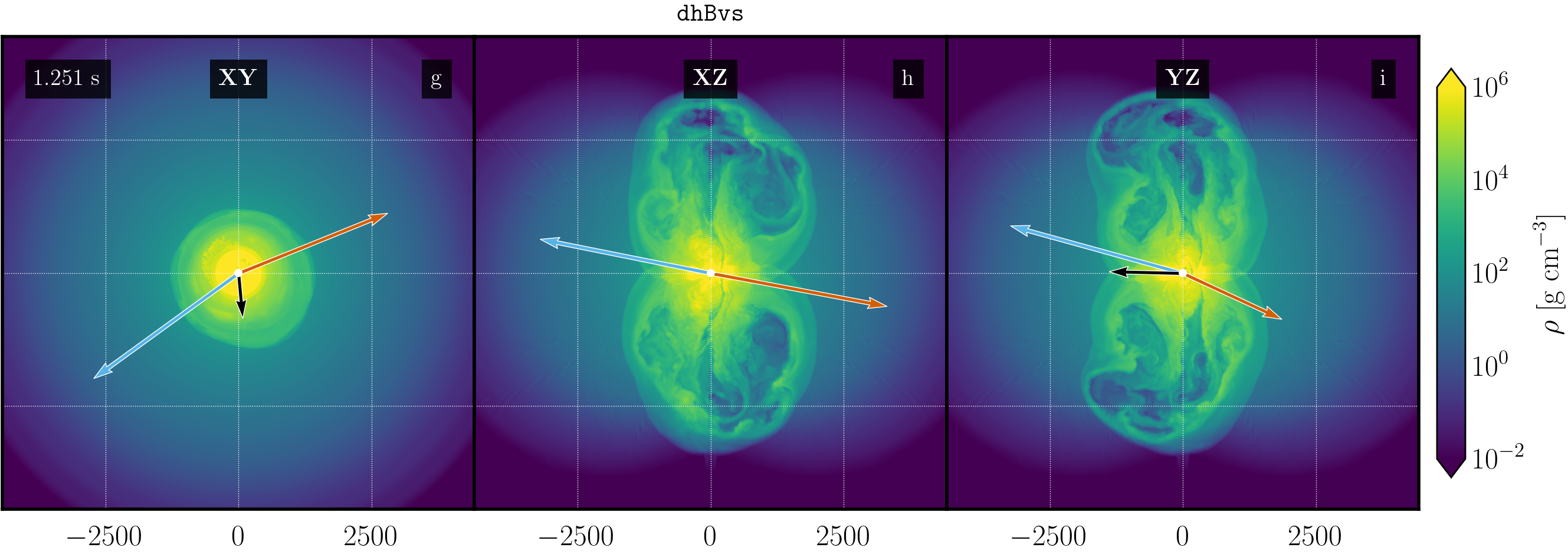}
\caption{Same as \fig{fiducial_explosion} but at the end of the direct-horizon (high-mass progenitor) models \texttt{dhBm} (top), \texttt{dhBs} (middle), and \texttt{dhBvs} (bottom). For the moderately magnetized model \texttt{dhBm}, disk winds aspherically expel a large fraction of the envelope, and a one-sided jet kicks the BH toward the northern hemisphere ([b] and [c]). The gravitational tugboat kick dominates and points in the opposite direction to the accretion kick. For the strongly magnetized model \texttt{dhBs}, highly asymmetric jets create an enormous explosion asymmetry across the $z=0$ plane ([e] and [f]), imparting a large $+z$ kick dominated by the gravity of
the slower ejecta. Remarkably, for the very strongly magnetized model
\texttt{dhBvs}, the jet outflows are roughly symmetric across the $z=0$ plane
([h] and [i]), but powerful magnetic-flux eruptions and deflected jets drive an equatorial kick away from a low-density cavity (g).}
    \label{fig:Bm_Bs_Bvs_explosion}
\end{figure*}

\section{Kick Results}\label{sec:results}
\subsection{Explosion Properties and Final Kicks}

Asymmetric explosions powered by the jets and disk produce natal kicks of hundreds of $\rm{km\,s^{-1}}$ in our simulations. The most strongly magnetized DH collapsar yields a kick of thousands of $\rm{km\,s^{-1}}$. To characterize the explosions responsible for these recoils, we first quantify the ejecta energy and momentum asymmetry before relating them to the resulting kick magnitudes
and orientations.

The amount of matter that ultimately escapes the progenitor is uncertain at finite simulation time, so we bracket the explosion properties using two criteria for unbound material: the relativistic Bernoulli criterion, $\mathcal{B}>0$, and the geodesic criterion, $u_t<-1$. The Bernoulli criterion is less restrictive and generally yields larger ejecta masses and energies, whereas the geodesic criterion provides a more conservative estimate. We therefore report diagnostics based on both criteria where applicable, adopting the Bernoulli values as our fiducial measurements
unless otherwise noted.

We define a diagnostic explosion energy that estimates the asymptotic
energy of the outflowing ejecta,
\begin{align}
    E_{\rm ej} \equiv
    \int D\mathcal{B}\sqrt{-g}\,
    \dd{r}\dd{\theta}\dd{\phi},
    \label{eq:e-diag}
\end{align}
where the integral is restricted to outward-moving fluid elements satisfying
the chosen unbound criterion.

We likewise quantify the anisotropy of the explosion using the momentum asymmetry parameter introduced by \citet{Janka2017},
\begin{align}
    \alpha_{\rm ej} \equiv
    \frac{p_{\rm ej}}
    {\int \sqrt{T^{ti}\tensor{T}{^t_i}}\sqrt{-g}\,
    \dd{r}\dd{\theta}\dd{\phi}},
    \label{eq:alpha-ej}
\end{align}
where the numerator is given by
\begin{align}
    p_{\rm ej}^2 =
    \sum_{i\in\{x,y,z\}}
    \left(p^{\rm ej}_{i}\right)^2.
\end{align}
The same unbound-matter criterion is applied consistently to the numerator and denominator. This quantity measures the magnitude of the net ejecta momentum relative to the scalar sum of the momentum carried in all directions, and therefore vanishes for a perfectly momentum-symmetric explosion. For the time evolution of the mass fraction processed by accretion or ejection, $f_{\rm mass}$, explosion energies, and asymmetry parameters, see Appendix~\ref{app:explosion}.

Each of our fiducial collapsars unbinds $>90\%$ ($>60\%$) of the stellar envelope according to the Bernoulli (geodesic) criterion by the end of the simulations, with small cumulative accreted masses $M_{\rm acc}\approx 0.24\,M_\odot$, $0.08\,M_\odot$, and $0.04\,M_\odot$ for \texttt{a1}, \texttt{a5}, and \texttt{a8}, respectively. The more powerful jets in the higher-spin models unbind the envelope more rapidly and leave less material available for accretion. In particular, \texttt{a8} has accreted or unbound 99\% of the envelope within $17.9\,\rm{s}$, whereas \texttt{a1} reaches 93\% only after $27.6\,\rm{s}$. Using the Bernoulli criterion, the peak explosion energy of \texttt{a1}, $\max E_{\rm ej}\approx10^{52}\,\rm erg$, is comparable to the characteristic kinetic energies of GRB-associated supernovae \citep{Mazzali14}, whereas the higher-spin models produce substantially more energetic explosions, with $\max E_{\rm ej}\sim3$--$5\times10^{52}\,\rm erg$. The explosion energy increases with spin across the fiducial models, broadly
tracking the jet power.

\begin{deluxetable}{crr|rr|rr}[b!]
    \tablecaption{Magnitude $v^{\rm BH}$ (rounded to the nearest $\rm km\,s^{-1}$) and angle $\theta^{\rm BH}$ (degrees) with respect to the BH spin axis of the BH kick velocity at $t = T_{\rm sim}$ and conservative estimates for their asymptotic values (denoted with the subscript ``$\inf$") using the Bernoulli (``Be") and geodesic (``geo" ) criteria.\label{tab:kick}}
    \tablehead{\colhead{Model} & \colhead{$v^{\rm BH}$} & \colhead{$\theta^{\rm BH}$} & \colhead{$v^{\rm BH}_{\inf,\,
    \rm Be}$} & \colhead{$\theta^{\rm BH}_{\inf,\,\rm Be}$} & \colhead{$v^{\rm BH}_{\inf,\,\rm geo}$} & \colhead{$\theta^{\rm BH}_{\inf,\,\rm geo}$}}
    \startdata
    \multicolumn{7}{c}{Fiducial} \\
    \midrule
    \texttt{a1} & 304 & 97  & 214 & 100 & 136 & 96  \\
    \texttt{a5} & 445 & 148 & 417 & 149 & 255 & 153 \\
    \texttt{a8} & 516 & 27  & 497 & 27  & 342 & 30  \\
    \midrule 
    \multicolumn{7}{c}{Modified Field} \\
    \midrule 
    \texttt{a5*} & 750 & 166 & 608 & 165 & 370 & 157 \\
    \midrule 
    \multicolumn{7}{c}{Direct Horizon} \\
    \midrule 
    \texttt{dhBm}  & 346 & 39 & 348  & 37 & 309  & 44 \\
    \texttt{dhBs}  & 569 & 35 & 530  & 35 & 354  & 49 \\
    \texttt{dhBvs} & 3691 & 89 & 3342 & 89 & 1856 & 89
    \enddata
\end{deluxetable}

\begin{deluxetable*}{crrr|rrr|rrr|rrrrr}
    \tablewidth{\textwidth}
    \tablecaption{Cartesian KS components of the kick velocity contributions due to momentum accretion, $v_i^{\rm acc}$, and gravity, $v_i^{\rm grav}$; components of the net kick velocity $v_i^{\rm BH}$; angle $\smash\theta^{\rm hyd}_{\rm EM}$ between the kick due to hydrodynamic and EM stress; angle $\smash\theta^{\rm grav}_{\rm acc}$ between $v_i^{\rm acc}$ and $v_i^{\rm grav}$; angle $\smash\theta^{\rm BH}_{\rm acc}$ between $v_i^{\rm acc}$ and $v_i^{\rm BH}$; and angle $\smash\theta^{\rm BH}_{\rm grav}$ between $v_i^{\rm grav}$ and $v_i^{\rm BH}$ at the end of the simulations. Velocities are shown in $\rm km\,s^{-1}$ and angles are shown in degrees.\label{tab:kick-xyz}}
    \tablehead{
    \colhead{Model} & \multicolumn{3}{c}{$v^{\rm acc}_i$} & \multicolumn{3}{c}{$v^{\rm grav}_i$} &
    \multicolumn{3}{c}{$v^{\rm BH}_i$} & $\theta^{\rm hyd}_{\rm EM}$ & $\theta^{\rm grav}_{\rm acc}$ & $\theta^{\rm BH}_{\rm acc}$ & $\theta^{\rm BH}_{\rm grav}$ \\ & \colhead{$x$} & \colhead{$y$} & \colhead{$z$} & \colhead{$x$} & \colhead{$y$} & \colhead{$z$} & \colhead{$x$} & \colhead{$y$} & \colhead{$z$} & & &
    }
    \startdata 
    \multicolumn{14}{c}{Fiducial} \\
    \midrule
    \texttt{a1} & $-193$ & $617$ & $-59$ & $465$ & $-486$ & $22$ & $272$ & $131$ & $-37$ & 55 & 154 & 81 & 72 \\
    \texttt{a5} & $106$ & $-4$ & $-59$ & $125$ & $-32$ & $-320$ & $231$ & $-36$ & $-379$ & 104 & 40 & 30 & 10 \\
    \texttt{a8} & $-12$ & $165$ & $-12$ & $-168$ & $-11$ & $470$ & $-180$ & $154$ & $458$ & 45 & 94 & 75 & 19 \\
    \midrule
    \multicolumn{14}{c}{Modified Field} \\
    \midrule
    \texttt{a5*} & $-337$ & $-381$ & $244$ & $180$ & $460$ & $-973$ & $-157$ & $79$ & $-729$ & 46 & 140 & 112 & 29 \\
    \midrule
    \multicolumn{14}{c}{Direct Horizon} \\
    \midrule 
    \texttt{dhBm} & $-1243$ & $-1466$ & $-1379$ & $1078$ & $1323$ & $1648$ & $-164$ & $-143$ & $269$ & 82 & 172 & 87 & 85 \\
    \texttt{dhBs} & $3442$ & $-1005$ & $-2534$ & $-3766$ & $956$ & $2998$ & $-324$ & $-50$ & $464$ & 122 & 177 & 154 & 23 \\
    \texttt{dhBvs} & $12195$ & $4917$ & $-2282$ & $-11854$ & $-8591$ & $2369$ & $341$ & $-3675$ & $87$ & 77 & 166 & 107 & 60 \\    
    \enddata
\end{deluxetable*}

Model \texttt{a1} achieves a maximum ejecta asymmetry of $\sim16.4\%$, whereas \texttt{a5} and \texttt{a8} reach only $3.4\%$ and $2.5\%$, respectively. Among the four low-mass models, \texttt{a5*} exhibits the largest asymmetry, reaching $\max\alpha_{\rm ej}=30\%$, associated with its weak, intermittent jets during the prolonged sub-MAD phase. This contrast shows that the large-scale ejecta contribution to the BH kick depends on the net ejecta momentum rather than on the total explosion energy or jet power alone: a sufficiently large ejecta asymmetry can produce a large recoil even when the jets are weaker and more intermittent.

Table~\ref{tab:kick} summarizes the inferred kick magnitudes and orientations. For all models except ``very strongly" magnetized DH collapsar \texttt{dhBvs}, the kicks range from a few to several hundreds of $\rm{km\,s^{-1}}$. The kick orientation depends sensitively on the explosion geometry, ranging from nearly polar to nearly equatorial.  The modified-field model \texttt{a5*} produces the largest kick among the low-mass collapsars despite its weaker average jet power, whereas the most strongly magnetized DH model (\texttt{dhBvs}) reaches nearly $4000\,{\rm km\,s^{-1}}$ owing to its exceptionally asymmetric outflow.

The origin of these different kick magnitudes is evident in the global explosion morphology. Figure \ref{fig:fiducial_explosion} shows the final rest-mass-density morphology of the three fiducial models. In \texttt{a1}, the equatorial slice (\fig{fiducial_explosion}[a]) reveals asymmetric outflows and anisotropic fallback, while the vertical slices (\fig{fiducial_explosion}[b],[c]) show an oblique, strongly asymmetric bipolar explosion across the equatorial plane. In \texttt{a5} and \texttt{a8}, the ejecta are more strongly concentrated toward the BH spin axis, reflecting their more powerful and initially better-aligned jets. Nevertheless, both models retain appreciable non-axisymmetry in the equatorial outflows and fallback (\fig{fiducial_explosion}[d],[g]), as well as in the relative propagation of the two jet-driven shocks (\fig{fiducial_explosion}[e],[f],[h],[i]). The large cavities within the roughly paraboloid-shaped outflows trace the history of jet wobbling and intermittency. These global asymmetries generate the large-scale momentum imbalance and gravitational tugboat contribution, which we analyze quantitatively in Sec.~\ref{subsec:mechanisms} and Sec.~\ref{subsec:evolution}.

The low-mass models therefore illustrate two distinct controls on the eventual BH kick. Increasing BH spin generally raises the jet power and explosion energy, but can also produce a more nearly bipolar explosion with a smaller fractional ejecta asymmetry. Conversely, the weaker and more intermittent jets of \texttt{a5*} produce a much larger momentum asymmetry. The resulting kick therefore need not vary monotonically with either BH spin or explosion energy.

The DH models exhibit the same qualitative trends as the fiducial collapsars, but at much higher accretion rates and jet powers. The moderately magnetized model \texttt{dhBm} remains sub-MAD ($\langle\phi_{\rm BH}\rangle\approx3$), producing asymmetric disk winds and a one-sided jet, whereas the more strongly magnetized model \texttt{dhBs} launches sustained jets at $\langle\phi_{\rm BH}\rangle\approx25$. Model \texttt{dhBvs} has the strongest magnetic fields and yields the most powerful jets, sustained by MAD-level flux $\langle\phi_{\rm BH}\rangle\approx40$, that readily break out of the star.

The corresponding explosion morphologies for the DH models are shown in \fig{Bm_Bs_Bvs_explosion}. The equatorial maps (\fig{Bm_Bs_Bvs_explosion}[a],[d],[g]) show magnetized winds expelling the outer layers of the progenitors. In \texttt{dhBm}, the vertical slices (\fig{Bm_Bs_Bvs_explosion}[b],[c]) reveal a one-sided jet that dominates the outflow. In \texttt{dhBs}, both jets ultimately escape the envelope, but the southern jet breaks out first, producing a pronounced explosion asymmetry across the equatorial plane (\fig{Bm_Bs_Bvs_explosion}[e],[f]). In contrast, the jet outflows in the very strongly magnetized model \texttt{dhBvs} are remarkably symmetric across the $z=0$ plane, but powerful one-sided magnetic flux eruptions and deflection of both jets toward the $+y$-direction create a strongly lopsided equatorial explosion; see \fig{Bm_Bs_Bvs_explosion}(g).

These results show that the BH recoil is determined not simply by the explosion energy or jet power, but by the degree and persistence of the large-scale ejecta momentum asymmetry. In the following subsection we examine how linear momentum is transferred to the BH through gravitational forces and horizon stresses.

\subsection{Kick Mechanisms}\label{subsec:mechanisms}

The global explosion morphologies discussed above show how large-scale asymmetries in the ejecta determine the magnitude and direction of the BH recoil. We now examine how this momentum is transferred to the BH by decomposing the kick into contributions from momentum accretion through the horizon and from the gravitational interaction with the asymmetric ejecta. Table~\ref{tab:kick-xyz} summarizes these contributions.

We find that both mechanisms can contribute substantially to the final recoil and often oppose each other. For the higher-spin low-mass models \texttt{a5}, \texttt{a5*}, and \texttt{a8}, however, the gravitational contribution is more closely aligned with the net kick and provides the dominant contribution, consistent with a persistent gravitational tugboat driven by the large-scale ejecta asymmetry.

In these models, the angles between the gravitational contribution and the final kick are $\theta_{\rm grav}^{\rm BH}=10^\circ$, $29^\circ$, and $19^\circ$, respectively, whereas the corresponding angles for momentum accretion are $30^\circ$, $112^\circ$, and $75^\circ$. In \texttt{a8}, for example, the dominant $z$-component of the kick is $v^{\rm BH}_z=v^{\rm acc}_z + v^{\rm grav}_z=(-12+470)\,\rm km\,s^{-1} =458\,\rm km\,s^{-1}$, illustrating the dominance of the gravitational contribution. For the lowest-spin model, \texttt{a1}, the momentum-accretion and gravitational contributions are more comparable in magnitude, and neither is closely aligned with the final kick. The $x$-component is $v^{\rm BH}_x = v^{\rm acc}_x + v^{\rm grav}_x =(-193+465)\,\rm km\,s^{-1} =272\,\rm km\,s^{-1}$, while the smaller $y$-component results from a strong cancellation, $v^{\rm BH}_y=(617-486)\,\rm km\,s^{-1}=131\,\rm km\,s^{-1}$. The resulting kick is predominantly in the equatorial plane with $v^{\rm BH}_z=-37\,\rm km\,s^{-1}$.

\begin{figure*}
    \centering
    \includegraphics[width=\linewidth]{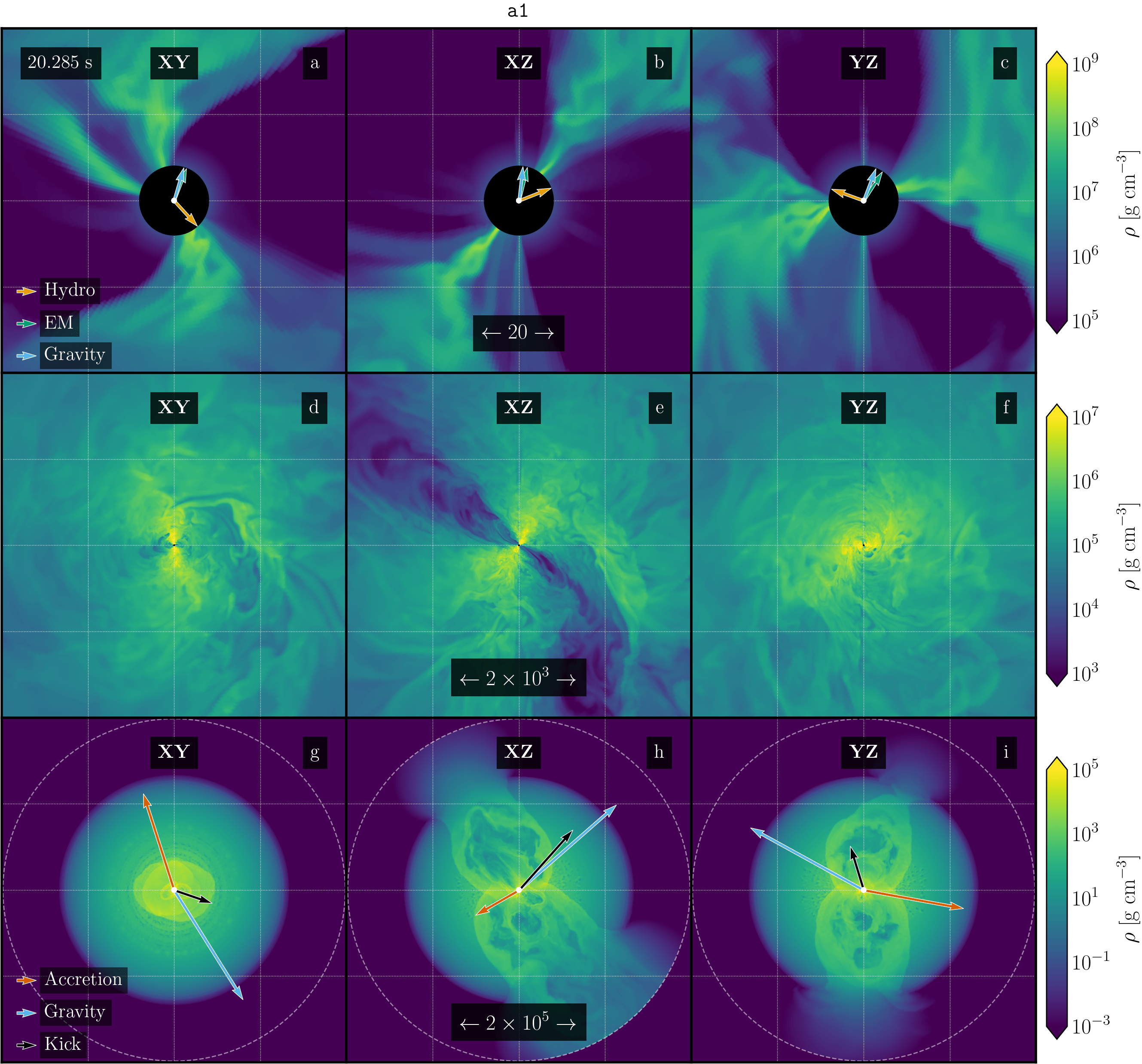}
    \caption{Slowly spinning BH model \texttt{a1} with $a=0.1$: equatorial ($xy$) and vertical ($xz$ and $yz$) slices of the rest mass density $\rho$ in Cartesian KS coordinates within the innermost $10\, r_g$ (top; [a]--[c]), $1000\,r_g$ (middle; [d]--[f]), and for the full simulation domain of $10^5\, r_g$ (bottom; [g]--[i]) taken at $t = 20.285\,\rm s$ after collapse. See \fig{a8_eruption} for an explanation of the arrows in the top three panels and \fig{fiducial_explosion} for the arrows in the bottom three panels. In this late snapshot, the disk angular momentum is nearly perpendicular to the BH spin axis in the innermost $\sim 100\,r_g$, and the jet asymmetry and wobbling are evident in the large-scale non-axisymmetric morphology of the accretion ([g], [h], and [i]). The southern jet shock is the first to break out of the star ([h] and [i]), and the gravitational kick points into the northern hemisphere, but not along the initial northern jet axis or the axis of the ejecta at the largest radii. Movies of all simulations discussed in this paper are available at \url{https://youtube.com/playlist?list=PLDkQSWz6UZNI&si=VMFtbJSVYpUrJk1r}.}
    \label{fig:a1_rho}
\end{figure*}

Snapshots of the 2D density profiles for \texttt{a1} taken at $t\approx20\,\rm{s}$ in \fig{a1_rho} illustrate asymmetries from the horizon to progenitor-star scales. The accretion disk and jet axes are seen to be highly tilted relative to the BH spin axis. The instantaneous hydrodynamic force makes a large angle with the EM, and gravitational forces, which happen to be aligned. At intermediate scales, the jets are deflected by asymmetrically infalling turbulent plasma, and on the largest scales the intersecting shock fronts within the overall dipolar explosion reveal the stochastic geometry produced by jet wobbling and sputtering. By this relatively late time, the southern outer shock has broken out of the star before the northern shock. These large-scale asymmetries help determine the direction of the net kick, whose time evolution we examine next.

The DH models exhibit a different balance of kick contributions. In the most weakly magnetized model, \texttt{dhBm}, the gravitational and kinetic contributions are comparable, reflecting the substantial momentum carried by the asymmetric disk winds and the one-sided jet. In the more strongly magnetized models \texttt{dhBs} and \texttt{dhBvs}, the gravitational contribution instead competes primarily with the EM stress. The resulting kicks are particularly large because the stronger magnetic fields drive more powerful, highly asymmetric outflows and magnetic-flux eruptions that produce substantial momentum transfer to the BH. Jets and disk winds powered by substantial accretion rates expel most of the progenitor envelope, allowing the BH to retain a large recoil. The final kick speeds are $v^{\rm BH}=346$, $569$, and $3691\,\rm km\,s^{-1}$ for \texttt{dhBm}, \texttt{dhBs}, and \texttt{dhBvs}, respectively. The first two models exhibit predominantly spin-aligned kicks, with $\theta^{\rm BH}=39^\circ$ and $35^\circ$, whereas \texttt{dhBvs} produces an almost entirely equatorial kick,
$\theta^{\rm BH}=89^\circ$.

\begin{figure*}
    \centering 
    \includegraphics[width=\linewidth]{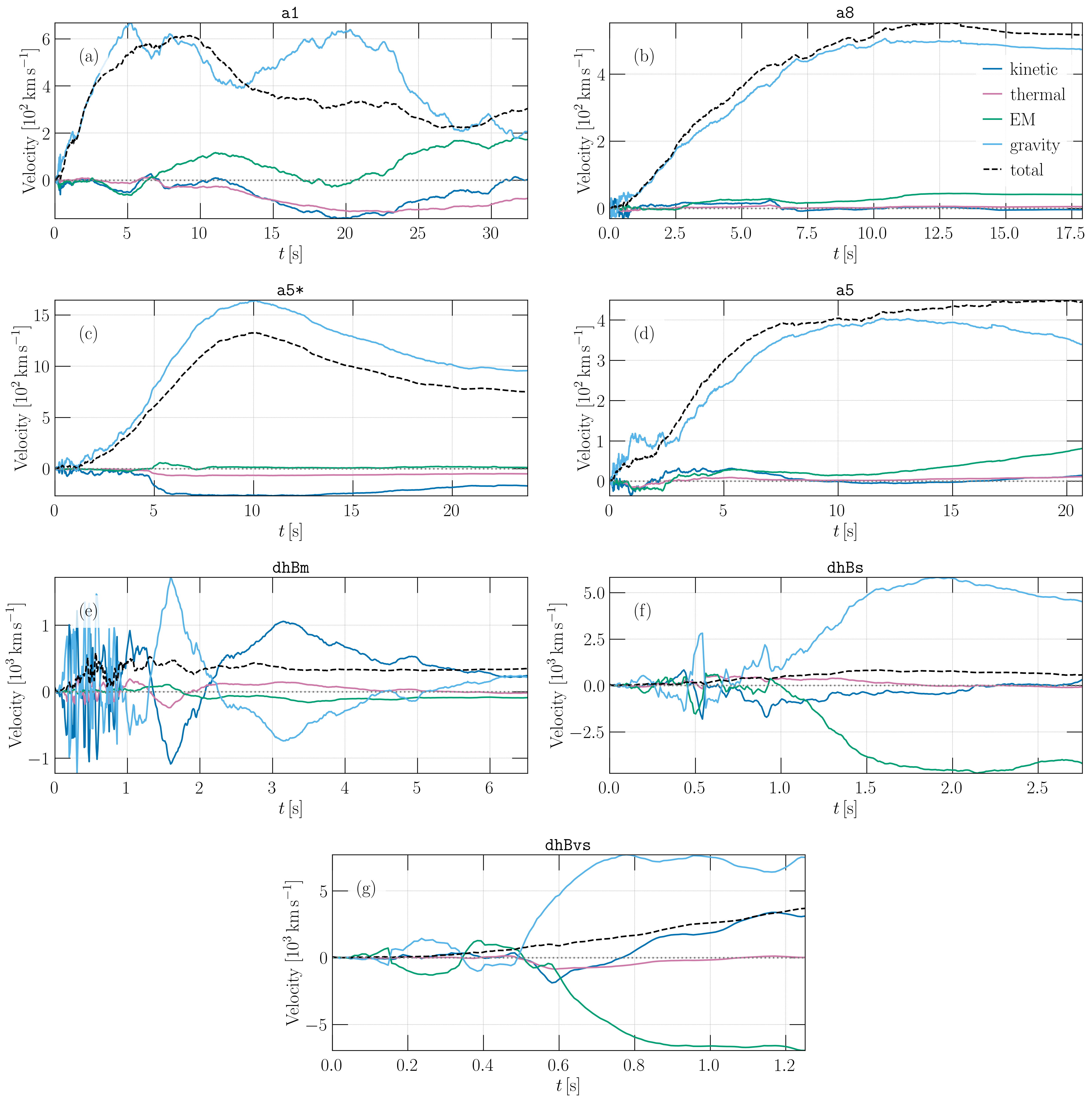}
    \caption{Contributions to the BH kick velocity from momentum accretion through the horizon, including kinetic, thermal, and electromagnetic (EM) stresses, and from gravitational interaction with the stellar envelope, as a function of coordinate time $t$. The solid curves show the projections $\sum_i v_i^{(x)} v_i^{\rm BH}/v^{\rm BH}$ (similar to, e.g., Fig.~6 of \citealt{Janka&Kreese24}), where $x\in\{\text{kinetic, thermal, EM, gravity}\}$. The dashed curve shows the total kick velocity. For the low-mass collapsar models (a)--(d), the kick is generally dominated by gravity at all times, with subdominant positive contributions from EM stress and negative contributions from kinetic stress. For \texttt{a1} (a), in which the disk undergoes the greatest wobbling, the EM and gravitational contributions are comparable by the end of the simulation. For the DH models, the balance changes with the magnetic-flux supply: in \texttt{dhBm} (e), kinetic and gravitational stresses make comparable contributions to the kick, whereas in the more strongly magnetized models \texttt{dhBs} (f) and \texttt{dhBvs} (g), the kick appears to be set by a competition between EM stress and gravity.}
    \label{fig:velocities}
\end{figure*}

The physical origin of the kick geometry varies across the DH models. In \texttt{dhBm}, the one-sided, partially deflected jet produces a substantial spin-aligned kick toward the northern hemisphere, $v^{\rm BH}_z\approx269\,\rm km\,s^{-1}$, while non-axisymmetric disk winds contribute equatorial components $v^{\rm BH}_x\approx-164\,\rm km\,s^{-1}$ and $v^{\rm BH}_y\approx-143\,\rm km\,s^{-1}$. In \texttt{dhBs}, the stronger southern jet plume dominates the momentum asymmetry across the equatorial plane and therefore the $z$-component of the kick. By contrast, in the very strongly magnetized model \texttt{dhBvs}, the bipolar jet outflows are nearly symmetric across the equatorial plane; instead, powerful magnetic-flux eruptions and the resulting jet deflections generate a strongly asymmetric equatorial outflow and an almost entirely equatorial BH kick.

\subsection{Time Evolution}\label{subsec:evolution}

The kick develops over the course of the explosion, with the relative contributions from gravitational interaction and momentum transfer through the horizon evolving as the jets, disk, and surrounding ejecta become increasingly asymmetric. Figure~\ref{fig:velocities} tracks these contributions for all models, separating the momentum-accretion term into kinetic, thermal, and EM stresses. The time evolution reveals a predominantly gravity-driven kick in the low-mass models, while the direct-horizon models show a stronger competition between gravitational and horizon-stress
contributions.

For the low-mass collapsar models, the kick is dominated by gravitational interaction with the stellar gas from the earliest times, with asymmetric EM and kinetic stresses providing secondary contributions and thermal stresses remaining subdominant. Model \texttt{a1} (\fig{velocities}[a]) is the main exception: by the end of the simulation, the EM and gravitational contributions are comparable. The gravitational contribution begins to deviate from the total kick at $t\approx12.5\,\rm{s}$, coincident with a period during which the ``$-$'' jet becomes more powerful than the ``$+$'' jet and the disk and jets undergo extreme tilting. During this interval, both the EM and kinetic contributions become more negative, suggesting that asymmetric inflow of highly magnetized plasma contributes appreciable momentum transfer to the BH. At $t \approx 20\,\rm{s}$, the EM and kinetic contributions become more positive again and the gravity contribution starts converging back toward the total kick. This corresponds roughly to a reversal of the earlier jet asymmetry and re-alignment of the disk and jet axes with the BH spin until $t \approx 30\,\rm{s}$ (\fig{angles}[a],[b]). The ``$+$" jet is generally dominant for the remainder of the evolution, and the movie of the density profile reveals that although the jet is launched along the BH spin, it is deflected strongly in the $-x$-direction beyond $\sim 100\,r_g$. As a result, the total kick and gravity kick directions both develop more positive $x$-components, re-orienting toward the equator.

Owing to very high jet power asymmetry at early times when the magnetic flux is accumulating but below saturation (see \fig{angles_a5}[a]), \texttt{a5*} reaches a kick speed $>1000\,\rm{km\,s^{-1}}$ at $t\approx 10\,\rm{s}$ before the kick smoothly falls to its final value of $750\,\rm{km\,s^{-1}}$. Gravity is almost wholly responsible for the evolution of the total kick (\fig{velocities}[c]), as the reversal of the jet asymmetry around $t\approx 10\,\rm{s}$ reverses the direction of the tugboat effect. 

For the DH collapsars, the time evolution reveals a stronger competition between gravitational and horizon-stress contributions (\fig{velocities}[e],[f],[g]). In \texttt{dhBm}, the kinetic and gravitational contributions are comparable and often oppose each other as the jet-power asymmetry changes sign. In the more strongly magnetized models \texttt{dhBs} and \texttt{dhBvs}, the EM stress instead becomes the principal competitor to gravity. These reversals and partial cancellations reflect the rapidly evolving magnetic and outflow geometry and allow the net kick to build to large values, albeit considerably smaller than individual momentum fluxes.

\begin{figure*}
    \centering
    \includegraphics[width=\linewidth]{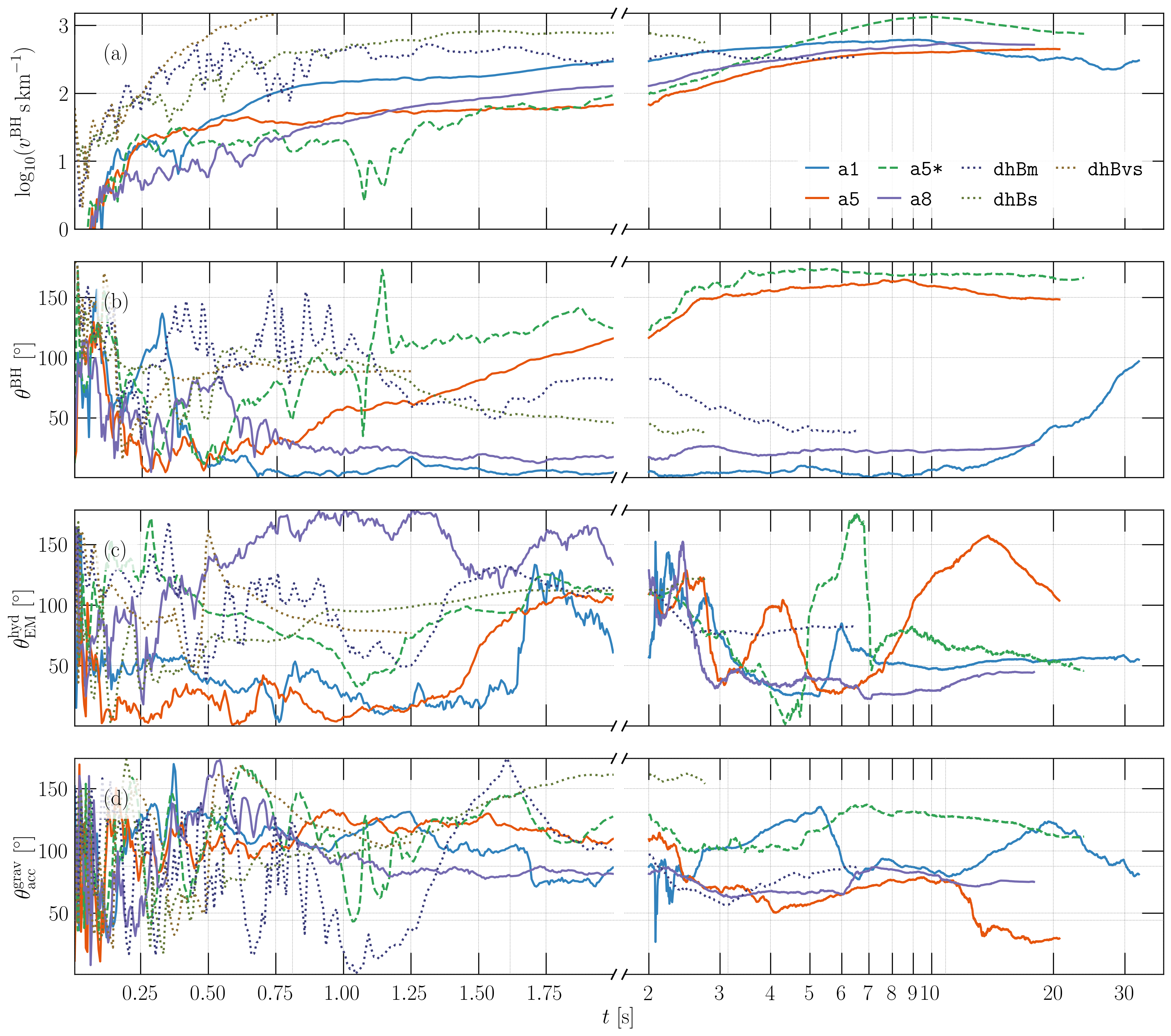}
    \caption{Time series of the kick speed $v^{\rm BH}$, spin-kick misalignment angle $\theta^{\rm BH}$, angular separation $\theta^{\rm hyd}_{\rm EM}$ between the hydrodynamic and electromagnetic contributions to the kick, and angular separation $\theta^{\rm grav}_{\rm acc}$ between the gravitational and momentum-accretion contributions. For visibility, we plot the time series on a linear scale until $t=2\,\rm{s}$ and a logarithmic scale thereafter.}
    \label{fig:kick_evolution}
\end{figure*}

Figure \ref{fig:kick_evolution}(a) shows that the low-mass kicks evolve relatively slowly by the end of the simulations, indicating that they are close to convergence as the progenitor mass is progressively accreted or ejected. Models \texttt{a5} and \texttt{a8} reach this converged state most rapidly, consistent with their more powerful jets. In contrast, \texttt{a1} initially develops a larger kick because its stronger outflow asymmetry compensates for its lower jet power, exceeding $500\,\rm km\,s^{-1}$ at $t\approx4\,\rm s$. The kick then begins to decline around $t\approx10\,\rm s$, ultimately settling below the final kick speeds
of \texttt{a5} and \texttt{a8}.

The kick-spin misalignment angle $\theta^{\rm BH}$ is highly variable at early times but converges within the first few seconds for most of the low-mass collapsars, except for \texttt{a1} (\fig{kick_evolution}[b]). For that model, $\theta^{\rm BH}$ continues to evolve toward its highly equatorial final value of $97^\circ$ as strongly asymmetric deflected jets drive net ejecta momentum in the $-x$-direction. This continued evolution highlights the need for long simulation times when estimating collapsar BH natal kicks. For \texttt{a5} and \texttt{a8}, the final kicks are more polar; in particular, the kick in \texttt{a5} is more nearly anti-aligned with the spin, whereas that in \texttt{a8} points more along the spin axis. For these models, the jetted outflows are more collimated due to greater jet power, and the comparative stability and spin-alignment of the jets at maximum power yields a steadier gravitational tugboat acceleration toward either pole. The smaller contributions from EM and kinetic stress compared with \texttt{a1} may be attributed to decreased inflow of plasma in the jet regions. For the modified field simulation, \texttt{a5*}, the gravitational tugboat pull of the southern ejecta causes the kick-spin angle to saturate to $\approx 166^\circ$ by $t \approx 3\,\rm{s}$, resulting in the most strongly polar kick of the simulations.

The angular separations between the horizon-stress contributions and between gravity and momentum accretion also evolve substantially at early times (\fig{kick_evolution}[c],[d]). For example, $\theta^{\rm hyd}_{\rm EM}$ reaches $\approx180^\circ$ before dropping to $45^\circ$ within $\approx3\,\rm{s}$ in \texttt{a8}. For \texttt{a1}, \texttt{a5*}, and \texttt{a8}, $\theta^{\rm hyd}_{\rm EM}$ converges to values between $45^\circ$ and $55^\circ$. In the fiducial $a=0.5$ model \texttt{a5}, $\theta^{\rm hyd}_{\rm EM}$ is still evolving at $t=T_{\rm sim}$ even though the kick speed has largely converged, but it trends toward the values reached by the other low-mass models. The angle $\theta^{\rm grav}_{\rm acc}$ evolves more slowly, consistent with the gravitational contribution being shaped by the long-range tugboat interaction. Its final values show substantially more dispersion than $\theta^{\rm hyd}_{\rm EM}$ and do not exhibit a clear dependence on BH spin. For DH models \texttt{dhBm} and \texttt{dhBs}, the kick has a dominant spin-aligned component, with $\theta^{\rm BH}=39^\circ$ and $35^\circ$, whereas the most strongly magnetized collapsar, \texttt{dhBvs}, yields an almost entirely equatorial kick, with $\theta^{\rm BH}=89^\circ$. In the latter case the kick is driven by magnetic flux eruptions that expel much of the star horizontally.
\section{Discussion}\label{sec:discussion}
\subsection{Kick Relationship to BH, Progenitor, and Explosion Properties}\label{subsec:properties}

\begin{figure*}
    \centering
    \includegraphics[width=\linewidth]{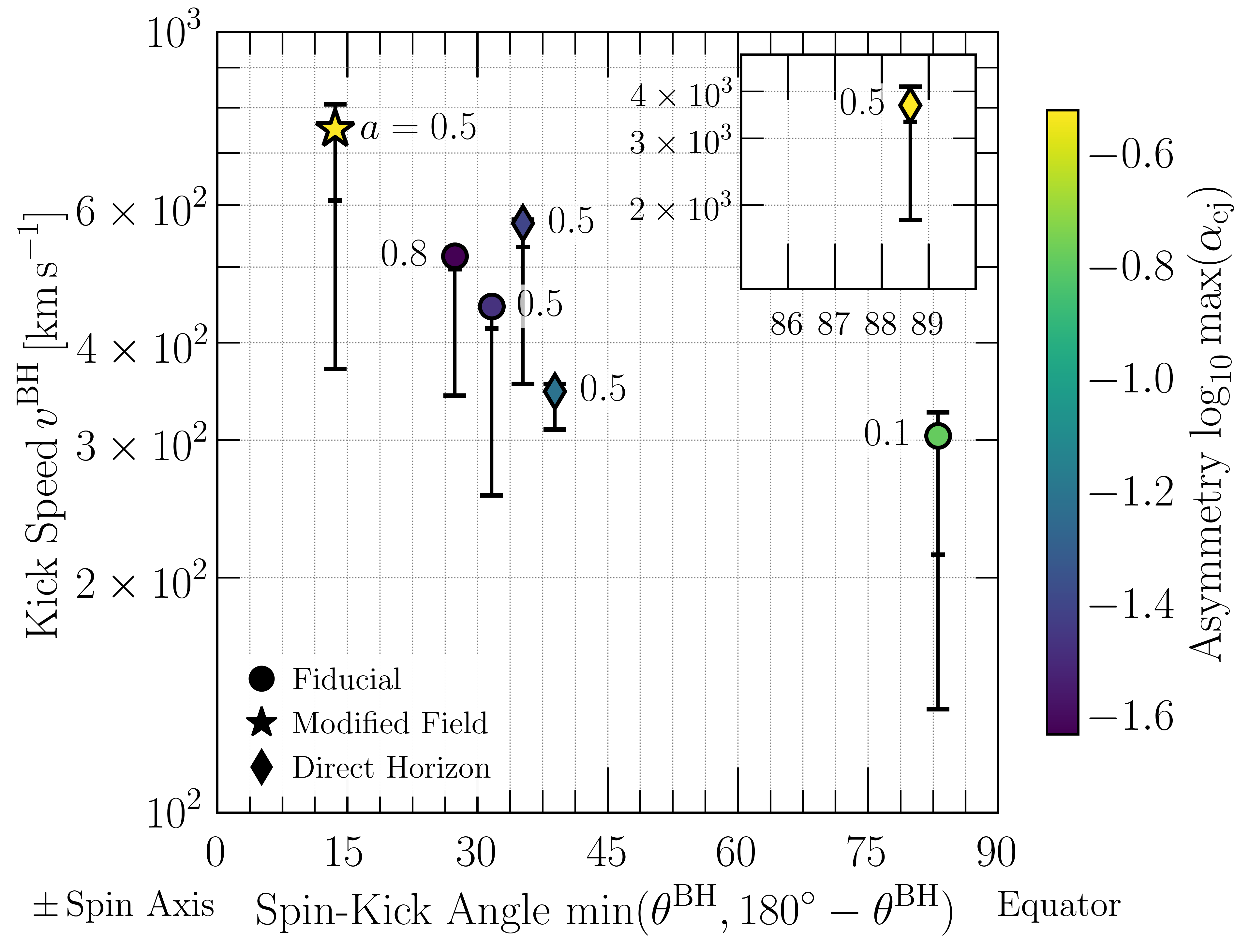}
    \caption{
    Summary of the natal BH kicks inferred in our collapsar simulations. Magnetized collapsars can impart large kicks, typically several hundred km s$^{-1}$. The most strongly magnetized direct-horizon model reaches several thousand km s$^{-1}$, illustrating the extreme kicks possible when powerful asymmetric outflows are sustained; however, this value should be regarded as an upper-end case because \citetalias{gottlieb_spinning_2025} showed that the DH models \texttt{dhBs} and \texttt{dhBvs} rapidly spin down their BHs. A self-consistent evolution of the jet power would likely reduce the largest kicks. The kick orientation varies substantially from model to model, ranging from nearly polar to nearly equatorial, reflecting the stochastic geometry of asymmetric jets, magnetic flux eruptions, and ejecta. Circles (fiducial models), stars (modified-field model), and diamonds (direct-horizon models) show the total kick speed, $v^{\rm BH}$, versus the acute kick angle with respect to the BH spin axis, $\min(\theta^{\rm BH},180^\circ-\theta^{\rm BH})$, for each simulation as measured at the end of the simulation ($t=T_{\rm sim}$).  Each model is labeled by the BH spin parameter $a$ and colored according to its maximum ejecta asymmetry parameter, $\max(\alpha_{\rm ej})$.  Vertical error bars correspond to estimated upper and lower bounds on the asymptotic kick, given by Eq.~(\ref{eq:v-high}) and Eq.~(\ref{eq:v-low}).  Of the two lower bounds for each model, the lower value corresponds to the geodesic criterion for unbound material.  For model \texttt{dhBm}, $v^{\rm BH}_{\infty,{\rm Be}}$ is not shown because it slightly exceeds the kick measured at $T_{\rm sim}$.
    }
    \label{fig:summary}
\end{figure*}

In \fig{summary}, we summarize how the kick varies with BH spin, magnetic-flux history, and progenitor model by plotting $v^{\rm BH}$ against the acute kick angle relative to the spin axis, $\min(\theta^{\rm BH},180^\circ-\theta^{\rm BH})$. Each point is labeled by the BH spin $a$ and colored by the peak ejecta momentum asymmetry, $\max(\alpha_{\rm ej})$. Among the three fiducial models (circles), which share the same progenitor magnetic-field structure, higher spin corresponds to stronger, more polar kicks, but smaller peak ejecta asymmetry.

The fiducial spin dependence aligns with the naive expectation that faster-spinning BHs should receive larger kicks. For comparable horizon-threading magnetic flux, the BZ jet power scales approximately quadratically with $a$ \citep{blandford_electromagnetic_1977}, so higher-spin BHs launch more powerful jets, drive more energetic explosions, and deposit a larger total momentum reservoir in the ejecta. However, the increase in kick speed is modest compared with the increase in jet power because stronger jets also promote spin-disk alignment. In the higher-spin models, stronger magneto-spin alignment and/or reduction of stochastic torques reduces early-time jet wobbling and produces a more nearly bipolar explosion, lowering the fractional ejecta momentum asymmetry. The same alignment also shapes the kick direction: for \texttt{a5} and \texttt{a8}, the jet-powered ejecta flow primarily along the $\pm z$-axis, so a few-percent momentum asymmetry yields a strong, mostly polar kick. Although our fiducial sequence contains only three spin values, it suggests that BH spin may influence both the magnitude and orientation of collapsar natal kicks.

The modified-field model, \texttt{a5*}, shows that this spin-based trend can be altered by the magnetic-flux accumulation history. Although \texttt{a5*} has the same BH spin and progenitor structure as \texttt{a5}, and a lower time-averaged jet power and horizon magnetic flux than \texttt{a8}, it reaches $v^{\rm BH}=750\,\rm km\,s^{-1}$, larger than any of the fiducial models. The greater kick magnitude reflects the much larger momentum asymmetry produced during its extended sub-MAD phase and subsequent transition back to the MAD state. In \texttt{a5*}, the kick grows most rapidly during this transition, before the jets reach their peak power, showing that collapsar kicks depend not only on the total magnetic flux supplied to the BH but also on how that flux is accumulated and reorganized in time. A moderately magnetized BH-disk system may therefore produce a larger kick than a more strongly magnetized one if it drives a sufficiently asymmetric outflow. The kick in \texttt{a5*} is the most polar of all the simulations, making an acute angle $14^\circ$ with the BH spin axis; however, different kick orientations could have arisen if the disk underwent more wobbling during the transition to MAD. 

Asymmetry alone is not enough to achieve a lasting kick: unless the outflow ultimately unbinds a large fraction of the progenitor, continued accretion will dilute the recoil by increasing the BH inertia. The late, more powerful ejection in \texttt{a5*} may therefore be essential for preserving, or ``freezing in,'' the kick generated during the earlier asymmetric phase. Dynamo amplification or reorganization of poloidal flux offers one possible route to this behavior.

\subsection{Implications for BBH Gravitational Wave Sources}
Coalescing BBH systems provide hints about their origins through their GW emission \citep[e.g.,][]{farr_distinguishing_2017}. The strong natal kicks we find in our simulations have implications for both major BBH formation channels, and hence the rates and properties of GW signatures of BBH mergers. Although our progenitors are motivated by the requirements for producing long GRBs, our simulations capture physical processes that may operate more broadly in engine-driven core collapse, including systems in which a jet is choked or otherwise fails to produce an observable GRB but still powers an asymmetric SN. In dense stellar systems, natal kicks determine whether newly formed BHs are retained long enough to undergo mass segregation, dynamical pairing, and subsequent mergers \citep[e.g.,][]{Antonini+16}, including the hierarchical assembly of intermediate-mass and supermassive BHs \citep{Sedda2021}. In isolated binaries, natal kicks alter the post-collapse orbit, including the separation, eccentricity, and spin--orbit misalignment, if the companions remain bound \citep[][and references therein]{Vigna-Gomez25}. This in turn affects the time to merger, and at the population level, the distribution of delay times between binary formation and GW emission as well as the overall local rate density $\mathcal{R}_0$ of BBH mergers.

\subsubsection{Dynamical Formation Channel}

In dynamically assembled BBH channels, natal kicks act primarily as a retention filter. Kicks exceeding the cluster escape velocity remove newly formed BHs before they can mass segregate, participate in dynamical encounters, form BBHs, or contribute to the growth of central massive BHs in nuclear star clusters. The kicks we obtain for our magnetized collapsar models, typically several hundred $\rm km\,s^{-1}$, are likely fast enough to eject their remnant BHs from most globular clusters, whose escape speeds are $v_{\rm esc} \lesssim 50\,\rm km\,s^{-1}$ \citep{belczynski_binaries}. In nuclear star clusters, where the potential wells are much deeper, the same kicks may not eject every remnant, but they would still reduce the retention fraction of BHs born through the magnetized collapsar channel and bias the population that remains available for subsequent mergers \citep{Antonini+16,Pavlik2018}. If the kick magnitude correlates with BH spin, as suggested by our fiducial models, this filtering could preferentially remove rapidly spinning collapsar remnants from dense stellar systems. The strength of this effect depends on the abundance of magnetized, rapidly rotating progenitors relative to other massive stars that collapse to BHs. Such filtering may be especially relevant if massive DH collapsars contribute to the population of BHs in or near the PISN mass gap \citep{Siegel+22,gottlieb_spinning_2025}.

\subsubsection{Isolated Binary Channel}

This binary context is especially relevant because rapidly rotating collapsar progenitors are themselves difficult to produce in isolation. Long-GRB central engines require massive stellar cores to retain large angular momentum until collapse, despite angular momentum transport by magnetic torques \citep[e.g.,][]{Kissin&Thompson18,Fuller+19,Skoutnev+24} and spin-down through post-main-sequence winds. In single stars, wind-driven angular momentum loss makes this difficult, especially at high metallicity. Binary evolution offers several viable alternatives: tidal spin-up in close binaries can transfer orbital angular momentum to the star \citep[e.g.,][]{Cantiello+07}, while subsequent mass transfer or stellar mergers may further enhance the core spin. An especially promising low-metallicity pathway is chemically homogeneous evolution \citep[e.g.,][]{Maeder87,Langer92}, in which rotationally induced mixing keeps the star compact and prevents envelope expansion, reducing angular momentum loss. Observations of long GRBs and broad-lined SNe Ic are consistent with BH central engines whose progenitors often reside in close binaries \citep{fryer_explaining_2025}.

Natal kicks are critical to the subsequent evolution of such binaries. For axisymmetric progenitors whose spins are initially aligned with the binary orbital angular momentum $\mathbf{L}$, the direction of the  kick component in the orbital plane is uniformly distributed. This in-plane component, perpendicular to $\mathbf{L}$, controls the immediate change in the orbital energy: a kick opposite the instantaneous orbital motion can tighten the binary and shorten the merger time, whereas a kick along the instantaneous orbital motion, or one that is too large, can help disrupt the system \citep{Vigna-Gomez25}. Symmetric mass loss from the explosion produces the Blaauw effect, which widens the orbit and increases the eccentricity, and can by itself unbind the binary if the lost mass fraction is sufficiently large \citep{blaauw,banagiri2023}. Because our collapsars can eject a large fraction of the collapsing star, the associated Blaauw effect can be substantial, although its importance for binary survival depends on the total pre-collapse binary mass. However, their natal kicks also contain in-plane components of order hundreds of $\rm km\,s^{-1}$ for aligned progenitor spins, so in favorable geometries the kick can counteract the Blaauw expansion and help the binary remain bound or merge within a Hubble time \citep{Hills1983}.

In a close binary, even natal kicks of hundreds of $\rm{km\,s^{-1}}$ need not cause disruption. For example, consider the binary scenario envisioned in \citetalias{gottlieb_spinning_2025} for the DH collapsar models. Suppose the progenitor He star, of mass $M_{\rm He} = 150 M_\odot$, were to exist in an equal-mass circular binary. For a semi-major axis $a_0 = 3\times 10^{12}\,\rm cm$, corresponding to an orbital period $\sim 2\,\rm days$ \citep[roughly the maximum for which massive binaries undergo efficient tidally induced mixing;][]{deMink2009,Marchant2016,Song2016}, the pre-explosion relative velocity $v_{\rm rel} = \sqrt{2GM_{\rm He}/a_0} = 1152\,\rm km\,s^{-1}$. For \texttt{dhBm}, the mass ejected is $M_{\rm ej} \approx 37M_\odot$, corresponding to $12\%$ of the total binary mass. The kick we find in our simulation, $v^{\rm BH} = 349\,\rm{km\,s^{-1}}$, is too low to unbind the system for any direction, i.e.~the survival probability $P_{\rm bound} = 100\%$ \citep{Hills1983}. For strongly magnetized model \texttt{dhBs}, the fractional mass ejected from the binary would be $23\%$, and a kick of $v^{\rm BH} = 569\,\rm{km\,s^{-1}}$ at an angle $\theta^{\rm BH} = 35^\circ$ relative to $\mathbf{L}$ corresponds to $P_{\rm bound} = 88\%$, assuming the probability distribution of the in-plane orientation of the kick is uniform. For a kick of the same magnitude purely in the plane of the orbit, $\theta^{\rm BH} = 90^\circ$, the survival probability is instead $P_{\rm bound} =68\%$.

The component of the kick parallel to $\mathbf{L}$ affects not only the probability of binary disruption and the post-collapse eccentricity, but also the tilt of the orbital plane, thereby generating spin--orbit misalignment and precession. These effects enter GW observations through the effective inspiral spin, $\chi_{\rm eff}$, and the precession spin, $\chi_{\rm p}$. Spin--orbit misalignment in LVK BBH mergers has been interpreted as possible evidence for strong BH natal kicks, $v^{\rm BH} \sim 10^2$--$10^3\,\rm km\,s^{-1}$, although such inferences depend sensitively on assumptions about the natal spin magnitudes and the prior binary evolution \citep[][and references therein]{wysocki_explaining_2018,callister_state_2021}. Our results provide a physical mechanism by which rapidly rotating collapsar progenitors can receive kicks in this range. Moreover, if higher-spinning BHs tend to receive more polar kicks, as in our fiducial models, then binaries in which one or both companions form through magnetized collapsars may exhibit correlations between $\chi_{\rm eff}$, $\chi_{\rm p}$, and the merger outcome. A comprehensive exploration of progenitor parameter space via simulations of both isolated collapse and binary evolution is needed to further investigate this possibility.

\citet{Boco2026} recently compared BBH merger efficiencies across several population synthesis codes using a standardized, observationally-motivated star formation rate density (SFRD), concluding that most state-of-the-art models overpredict $\mathcal{R}_0$ by up to an order of magnitude. The discrepancy arises from the low-metallicity tail of the SFRD \citep{Sgalletta+2025}, corresponding to low-mass galaxies and starbursts. One possible remedy to this tension among many is to adopt higher natal BH kicks than are commonly assumed; this can both reduce the fraction of surviving binaries and steepen the delay time distribution of those that survive \citep[see, however,][for a discussion of other resolutions, such as adjustments to mass transfer, angular momentum loss, etc.]{Broekgaarden2026}. Our results lend plausibility to this approach by demonstrating that BHs formed from the collapse of possible long GRB progenitors, which favor low-metallicity environments \citep{Fruchter2006, Palmerio2019}, can receive such natal kicks.

\subsection{Implications for Black Hole X-Ray Binaries}

In addition to imprinting signatures on compact-object mergers, BH natal kicks also influence the observable properties of Galactic black hole X-ray binaries \citep[BHXBs; for analogous discussion of the impact of kicks on NS binaries, see][]{Brandt&Podsiadlowski1995}. The kick magnitude primarily determines the systemic velocity of the binary relative to the local standard of rest. The peculiar velocities of Galactic BHXBs therefore provide indirect constraints on BH natal kicks. Using very long baseline interferometry and Gaia data, \citet{atri_potential_2019} estimated the probability distributions of so-called ``potential kicks,'' defined as the peculiar velocity of the system at the time of crossing the Galactic plane, for 16 BHXBs. They found that 75\% of these systems have potential kicks exceeding $70\,{\rm km\,s^{-1}}$, larger than the Blaauw recoil expected for the maximum mass loss that would still leave the binaries bound, thereby suggesting a contribution from intrinsic natal kicks. For the low-mass BHXB system H 1705-250, \citet{DashwoodBrown2024} infer a natal kick of $\approx 300\,\rm{km\,s^{-1}}$. The natal kicks of BHXBs can also be constrained from their heights above the Galactic plane \citep[e.g.,][]{Mandel2016,repetto_galactic_2017}. 

Observations of BHXBs are necessarily biased toward binaries that survive the explosion and subsequently evolve into Roche-lobe overflow, disfavoring systems that receive kicks large enough to disrupt the binary or widen it sufficiently to prevent mass transfer. Consequently, the observed distribution of BHXB peculiar velocities is expected to underestimate the full natal-kick distribution. Because our collapsar explosions eject several solar masses while simultaneously imparting intrinsic BH natal kicks, they naturally predict large systemic velocities, with a substantial contribution arising from the accompanying Blaauw recoil.

In addition to their magnitude, the direction of BH natal kicks influences the degree of spin--orbit misalignment in BHXBs. Spin--orbit misalignment can be inferred through polarimetry of relativistic jets or the inner accretion flow. For example, \citet{Poutanen2022} placed a lower limit of $40^\circ$ on the spin--orbit misalignment angle in the system MAXI~J1820+070 (note, however, that this system is much less massive than the progenitors considered in this work). In evolutionary channels that promote aligned stellar spins prior to collapse (e.g., through tidal interactions), natal kicks with a component perpendicular to the orbital plane are required to generate such misalignment. Although our slowly spinning fiducial model \texttt{a1} yields a kick almost entirely within the equatorial plane, our more rapidly rotating models exhibit substantial out-of-plane kick components, naturally producing spin--orbit misalignment in binaries whose spins were initially aligned. If this trend proves generic, BHXBs formed through the collapsar channel may preferentially exhibit larger spin--orbit misalignments for rapidly rotating BHs than for more slowly rotating ones. Testing for such correlations would provide a novel observational probe of magnetized collapsar formation. In contrast with the comparatively low spins, $a \lesssim 0.25$, inferred for the components of $85\%$ of merging LVK binaries \citep{LVK2026}, spin measurements in Galactic BHXBs preferentially probe a higher-spin BH population, with many systems inferred to have $a \gtrsim 0.7$ \citep{Draghis+2024}. Collapsars that launch strongly asymmetric polar outflows without significantly spinning down their BHs, for example in a ``failed jet'' scenario, could contribute to this population while also producing the large systemic velocities and spin--orbit misalignments observed in some BHXBs.
\subsection{Implications for Dark Lenses}
Several quiescent BH candidates have been identified through  astrometric microlensing \citep{Wyrzykowski+2020}, and forward modeling \citep{Vigna-Gomez&Ramirez-Ruiz2023} supports a binary origin for the confirmed system OB110462 \citep{Lam+2023}. The mass distribution of these dark lenses disfavors the putative lower mass gap unless appreciable BH natal kicks of $20$--$80\,\rm{km\,s^{-1}}$ are assumed \citep{Wyrzykowski+2020}, and the demographics of BBH mergers from GWTC-5.0 further reduce observational support for the gap \citep{LVK2026}. Our results suggest rapidly rotating progenitors can supply kicks of hundreds of $\rm{km\,s^{-1}}$ given sufficiently strong magnetic fields, but reduced rotation or magnetic flux may yield weaker ``failed-jet" explosions responsible for kicks in the range $\sim 10$--$100\,\rm{km\,s^{-1}}$. Microlensing measurements can also more directly constrain natal kicks assuming an isolated binary origin \citep{andrews_constraining_2022}. The Roman Galactic Bulge Time Domain Survey is expected to discover additional isolated BHs \citep{Lam+2026}, and massive BHs with high peculiar velocities will provide possible signatures of the collapsar kick mechanism we study here.

\subsection{Context and Limitations}
Recent work has demonstrated that BH natal kicks can arise through several channels of core collapse powered primarily by asymmetric neutrino-driven explosions. Our results identify a complementary channel operating in rapidly rotating, magnetized collapsars, in which the recoil is generated primarily after BH formation by asymmetric jet-driven outflows. 

The most extensive recent study is that of \citet{burrows_black_2023,Burrows+2024b,Burrows+2025} (hereafter B23,24,25), which identify multiple channels of stellar-mass BH formation in long-duration three-dimensional neutrino-radiation hydrodynamic simulations of non-rotating, solar-metallicity progenitors. In the ``silent'' (or ``failed explosion'') channel of \citetalias{Burrows+2025}, the bounce shock is never revived and the BH forms quiescently, receiving only a weak kick, $\lesssim10\,{\rm km\,s^{-1}}$, from anisotropic neutrino emission during the pre-collapse proto-neutron-star (PNS) phase. \citet{Janka&Kreese24} (hereafter J24) also studied non-rotating or slowly rotating neutrino-hydrodynamic models and found similarly small neutrino kicks for BHs formed without asymmetric mass ejection. Our simulations suggest that rapid rotation and magnetic fields fundamentally alter the outcome of the otherwise ``silent'' BH formation channel by powering sustained asymmetric outflows after BH formation. Thus, even if the PNS receives only a negligible kick, the BH itself may ultimately acquire a substantial recoil. 

At the opposite extreme, two of the most compact progenitors of \citetalias{Burrows+2025} undergo highly energetic, asymmetric neutrino-driven explosions that impart large kicks during the PNS/explosion phase, leaving relatively low-mass BHs ($M_{\rm BH}\sim2.5$--$3.5\,M_\odot$). Fallback reduces these kicks by at most a factor of $\sim2$, yielding final BH recoil velocities of $\sim500$--$1300\,{\rm km\,s^{-1}}$. In the $40\,M_\odot$ model, stochastic angular momentum in the accretion flow produces a transient disk-like structure around the PNS, leading the authors to speculate that collapsar-like outflows might arise if sufficiently strong magnetic fields were present. While delayed BH formation may provide an additional source of recoil, it is not clear that such systems would enter the same long-lived magnetized collapsar regime modeled in this work. 
 
We initialize our simulations with a central BH embedded in a freely falling progenitor, up to small symmetry-breaking perturbations. We therefore do not follow the preceding PNS phase, including the kick the PNS may receive from neutrinos or hydrodynamic instabilities as \citetalias{Burrows+2025} and \citetalias{Janka&Kreese24} explore. Another possibility is that a highly magnetized millisecond proto-magnetar produces an energetic explosion, including sub-relativistic equatorial winds and relativistic, highly magnetized jets collimated by the former \citep{Prasanna+2023,Desai+2026}, prior to collapsing to a BH \citep[e.g.,][]{Margalit+22}. In this case the collapsar evolution and kick may be significantly modified by the disturbed star.

The absence of neutrino cooling affects the structure and dynamics of our accretion disks. In collapsars, neutrino emission can cool the dense inner disk and reduce its scale height, thereby modifying the hydrodynamic and EM angular-momentum fluxes through the horizon \citep{Issa2026}. The direct angular-momentum flux carried by neutrinos is small, but their cooling effect changes the disk thickness and hence the balance between accretion-driven spin-up and jet-driven spin-down torques. These structural changes may also affect the accumulation and eruption of magnetic flux in the MAD state. Lastly, because magneto-spin alignment is most effective for geometrically thick, radiatively inefficient disks \citep{magneto-spin}, thinner neutrino-cooled disks may wobble more readily even around rapidly spinning BHs. Neutrino cooling could therefore weaken, or at least modify, the spin dependence of the natal kicks found in our adiabatic simulations.

Our simulations are performed in a fixed Kerr spacetime, and therefore do not self-consistently evolve the BH mass, spin, or position in the progenitor frame. This limitation is especially important for the DH collapsar models, in which the accreted mass can exceed the initial BH mass. Dynamical BH growth and spin evolution could alter both the jet power and the geometry of the outflow. Angular momentum extraction by jets competes with spin-up by disk-fed accretion to regulate the BH spin \citep{jacquemin-ide_collapsar_2024}.  Because the BZ jet power depends sensitively on spin, accounting for BH spin-down (spin-up) would generally reduce (enhance) the momentum available to the ejecta and hence the kick. This caveat is particularly important for the strongly magnetized DH collapsar models, for which \citetalias{gottlieb_spinning_2025} found rapid spin-down to $a \approx 0$ in less than a second. Self-consistent evolution of the spin direction may also matter. In wobbling or tilted disks, the BH spin axis need not remain fixed and could contribute directly to a spin--orbit misalignment. The kick--spin angles reported here should thus be interpreted relative to the imposed initial spin axis.

We also neglect gas self-gravity. This can lead to an overestimate of the unbound mass and explosion energy by making the stellar envelope easier to eject. Including self-gravity would deepen the potential well of the star and may therefore reduce the asymptotic ejecta momentum and the resulting kick. At the same time, allowing the BH to move through the stellar potential would introduce dynamical friction and change the gravitational coupling between the BH and asymmetric ejecta. Fully dynamical-spacetime simulations including self-gravity will be needed to determine how these effects modify the late-time recoil.

\section{Conclusion}\label{sec:conclusion}
We have presented the first numerical study of BH natal kicks from magnetized collapsars using long-duration 3D GRMHD simulations. Our models follow rapidly rotating He cores collapsing onto central BHs with fixed Kerr spin parameters. Four simulations are ``low-mass'' collapsars with $M_\star = 14M_\sun$ and three simulations are massive direct-horizon collapsars, motivated by \citetalias{gottlieb_spinning_2025}, whose progenitor He-core masses lie above the PISN gap and can leave behind spinning BHs within the gap.

We summarize our main conclusions as follows.

\begin{itemize}
    \item \emph{Kick magnitudes:} Our lower-mass progenitors, corresponding roughly to ZAMS masses $\sim 25$--$40 M_\odot$, represent promising collapsar engines for long GRBs and their associated broad-lined Type Ic SNe. If the BH is slowly- or moderately-spinning, and is threaded by a strong magnetic field, the explosion energies are broadly consistent with those inferred for observed GRB/SN Ic-BL systems. These models can kick their BHs to speeds of $\mathcal{O}({\rm few}\times100\,{\rm km\,s^{-1}})$ when asymmetric jets and disk outflows unbind a large fraction of the stellar envelope, allowing the ejecta to carry net linear momentum to infinity. Magnetic-flux eruptions are expected in magnetized jet-launching disks and can contribute to the recoil by producing non-axisymmetric inner outflows; however, they affect the asymptotic natal kick only insofar as their momentum is communicated to material that ultimately escapes the star.
    
    \item \emph{Spin--kick correlations:} At a given progenitor magnetic-field structure, we find a positive association between the BH spin parameter $a$ and the estimated natal kick speed. This trend is partially self-regulated: higher spin increases the BZ jet power and explosion energy, but stronger jets also promote magneto-spin alignment of the accretion disk, reducing jet wobbling and the fractional ejecta momentum asymmetry. The same alignment tends to orient the kick more closely along the spin axis for higher-spinning BHs.
    
    \item \emph{Effects of magnetic field:} The BH kick depends not only on the amount of magnetic flux available to the BH, but also on how that flux is accumulated and reorganized in time. An extended sub-MAD phase and a subsequent transition to MAD produce a larger ejecta momentum asymmetry and the largest kick among our low-mass collapsars, despite having a lower time-averaged horizon flux and explosion energy than persistent MAD models. The subsequent MAD state launches more powerful ejection, which helps preserve the earlier recoil by expelling much of the remaining envelope before continued accretion can dilute the kick.
    
    \item \emph{Kick mechanisms:} Consistent with numerical studies of NS kicks \citep[e.g.,][]{scheck_multidimensional_2006,wongwathanarat_three-dimensional_2013,Janka&Kreese24}, the largest contribution to the recoil in most models comes from the gravitational interaction between the BH and the asymmetric stellar ejecta. Asymmetric, misaligned, and wobbling jets carve out a disordered ejecta distribution, producing a long-range tugboat acceleration toward the slower-moving ejecta. Momentum accretion through the BH horizon provides an additional contribution, dominated by asymmetric kinetic and electromagnetic stresses. The relative importance and orientation of these horizon-stress and gravitational contributions vary with the magnetization and geometry of the accretion flow.
    
    \item \emph{Massive DH collapsars}, which may yield BHs in the upper PISN mass gap with a range of spins \citepalias{gottlieb_spinning_2025}, also receive powerful natal kicks. If the magnetic field in the progenitor is moderate, a one-sided jet kicks the BH to hundreds of $\rm km\,s^{-1}$, misaligned with the BH spin. In a sufficiently close binary, such a kick can counteract the Blaauw expansion due to mass loss and reduce the merger time without disrupting the system, illustrating how the collapsar channel for producing rapidly spinning BHs in the PISN gap could remain viable. Stronger progenitor magnetic fields lead to considerably stronger kicks that would almost certainly disrupt any binary system or eject the BH from its host cluster. However, because \citetalias{gottlieb_spinning_2025} found that similarly strongly magnetized DH collapsars rapidly spin down their BHs, the largest kicks in our fixed-spin DH models should be interpreted as upper-end estimates.
\end{itemize}

The most astrophysically relevant kicks may be those produced by slowly and moderately spinning collapsars. Typical long-GRB luminosities favor low to moderate BH spins, $0.1 \lesssim a \lesssim 0.5$, depending on jet wobbling and radiative efficiency \citep{gottlieb_collapsar_2023}; moreover, highly spinning GRB engines may rapidly spin down through magnetic torques \citep{jacquemin-ide_collapsar_2024,Issa2026}. In this regime, comparatively weak jets may fail to maintain disk-spin alignment, allowing disk wobbling, jet obliquity, and jet-power asymmetry to produce kicks of several hundred $\rm km\,s^{-1}$. More broadly, however, the kick mechanism studied here need not be restricted to classical long GRBs. If a significant fraction of stripped-envelope SNe require central-engine activity, as argued from stripped-envelope SN light curves \citep{Afsariardchi2021,Sollerman2022,Rodriguez2024}, then asymmetric engine-driven outflows may accompany a broader class of massive stellar deaths. Such outflows may fail to sustain a relativistic GRB jet after breakout, yet expel enough stellar material asymmetrically to power luminous stripped-envelope SNe and impart substantial BH natal kicks.

Our simulations broadly suggest that BH kicks of several hundred $\rm km\,s^{-1}$ can be achieved when a powerful BH--disk system expels a large fraction of the stellar envelope with a few-percent net momentum asymmetry and sub-relativistic velocities. In our models this corresponds to jet powers $P_{\rm jet}\gtrsim10^{50}\,\rm erg\,s^{-1}$, but the outflow need not produce an observable GRB. Future work should explore a wider progenitor parameter space, including angular momentum, magnetic-field strength and geometry, compactness, neutrino cooling, self-gravity, and dynamical BH mass, spin, and position evolution. Such simulations will be needed to determine how often magnetized collapsars produce large BH natal kicks, and how those kicks shape BBH formation across different masses and spins.

\begin{acknowledgments}
We thank Yuri Levin, Chris Fragile, Danat Issa, Nicholas Kaaz, Yoonsoo Kim, and H. T. Janka for helpful conversations. We thank Alejandro Vigna-G\'omez for useful comments on our manuscript. S.E.L. and B.D.M. are supported in part by the National Science Foundation (grant AST-2406637), the Simons Foundation (grant 727700), and Columbia University through the Research Stabilization Fund. Computations for this work were done in part at facilities supported by the Scientific Computing Core (SCC) at the Flatiron Institute, which is supported by the Simons Foundation. S.E.L. thanks Yan-Fei Jiang for sponsoring his access to SCC resources. This research used resources of the Argonne Leadership Computing Facility, a U.S. Department of Energy (DOE) Office of Science user facility at Argonne National Laboratory and is based on research supported by the U.S. DOE Office of Science-Advanced Scientific Computing Research Program, under Contract No. DE-AC02-06CH11357 (NeutronStarRemnants project).
\end{acknowledgments}

\bibliography{references}{}

@article{Afsariardchi2021,
  author  = {Afsariardchi, Niloufar and Drout, Maria R. and Khatami, David K. and Matzner, Christopher D. and Moon, Dae-Sik and Ni, Yuan Qi},
  title   = {The Nickel Mass Distribution of Stripped-envelope Supernovae: Implications for Additional Power Sources},
  journal = {The Astrophysical Journal},
  year    = {2021},
  volume  = {918},
  pages   = {89},
  eprint  = {2009.06683},
  archivePrefix = {arXiv},
  primaryClass = {astro-ph.HE}
}

@article{Sollerman2022,
  author  = {Sollerman, J. and Yang, S. and Perley, D. and Schulze, S. and Fremling, C. and Kasliwal, M. and Shin, K. and Racine, B.},
  title   = {On the maximum luminosities of normal stripped-envelope supernovae: brighter than explosion models allow},
  journal = {Astronomy \& Astrophysics},
  year    = {2022},
  volume  = {657},
  pages   = {A64},
  doi     = {10.1051/0004-6361/202142049},
  eprint  = {2109.14339},
  archivePrefix = {arXiv},
  primaryClass = {astro-ph.SR}
}

@article{Rodriguez2024,
  author  = {Rodr{\'i}guez, {\'O}smar and Nakar, Ehud and Maoz, Dan},
  title   = {Stripped-envelope supernova light curves argue for central engine activity},
  journal = {Nature},
  year    = {2024},
  volume  = {628},
  pages   = {733--735},
  doi     = {10.1038/s41586-024-07262-x},
  eprint  = {2404.10846},
  archivePrefix = {arXiv},
  primaryClass = {astro-ph.HE}
}

@article{Mazzali14,
  author        = {Mazzali, P. and MacFadyen, A. and Woosley, S. and Pian, E. and Tanaka, M.},
  title         = {An upper limit to the energy of gamma-ray bursts indicates that GRB/SNe are powered by magnetars},
  journal       = {Monthly Notices of the Royal Astronomical Society},
  volume        = {443},
  number        = {1},
  pages         = {67--71},
  year          = {2014},
  doi           = {10.1093/mnras/stu1124},
  eprint        = {1406.1209},
  archivePrefix = {arXiv},
  primaryClass  = {astro-ph.HE}
}

@ARTICLE{Wanderman&Piran10,
       author = {{Wanderman}, David and {Piran}, Tsvi},
        title = "{The luminosity function and the rate of Swift's gamma-ray bursts}",
      journal = {\mnras},
         year = 2010,
        month = aug,
       volume = {406},
       number = {3},
        pages = {1944-1958},
          doi = {10.1111/j.1365-2966.2010.16787.x},
archivePrefix = {arXiv},
       eprint = {0912.0709},
 primaryClass = {astro-ph.HE},
       adsurl = {https://ui.adsabs.harvard.edu/abs/2010MNRAS.406.1944W}
}

@ARTICLE{Oppenheimer&Synder39,
       author = {{Oppenheimer}, J.~R. and {Snyder}, H.},
        title = "{On Continued Gravitational Contraction}",
      journal = {Physical Review},
         year = 1939,
        month = sep,
       volume = {56},
       number = {5},
        pages = {455-459},
          doi = {10.1103/PhysRev.56.455},
       adsurl = {https://ui.adsabs.harvard.edu/abs/1939PhRv...56..455O}
}

@article{Bini2026,
    author = "Bini, Sophie and Kr{\'o}l, Krzysztof and Chatziioannou, Katerina and Isi, Maximiliano",
    title = "{Impact of waveform systematics and Gaussian noise on the interpretation of GW231123}",
    eprint = "2601.09678",
    archivePrefix = "arXiv",
    primaryClass = "gr-qc",
    doi = "10.1103/gxjb-23lv",
    journal = "Phys. Rev. D",
    volume = "113",
    number = "8",
    pages = "083036",
    year = "2026"
}

@ARTICLE{Janka&Kreese24,
       author = {{Janka}, Hans-Thomas and {Kresse}, Daniel},
        title = "{Interplay between neutrino kicks and hydrodynamic kicks of neutron stars and black holes}",
      journal = {\apss},
         year = 2024,
        month = aug,
       volume = {369},
       number = {8},
          eid = {80},
        pages = {80},
          doi = {10.1007/s10509-024-04343-1},
archivePrefix = {arXiv},
       eprint = {2401.13817},
 primaryClass = {astro-ph.HE},
       adsurl = {https://ui.adsabs.harvard.edu/abs/2024Ap&SS.369...80J}
}

@ARTICLE{Janka25,
       author = {{Janka}, Hans-Thomas},
        title = "{Long-Term Multidimensional Models of Core-Collapse Supernovae: Progress and Challenges}",
      journal = {Annual Review of Nuclear and Particle Science},
         year = 2025,
        month = sep,
       volume = {75},
       number = {1},
        pages = {425-461},
          doi = {10.1146/annurev-nucl-121423-100945},
archivePrefix = {arXiv},
       eprint = {2502.14836},
 primaryClass = {astro-ph.HE},
       adsurl = {https://ui.adsabs.harvard.edu/abs/2025ARNPS..75..425J}
}

@ARTICLE{Woosley&Bloom06,
       author = {{Woosley}, S.~E. and {Bloom}, J.~S.},
        title = "{The Supernova Gamma-Ray Burst Connection}",
      journal = {\araa},
         year = 2006,
        month = sep,
       volume = {44},
       number = {1},
        pages = {507-556},
          doi = {10.1146/annurev.astro.43.072103.150558},
archivePrefix = {arXiv},
       eprint = {astro-ph/0609142},
 primaryClass = {astro-ph},
       adsurl = {https://ui.adsabs.harvard.edu/abs/2006ARA&A..44..507W}
}

@ARTICLE{Komissarov2009,
       author = {{Komissarov}, Serguei S. and {Barkov}, Maxim V.},
        title = "{Activation of the Blandford-Znajek mechanism in collapsing stars}",
      journal = {\mnras},
         year = 2009,
        month = aug,
       volume = {397},
       number = {3},
        pages = {1153-1168},
          doi = {10.1111/j.1365-2966.2009.14831.x},
archivePrefix = {arXiv},
       eprint = {0902.2881},
 primaryClass = {astro-ph.HE},
       adsurl = {https://ui.adsabs.harvard.edu/abs/2009MNRAS.397.1153K}
}

@ARTICLE{Issa2025,
       author = {{Issa}, Danat and {Gottlieb}, Ore and {Metzger}, Brian D. and {Jacquemin-Ide}, Jonatan and {Liska}, Matthew and {Foucart}, Francois and {Halevi}, Goni and {Tchekhovskoy}, Alexander},
        title = "{Magnetically Driven Neutron-rich Ejecta Unleashed: Global 3D Neutrino─General Relativistic Magnetohydrodynamic Simulations of Collapsars Probe the Conditions for r-process Nucleosynthesis}",
      journal = {\apjl},
         year = 2025,
        month = jun,
       volume = {985},
       number = {2},
          eid = {L26},
        pages = {L26},
          doi = {10.3847/2041-8213/adc694},
archivePrefix = {arXiv},
       eprint = {2410.02852},
 primaryClass = {astro-ph.HE},
       adsurl = {https://ui.adsabs.harvard.edu/abs/2025ApJ...985L..26I}
}

@ARTICLE{Antoni&Quataert23,
       author = {{Antoni}, Andrea and {Quataert}, Eliot},
        title = "{Numerical simulations of the random angular momentum in convection - II. Delayed explosions of red supergiants following 'failed' supernovae}",
      journal = {\mnras},
         year = 2023,
        month = oct,
       volume = {525},
       number = {1},
        pages = {1229-1245},
          doi = {10.1093/mnras/stad2328},
archivePrefix = {arXiv},
       eprint = {2301.05237},
 primaryClass = {astro-ph.HE},
       adsurl = {https://ui.adsabs.harvard.edu/abs/2023MNRAS.525.1229A}
}

@ARTICLE{Woosley&Heger06,
       author = {{Woosley}, S.~E. and {Heger}, A.},
        title = "{The Progenitor Stars of Gamma-Ray Bursts}",
      journal = {\apj},
         year = 2006,
        month = feb,
       volume = {637},
       number = {2},
        pages = {914-921},
          doi = {10.1086/498500},
archivePrefix = {arXiv},
       eprint = {astro-ph/0508175},
 primaryClass = {astro-ph},
       adsurl = {https://ui.adsabs.harvard.edu/abs/2006ApJ...637..914W}
}

@ARTICLE{Siegel+22,
       author = {{Siegel}, Daniel M. and {Agarwal}, Aman and {Barnes}, Jennifer and {Metzger}, Brian D. and {Renzo}, Mathieu and {Villar}, V. Ashley},
        title = "{``Super-kilonovae'' from Massive Collapsars as Signatures of Black Hole Birth in the Pair-instability Mass Gap}",
      journal = {\apj},
         year = 2022,
        month = dec,
       volume = {941},
       number = {1},
          eid = {100},
        pages = {100},
          doi = {10.3847/1538-4357/ac8d04},
archivePrefix = {arXiv},
       eprint = {2111.03094},
 primaryClass = {astro-ph.HE},
       adsurl = {https://ui.adsabs.harvard.edu/abs/2022ApJ...941..100S}
}

@ARTICLE{Mandel2016,
       author = {{Mandel}, Ilya},
        title = "{Estimates of black hole natal kick velocities from observations of low-mass X-ray binaries}",
      journal = {\mnras},
         year = 2016,
        month = feb,
       volume = {456},
       number = {1},
        pages = {578-581},
          doi = {10.1093/mnras/stv2733},
archivePrefix = {arXiv},
       eprint = {1510.03871},
 primaryClass = {astro-ph.HE},
       adsurl = {https://ui.adsabs.harvard.edu/abs/2016MNRAS.456..578M}
}

@ARTICLE{Brandt&Podsiadlowski1995,
       author = {{Brandt}, Niel and {Podsiadlowski}, Philipp},
        title = "{The effects of high-velocity supernova kicks on the orbital properties and sky distributions of neutron-star binaries}",
      journal = {\mnras},
         year = 1995,
        month = may,
       volume = {274},
       number = {2},
        pages = {461-484},
          doi = {10.1093/mnras/274.2.461},
       adsurl = {https://ui.adsabs.harvard.edu/abs/1995MNRAS.274..461B}
}

@ARTICLE{Vigna-Gomez&Ramirez-Ruiz2023,
       author = {{Vigna-G{\'o}mez}, Alejandro and {Ramirez-Ruiz}, Enrico},
        title = "{A Binary Origin for the First Isolated Stellar-mass Black Hole Detected with Astrometric Microlensing}",
      journal = {\apjl},
         year = 2023,
        month = mar,
       volume = {946},
       number = {1},
          eid = {L2},
        pages = {L2},
          doi = {10.3847/2041-8213/acc076},
archivePrefix = {arXiv},
       eprint = {2203.08478},
 primaryClass = {astro-ph.GA},
       adsurl = {https://ui.adsabs.harvard.edu/abs/2023ApJ...946L...2V}
}

@ARTICLE{Vigna-Gomez+2024,
       author = {{Vigna-G{\'o}mez}, Alejandro and {Willcox}, Reinhold and {Tamborra}, Irene and {Mandel}, Ilya and {Renzo}, Mathieu and {Wagg}, Tom and {Janka}, Hans-Thomas and {Kresse}, Daniel and {Bodensteiner}, Julia and {Shenar}, Tomer and et al.},
        title = "{Constraints on Neutrino Natal Kicks from Black-Hole Binary VFTS 243}",
      journal = {\prl},
         year = 2024,
        month = may,
       volume = {132},
       number = {19},
          eid = {191403},
        pages = {191403},
          doi = {10.1103/PhysRevLett.132.191403},
archivePrefix = {arXiv},
       eprint = {2310.01509},
 primaryClass = {astro-ph.HE},
       adsurl = {https://ui.adsabs.harvard.edu/abs/2024PhRvL.132s1403V}
}

@ARTICLE{Lam+2023,
       author = {{Lam}, Casey Y. and {Lu}, Jessica R.},
        title = "{A Reanalysis of the Isolated Black Hole Candidate OGLE-2011-BLG-0462/MOA-2011-BLG-191}",
      journal = {\apj},
         year = 2023,
        month = oct,
       volume = {955},
       number = {2},
          eid = {116},
        pages = {116},
          doi = {10.3847/1538-4357/aced4a},
archivePrefix = {arXiv},
       eprint = {2308.03302},
 primaryClass = {astro-ph.SR},
       adsurl = {https://ui.adsabs.harvard.edu/abs/2023ApJ...955..116L}
}

@ARTICLE{Wyrzykowski+2020,
       author = {{Wyrzykowski}, {\L}ukasz and {Mandel}, Ilya},
        title = "{Constraining the masses of microlensing black holes and the mass gap with Gaia DR2}",
      journal = {\aap},
         year = 2020,
        month = apr,
       volume = {636},
          eid = {A20},
        pages = {A20},
          doi = {10.1051/0004-6361/201935842},
archivePrefix = {arXiv},
       eprint = {1904.07789},
 primaryClass = {astro-ph.SR},
       adsurl = {https://ui.adsabs.harvard.edu/abs/2020A&A...636A..20W}
}

@ARTICLE{LVK2026,
       author = {{LIGO--Virgo--KAGRA Collaboration}},
        title = "{GWTC-5.0: Population Properties of Merging Compact Binaries}",
      journal = {arXiv e-prints},
         year = 2026,
        month = may,
          eid = {arXiv:2605.27226},
        pages = {arXiv:2605.27226},
          doi = {10.48550/arXiv.2605.27226},
archivePrefix = {arXiv},
       eprint = {2605.27226},
 primaryClass = {astro-ph.HE},
       adsurl = {https://ui.adsabs.harvard.edu/abs/2026arXiv260527226T}
}

@ARTICLE{Vivanco+2026,
       author = {{Vivanco C{\'a}diz}, Felipe and {Artale}, M. Celeste and {Masetti}, Nicola and {Escobar}, Gaston and {Iorio}, Giuliano and {Kumar}, Ankit and {Liu}, Boyuan and {Mapelli}, Michela},
        title = "{Linking high-mass X-ray binaries to binary compact object mergers in Milky Way-like galaxies}",
      journal = {arXiv e-prints},
         year = 2026,
        month = aug,
          eid = {arXiv:2608.23945},
        pages = {arXiv:2608.23945},
          doi = {10.48550/arXiv.2608.23945},
archivePrefix = {arXiv},
       eprint = {2608.23945},
 primaryClass = {astro-ph.HE},
       adsurl = {https://ui.adsabs.harvard.edu/abs/2026arXiv260823945V}
}

@ARTICLE{Prasanna+2023,
       author = {{Prasanna}, Tejas and {Coleman}, Matthew S.~B. and {Raives}, Matthias J. and {Thompson}, Todd A.},
        title = "{The early evolution of magnetar rotation - II. Rapidly rotating magnetars: implications for gamma-ray bursts and superluminous supernovae}",
      journal = {\mnras},
         year = 2023,
        month = dec,
       volume = {526},
       number = {2},
        pages = {3141-3155},
          doi = {10.1093/mnras/stad2948},
archivePrefix = {arXiv},
       eprint = {2305.16412},
 primaryClass = {astro-ph.HE},
       adsurl = {https://ui.adsabs.harvard.edu/abs/2023MNRAS.526.3141P}
}

@ARTICLE{DashwoodBrown2024,
       author = {{Dashwood Brown}, Cordelia and {Gandhi}, Poshak and {Zhao}, Yue},
        title = "{On the natal kick of the black hole X-ray binary H 1705-250}",
      journal = {\mnras},
         year = 2024,
        month = jan,
       volume = {527},
       number = {1},
        pages = {L82-L87},
          doi = {10.1093/mnrasl/slad151},
archivePrefix = {arXiv},
       eprint = {2310.11492},
 primaryClass = {astro-ph.HE},
       adsurl = {https://ui.adsabs.harvard.edu/abs/2024MNRAS.527L..82D}
}

@ARTICLE{Burdge_2024,
       author = {{Burdge}, Kevin B. and {El-Badry}, Kareem and {Kara}, Erin and {Canizares}, Claude and {Chakrabarty}, Deepto and {Frebel}, Anna and {Millholland}, Sarah C. and {Rappaport}, Saul and {Simcoe}, Rob and {Vanderburg}, Andrew},
        title = "{The black hole low-mass X-ray binary V404 Cygni is part of a wide triple}",
      journal = {\nat},
         year = 2024,
        month = nov,
       volume = {635},
       number = {8038},
        pages = {316-320},
          doi = {10.1038/s41586-024-08120-6},
archivePrefix = {arXiv},
       eprint = {2404.03719},
 primaryClass = {astro-ph.HE},
       adsurl = {https://ui.adsabs.harvard.edu/abs/2024Natur.635..316B}
}

@ARTICLE{Shenar+2022,
       author = {{Shenar}, Tomer and {Sana}, Hugues and {Mahy}, Laurent and {El-Badry}, Kareem and {Marchant}, Pablo and {Langer}, Norbert and {Hawcroft}, Calum and {Fabry}, Matthias and {Sen}, Koushik and {Almeida}, Leonardo A. and {Abdul-Masih}, Michael and {Bodensteiner}, Julia and {Crowther}, Paul A. and {Gieles}, Mark and {Gromadzki}, Mariusz and {H{\'e}nault-Brunet}, Vincent and {Herrero}, Artemio and {de Koter}, Alex and {Iwanek}, Patryk and {Koz{\l}owski}, Szymon and {Lennon}, Daniel J. and {Ma{\'\i}z Apell{\'a}niz}, Jes{\'u}s and {Mr{\'o}z}, Przemys{\l}aw and {Moffat}, Anthony F.~J. and {Picco}, Annachiara and {Pietrukowicz}, Pawe{\l} and {Poleski}, Rados{\l}aw and {Rybicki}, Krzysztof and {Schneider}, Fabian R.~N. and {Skowron}, Dorota M. and {Skowron}, Jan and {Soszy{\'n}ski}, Igor and {Szyma{\'n}ski}, Micha{\l} K. and {Toonen}, Silvia and {Udalski}, Andrzej and {Ulaczyk}, Krzysztof and {Vink}, Jorick S. and {Wrona}, Marcin},
        title = "{An X-ray-quiet black hole born with a negligible kick in a massive binary within the Large Magellanic Cloud}",
      journal = {Nature Astronomy},
         year = 2022,
        month = jul,
       volume = {6},
        pages = {1085-1092},
          doi = {10.1038/s41550-022-01730-y},
archivePrefix = {arXiv},
       eprint = {2207.07675},
 primaryClass = {astro-ph.HE},
       adsurl = {https://ui.adsabs.harvard.edu/abs/2022NatAs...6.1085S}
}

@ARTICLE{Desai+2026,
       author = {{Desai}, Dhruv K. and {Combi}, Luciano and {Siegel}, Daniel M. and {Metzger}, Brian D.},
        title = "{Relativistic jets from millisecond proto-magnetars}",
      journal = {arXiv e-prints},
         year = 2026,
        month = jan,
          eid = {arXiv:2601.07918},
        pages = {arXiv:2601.07918},
          doi = {10.48550/arXiv.2601.07918},
archivePrefix = {arXiv},
       eprint = {2601.07918},
 primaryClass = {astro-ph.HE},
       adsurl = {https://ui.adsabs.harvard.edu/abs/2026arXiv260107918D}
}

@ARTICLE{Margalit+22,
       author = {{Margalit}, Ben and {Jermyn}, Adam S. and {Metzger}, Brian D. and {Roberts}, Luke F. and {Quataert}, Eliot},
        title = "{Angular-momentum Transport in Proto-neutron Stars and the Fate of Neutron Star Merger Remnants}",
      journal = {\apj},
         year = 2022,
        month = nov,
       volume = {939},
       number = {1},
          eid = {51},
        pages = {51},
          doi = {10.3847/1538-4357/ac8b01},
archivePrefix = {arXiv},
       eprint = {2206.10645},
 primaryClass = {astro-ph.HE},
       adsurl = {https://ui.adsabs.harvard.edu/abs/2022ApJ...939...51M}
}

@article{woosley_gamma-ray_1993,
	title = {Gamma-ray bursts from stellar mass accretion disks around black holes},
	volume = {405},
	issn = {0004-637X, 1538-4357},
	url = {http://adsabs.harvard.edu/doi/10.1086/172359},
	doi = {10.1086/172359},
	language = {en},
	urldate = {2025-06-09},
	journal = {ApJ},
	author = {Woosley, S. E.},
	month = mar,
	year = {1993},
	pages = {273},
}

@ARTICLE{Sana_2012,
       author = {{Sana}, H. and {de Mink}, S.~E. and {de Koter}, A. and {Langer}, N. and {Evans}, C.~J. and {Gieles}, M. and {Gosset}, E. and {Izzard}, R.~G. and {Le Bouquin}, J.-B. and {Schneider}, F.~R.~N.},
        title = "{Binary Interaction Dominates the Evolution of Massive Stars}",
      journal = {Science},
         year = 2012,
        month = jul,
       volume = {337},
       number = {6093},
        pages = {444},
          doi = {10.1126/science.1223344},
archivePrefix = {arXiv},
       eprint = {1207.6397},
 primaryClass = {astro-ph.SR},
       adsurl = {https://ui.adsabs.harvard.edu/abs/2012Sci...337..444S}
}

@article{Chan:2017tdg,
    author = {Chan, Conrad and M{\"u}ller, Bernhard and Heger, Alexander and Pakmor, R{\"u}diger and Springel, Volker},
    title = "{Black hole formation and fallback during the supernova explosion of a $40 \,\mathrm{M}_\odot$ star}",
    eprint = "1710.00838",
    archivePrefix = "arXiv",
    primaryClass = "astro-ph.SR",
    doi = "10.3847/2041-8213/aaa28c",
    journal = "Astrophys. J. Lett.",
    volume = "852",
    number = "1",
    pages = "L19",
    year = "2018"
}

@ARTICLE{Tchekhovskoy2011,
       author = {{Tchekhovskoy}, Alexander and {Narayan}, Ramesh and {McKinney}, Jonathan C.},
        title = "{Efficient generation of jets from magnetically arrested accretion on a rapidly spinning black hole}",
      journal = {\mnras},
         year = 2011,
        month = nov,
       volume = {418},
       number = {1},
        pages = {L79-L83},
          doi = {10.1111/j.1745-3933.2011.01147.x},
archivePrefix = {arXiv},
       eprint = {1108.0412},
 primaryClass = {astro-ph.HE},
       adsurl = {https://ui.adsabs.harvard.edu/abs/2011MNRAS.418L..79T}
}

@ARTICLE{Gottlieb2024,
       author = {{Gottlieb}, Ore and {Renzo}, Mathieu and {Metzger}, Brian D. and {Goldberg}, Jared A. and {Cantiello}, Matteo},
        title = "{She's Got Her Mother's Hair: Unveiling the Origin of Black Hole Magnetic Fields through Stellar to Collapsar Simulations}",
      journal = {\apjl},
         year = 2024,
        month = nov,
       volume = {976},
       number = {1},
          eid = {L13},
        pages = {L13},
          doi = {10.3847/2041-8213/ad8563},
archivePrefix = {arXiv},
       eprint = {2407.16745},
 primaryClass = {astro-ph.HE},
       adsurl = {https://ui.adsabs.harvard.edu/abs/2024ApJ...976L..13G}
}

@ARTICLE{Draghis+2024,
       author = {{Draghis}, Paul A. and {Miller}, Jon M. and {Costantini}, Elisa and {Gallo}, Luigi C. and {Reynolds}, Mark and {Tomsick}, John A. and {Zoghbi}, Abderahmen},
        title = "{Systematically Revisiting All NuSTAR Spins of Black Holes in X-Ray Binaries}",
      journal = {\apj},
         year = 2024,
        month = jul,
       volume = {969},
       number = {1},
          eid = {40},
        pages = {40},
          doi = {10.3847/1538-4357/ad43ea},
archivePrefix = {arXiv},
       eprint = {2311.16225},
 primaryClass = {astro-ph.HE},
       adsurl = {https://ui.adsabs.harvard.edu/abs/2024ApJ...969...40D}
}

@article{narayan_magnetically_2003,
	title = {Magnetically {Arrested} {Disk}: an {Energetically} {Efficient} {Accretion} {Flow}},
	volume = {55},
	issn = {0004-6264, 2053-051X},
	shorttitle = {Magnetically {Arrested} {Disk}},
	url = {https://academic.oup.com/pasj/article/55/6/L69/2055460},
	doi = {10.1093/pasj/55.6.L69},
	language = {en},
	number = {6},
	urldate = {2025-01-27},
	journal = {PASJ},
	author = {Narayan, Ramesh and Igumenshchev, Igor V. and Abramowicz, Marek A.},
	month = dec,
	year = {2003},
	pages = {L69--L72},
}

@article{atri_potential_2019,
	title = {Potential kick velocity distribution of black hole {X}-ray binaries and implications for natal kicks},
	volume = {489},
	copyright = {https://academic.oup.com/journals/pages/open\_access/funder\_policies/chorus/standard\_publication\_model},
	issn = {0035-8711, 1365-2966},
	url = {https://academic.oup.com/mnras/article/489/3/3116/5556959},
	doi = {10.1093/mnras/stz2335},
	language = {en},
	number = {3},
	urldate = {2024-09-04},
	journal = {Monthly Notices of the Royal Astronomical Society},
	author = {Atri, P and Miller-Jones, J C A and Bahramian, A and Plotkin, R M and Jonker, P G and Nelemans, G and Maccarone, T J and Sivakoff, G R and Deller, A T and Chaty, S and Torres, M A P and Horiuchi, S and McCallum, J and Natusch, T and Phillips, C J and Stevens, J and Weston, S},
	month = nov,
	year = {2019},
	pages = {3116--3134},
}

@ARTICLE{Chan2026,
       author = {{Chan}, Ho-Sang and {Gottlieb}, Ore and {Jacquemin-Ide}, Jonatan and {Cantiello}, Matteo and {Renzo}, Mathieu},
        title = "{Jets from Scratch: A 3D Dynamo Origin of Long Gamma-Ray Burst Jets}",
      journal = {arXiv e-prints},
         year = 2026,
        month = may,
          eid = {arXiv:2605.12931},
        pages = {arXiv:2605.12931},
          doi = {10.48550/arXiv.2605.12931},
archivePrefix = {arXiv},
       eprint = {2605.12931},
 primaryClass = {astro-ph.HE},
       adsurl = {https://ui.adsabs.harvard.edu/abs/2026arXiv260512931C}
}

@article{Ka2015,
  title = {Properties of hypermassive neutron stars formed in mergers of spinning binaries},
  author = {Kastaun, Wolfgang and Galeazzi, Filippo},
  journal = {Phys. Rev. D},
  volume = {91},
  issue = {6},
  pages = {064027},
  numpages = {17},
  year = {2015},
  month = {Mar},
  publisher = {American Physical Society},
  doi = {10.1103/PhysRevD.91.064027},
  url = {https://link.aps.org/doi/10.1103/PhysRevD.91.064027}
}

@article{Vi+2020,
  title = {Unequal mass binary neutron star simulations with neutrino transport: Ejecta and neutrino emission},
  author = {Vincent, Trevor and Foucart, Francois and Duez, Matthew D. and Haas, Roland and Kidder, Lawrence E. and Pfeiffer, Harald P. and Scheel, Mark A.},
  journal = {Phys. Rev. D},
  volume = {101},
  issue = {4},
  pages = {044053},
  numpages = {21},
  year = {2020},
  month = {Feb},
  publisher = {American Physical Society},
  doi = {10.1103/PhysRevD.101.044053},
  url = {https://link.aps.org/doi/10.1103/PhysRevD.101.044053}
}

@article{Janka2017,
	title = {Neutron {Star} {Kicks} by the {Gravitational} {Tug}-boat {Mechanism} in {Asymmetric} {Supernova} {Explosions}: {Progenitor} and {Explosion} {Dependence}},
	volume = {837},
	issn = {0004-637X, 1538-4357},
	shorttitle = {Neutron {Star} {Kicks} by the {Gravitational} {Tug}-boat {Mechanism} in {Asymmetric} {Supernova} {Explosions}},
	url = {https://iopscience.iop.org/article/10.3847/1538-4357/aa618e},
	doi = {10.3847/1538-4357/aa618e},
	language = {en},
	number = {1},
	urldate = {2024-05-15},
	journal = {The Astrophysical Journal},
	author = {Janka, Hans-Thomas},
	month = mar,
	year = {2017},
	pages = {84},
}

@ARTICLE{Issa2026,
       author = {{Issa}, Danat and {Lowell}, Beverly and {Jacquemin-Ide}, Jonatan and {Liska}, Matthew and {Tchekhovskoy}, Alexander},
        title = "{Collapsar black hole spin evolution in 3D neutrino transport GRMHD simulations}",
      journal = {\prd},
         year = 2026,
        month = apr,
       volume = {113},
       number = {8},
          eid = {083020},
        pages = {083020},
          doi = {10.1103/fzhd-hcpp},
archivePrefix = {arXiv},
       eprint = {2502.08732},
 primaryClass = {astro-ph.HE},
       adsurl = {https://ui.adsabs.harvard.edu/abs/2026PhRvD.113h3020I}
}

@article{lowell_rapid_2024,
	title = {Rapid {Black} {Hole} {Spin}-down by {Thick} {Magnetically} {Arrested} {Disks}},
	volume = {960},
	issn = {0004-637X, 1538-4357},
	url = {https://iopscience.iop.org/article/10.3847/1538-4357/ad09af},
	doi = {10.3847/1538-4357/ad09af},
	language = {en},
	number = {1},
	urldate = {2025-06-24},
	journal = {The Astrophysical Journal},
	author = {Lowell, Beverly and Jacquemin-Ide, Jonatan and Tchekhovskoy, Alexander and Duncan, Alex},
	month = jan,
	year = {2024},
	pages = {82},
}

@ARTICLE{Hills1983,
       author = {{Hills}, J.~G.},
        title = "{The effects of sudden mass loss and a random kick velocity produced in a supernova explosion on the dynamics of a binary star of arbitrary orbital eccentricity. Applications to X-ray binaries and to the binarypulsars.}",
      journal = {\apj},
         year = 1983,
        month = apr,
       volume = {267},
        pages = {322-333},
          doi = {10.1086/160871},
       adsurl = {https://ui.adsabs.harvard.edu/abs/1983ApJ...267..322H}
}

@article{oshaughnessy_inferences_2017,
	title = {Inferences about {Supernova} {Physics} from {Gravitational}-{Wave} {Measurements}: {GW151226} {Spin} {Misalignment} as an {Indicator} of {Strong} {Black}-{Hole} {Natal} {Kicks}},
	volume = {119},
	copyright = {http://link.aps.org/licenses/aps-default-license},
	issn = {0031-9007, 1079-7114},
	shorttitle = {Inferences about {Supernova} {Physics} from {Gravitational}-{Wave} {Measurements}},
	url = {http://link.aps.org/doi/10.1103/PhysRevLett.119.011101},
	doi = {10.1103/PhysRevLett.119.011101},
	language = {en},
	number = {1},
	urldate = {2024-09-04},
	journal = {Phys. Rev. Lett.},
	author = {O’Shaughnessy, Richard and Gerosa, Davide and Wysocki, Daniel},
	month = jul,
	year = {2017},
	pages = {011101},
}

@article{andrews_constraining_2022,
	title = {Constraining {Black} {Hole} {Natal} {Kicks} with {Astrometric} {Microlensing}},
	volume = {930},
	issn = {0004-637X, 1538-4357},
	url = {https://iopscience.iop.org/article/10.3847/1538-4357/ac66d6},
	doi = {10.3847/1538-4357/ac66d6},
	language = {en},
	number = {2},
	urldate = {2025-05-21},
	journal = {ApJ},
	author = {Andrews, Jeff J. and Kalogera, Vicky},
	month = may,
	year = {2022},
	pages = {159},
}

@article{jacquemin-ide_magnetorotational_2024,
	title = {Magnetorotational dynamo can generate large-scale vertical magnetic fields in {3D} {GRMHD} simulations of accreting black holes},
	volume = {532},
	copyright = {https://creativecommons.org/licenses/by/4.0/},
	issn = {0035-8711, 1365-2966},
	url = {https://academic.oup.com/mnras/article/532/2/1522/7701801},
	doi = {10.1093/mnras/stae1538},
	language = {en},
	number = {2},
	urldate = {2025-07-23},
	journal = {Monthly Notices of the Royal Astronomical Society},
	author = {Jacquemin-Ide, Jonatan and Rincon, François and Tchekhovskoy, Alexander and Liska, Matthew},
	month = jul,
	year = {2024},
	note = {Publisher: Oxford University Press (OUP)},
	pages = {1522--1545},
}

@ARTICLE{stockinger2020,
       author = {{Stockinger}, G. and {Janka}, H.-T. and {Kresse}, D. and {Melson}, T. and {Ertl}, T. and {Gabler}, M. and {Gessner}, A. and {Wongwathanarat}, A. and {Tolstov}, A. and {Leung}, S.-C. and {Nomoto}, K. and {Heger}, A.},
        title = "{Three-dimensional models of core-collapse supernovae from low-mass progenitors with implications for Crab}",
      journal = {\mnras},
         year = 2020,
        month = aug,
       volume = {496},
       number = {2},
        pages = {2039-2084},
          doi = {10.1093/mnras/staa1691},
archivePrefix = {arXiv},
       eprint = {2005.02420},
 primaryClass = {astro-ph.HE},
       adsurl = {https://ui.adsabs.harvard.edu/abs/2020MNRAS.496.2039S}
}

@article{nagarajan_mixed_2025,
	title = {Mixed {Origins}: {Strong} {Natal} {Kicks} for {Some} {Black} {Holes} and {None} for {Others}},
	volume = {137},
	issn = {0004-6280, 1538-3873},
	shorttitle = {Mixed {Origins}},
	url = {https://iopscience.iop.org/article/10.1088/1538-3873/adb6d6},
	doi = {10.1088/1538-3873/adb6d6},
	language = {en},
	number = {3},
	urldate = {2025-05-21},
	journal = {PASP},
	author = {Nagarajan, Pranav and El-Badry, Kareem},
	month = mar,
	year = {2025},
	pages = {034203},
}

@article{macfadyen_collapsars_1999,
	title = {Collapsars: {Gamma}‐{Ray} {Bursts} and {Explosions} in “{Failed} {Supernovae}”},
	volume = {524},
	issn = {0004-637X, 1538-4357},
	shorttitle = {Collapsars},
	url = {https://iopscience.iop.org/article/10.1086/307790},
	doi = {10.1086/307790},
	language = {en},
	number = {1},
	urldate = {2025-06-09},
	journal = {ApJ},
	author = {MacFadyen, A. I. and Woosley, S. E.},
	month = oct,
	year = {1999},
	pages = {262--289},
}

@ARTICLE{Gottlieb2022b,
       author = {{Gottlieb}, Ore and {Liska}, Matthew and {Tchekhovskoy}, Alexander and {Bromberg}, Omer and {Lalakos}, Aretaios and {Giannios}, Dimitrios and {M{\"o}sta}, Philipp},
        title = "{Black Hole to Photosphere: 3D GRMHD Simulations of Collapsars Reveal Wobbling and Hybrid Composition Jets}",
      journal = {\apjl},
         year = 2022,
        month = jul,
       volume = {933},
       number = {1},
          eid = {L9},
        pages = {L9},
          doi = {10.3847/2041-8213/ac7530},
archivePrefix = {arXiv},
       eprint = {2204.12501},
 primaryClass = {astro-ph.HE},
       adsurl = {https://ui.adsabs.harvard.edu/abs/2022ApJ...933L...9G}
}

@ARTICLE{Burrows+2025,
       author = {{Burrows}, Adam and {Wang}, Tianshu and {Vartanyan}, David},
        title = "{Channels of Stellar-mass Black Hole Formation}",
      journal = {\apj},
         year = 2025,
        month = jul,
       volume = {987},
       number = {2},
          eid = {164},
        pages = {164},
          doi = {10.3847/1538-4357/addd04},
archivePrefix = {arXiv},
       eprint = {2412.07831},
 primaryClass = {astro-ph.SR},
       adsurl = {https://ui.adsabs.harvard.edu/abs/2025ApJ...987..164B}
}

@article{burrows_black_2023,
	title = {Black {Hole} {Formation} {Accompanied} by the {Supernova} {Explosion} of a 40 {M}$_{\textrm{⊙}}$ {Progenitor} {Star}},
	volume = {957},
	issn = {0004-637X, 1538-4357},
	url = {https://iopscience.iop.org/article/10.3847/1538-4357/acfc1c},
	doi = {10.3847/1538-4357/acfc1c},
	language = {en},
	number = {2},
	urldate = {2024-05-15},
	journal = {ApJ},
	author = {Burrows, Adam and Vartanyan, David and Wang, Tianshu},
	month = nov,
	year = {2023},
	pages = {68},
}

@article{gottlieb_spinning_2025,
	title = {Spinning into the {Gap}: {Direct}-horizon {Collapse} as the {Origin} of {GW231123} from {End}-to-end {General}-relativistic {Magnetohydrodynamic} {Simulations}},
	volume = {993},
	issn = {2041-8205},
	shorttitle = {Spinning into the {Gap}},
	url = {https://doi.org/10.3847/2041-8213/ae0d81},
	doi = {10.3847/2041-8213/ae0d81},
	language = {en},
	number = {2},
	urldate = {2025-11-12},
	journal = {The Astrophysical Journal Letters},
	author = {Gottlieb, Ore and Metzger, Brian D. and Issa, Danat and Li, Sean E. and Renzo, Mathieu and Isi, Maximiliano},
	month = nov,
	year = {2025},
	note = {Publisher: The American Astronomical Society},
	pages = {L54},
}

@ARTICLE{Antonini+16,
       author = {{Antonini}, Fabio and {Rasio}, Frederic A.},
        title = "{Merging Black Hole Binaries in Galactic Nuclei: Implications for Advanced-LIGO Detections}",
      journal = {\apj},
         year = 2016,
        month = nov,
       volume = {831},
       number = {2},
          eid = {187},
        pages = {187},
          doi = {10.3847/0004-637X/831/2/187},
archivePrefix = {arXiv},
       eprint = {1606.04889},
 primaryClass = {astro-ph.HE},
       adsurl = {https://ui.adsabs.harvard.edu/abs/2016ApJ...831..187A}
}

@ARTICLE{Sgalletta+2025,
       author = {{Sgalletta}, Cecilia and {Mapelli}, Michela and {Boco}, Lumen and {Santoliquido}, Filippo and {Artale}, M. Celeste and {Iorio}, Giuliano and {Lapi}, Andrea and {Spera}, Mario},
        title = "{The more accurately the metal-dependent star formation rate is modeled, the larger the predicted excess of binary black hole mergers}",
      journal = {\aap},
         year = 2025,
        month = jun,
       volume = {698},
          eid = {A144},
        pages = {A144},
          doi = {10.1051/0004-6361/202452757},
archivePrefix = {arXiv},
       eprint = {2410.21401},
 primaryClass = {astro-ph.HE},
       adsurl = {https://ui.adsabs.harvard.edu/abs/2025A&A...698A.144S}
}

@ARTICLE{Mapelli+2026,
       author = {{Mapelli}, Michela and {Sgalletta}, Cecilia and {M{\"u}ller-Horn}, Johanna and {Iorio}, Giuliano and {Rinaldi}, Stefano and {Burt}, Christian and {Mar{\'\i}n Pina}, Daniel and {Romagnolo}, Amedeo},
        title = "{The role of accretion efficiency, natal kicks, and angular momentum transport in the formation of the Gaia black holes}",
      journal = {arXiv e-prints},
         year = 2026,
        month = apr,
          eid = {arXiv:2604.12839},
        pages = {arXiv:2604.12839},
          doi = {10.48550/arXiv.2604.12839},
archivePrefix = {arXiv},
       eprint = {2604.12839},
 primaryClass = {astro-ph.HE},
       adsurl = {https://ui.adsabs.harvard.edu/abs/2026arXiv260412839M}
}

@ARTICLE{Lam+2026,
       author = {{Lam}, Casey Y. and {Lu}, Jessica R. and {El-Badry}, Kareem and {Seth}, Anil and {Simon}, Joshua D. and {Abrams}, Natasha S. and {Bellini}, Andrea and {Hosek}, Jr., Matthew W. and {McGill}, Peter},
        title = "{Black hole astrometric binaries in the Roman Galactic Bulge Time Domain Survey}",
      journal = {arXiv e-prints},
         year = 2026,
        month = aug,
          eid = {arXiv:2608.24998},
        pages = {arXiv:2608.24998},
          doi = {10.48550/arXiv.2608.24998},
archivePrefix = {arXiv},
       eprint = {2608.24998},
 primaryClass = {astro-ph.SR},
       adsurl = {https://ui.adsabs.harvard.edu/abs/2026arXiv260824998L}
}

@ARTICLE{Shibata2025,
       author = {{Shibata}, Masaru and {Fujibayashi}, Sho and {Wanajo}, Shinya and {Ioka}, Kunihito and {Lam}, Alan Tsz-Lok and {Sekiguchi}, Yuichiro},
        title = "{Self-consistent scenario for jet and stellar explosions in collapsar: General relativistic magnetohydrodynamics simulation with a dynamo}",
      journal = {\prd},
         year = 2025,
        month = jun,
       volume = {111},
       number = {12},
          eid = {123017},
        pages = {123017},
          doi = {10.1103/msy2-fwhx},
archivePrefix = {arXiv},
       eprint = {2502.02077},
 primaryClass = {astro-ph.HE},
       adsurl = {https://ui.adsabs.harvard.edu/abs/2025PhRvD.111l3017S}
}

@ARTICLE{Nordhaus2010,
       author = {{Nordhaus}, J. and {Brandt}, T.~D. and {Burrows}, A. and {Livne}, E. and {Ott}, C.~D.},
        title = "{Theoretical support for the hydrodynamic mechanism of pulsar kicks}",
      journal = {\prd},
         year = 2010,
        month = nov,
       volume = {82},
       number = {10},
          eid = {103016},
        pages = {103016},
          doi = {10.1103/PhysRevD.82.103016},
archivePrefix = {arXiv},
       eprint = {1010.0674},
 primaryClass = {astro-ph.HE},
       adsurl = {https://ui.adsabs.harvard.edu/abs/2010PhRvD..82j3016N}
}

@ARTICLE{Bopp2025,
       author = {{Bopp}, Justin and {Gottlieb}, Ore},
        title = "{Fast Transients from Magnetic Disks around Nonspinning Collapsar Black Holes}",
      journal = {\apjl},
         year = 2025,
        month = apr,
       volume = {982},
       number = {2},
          eid = {L56},
        pages = {L56},
          doi = {10.3847/2041-8213/adbdcd},
archivePrefix = {arXiv},
       eprint = {2410.22401},
 primaryClass = {astro-ph.HE},
       adsurl = {https://ui.adsabs.harvard.edu/abs/2025ApJ...982L..56B}
}

@ARTICLE{Gottlieb2024b,
       author = {{Gottlieb}, Ore and {Levinson}, Amir and {Levin}, Yuri},
        title = "{In LIGO's Sight? Vigorous Coherent Gravitational Waves from Cooled Collapsar Disks}",
      journal = {\apjl},
         year = 2024,
        month = sep,
       volume = {972},
       number = {1},
          eid = {L4},
        pages = {L4},
          doi = {10.3847/2041-8213/ad697c},
archivePrefix = {arXiv},
       eprint = {2406.19452},
 primaryClass = {astro-ph.HE},
       adsurl = {https://ui.adsabs.harvard.edu/abs/2024ApJ...972L...4G}
}

@article{halevi_black_2025,
	title = {A {Black} {Hole} {Is} {Born}: {3D} {General}-relativistic {Magnetohydrodynamic} {Simulation} of {Black} {Hole} {Formation} from {Core} {Collapse}},
	volume = {991},
	issn = {2041-8205},
	shorttitle = {A {Black} {Hole} {Is} {Born}},
	url = {https://doi.org/10.3847/2041-8213/ae08a6},
	doi = {10.3847/2041-8213/ae08a6},
	language = {en},
	number = {2},
	urldate = {2025-11-13},
	journal = {The Astrophysical Journal Letters},
	author = {Halevi, Goni and Shankar, Swapnil and Mösta, Philipp and Haas, Roland and Schnetter, Erik},
	month = sep,
	year = {2025},
	note = {Publisher: The American Astronomical Society},
	pages = {L51},
}

@ARTICLE{Poutanen2022,
       author = {{Poutanen}, Juri and {Veledina}, Alexandra and {Berdyugin}, Andrei V. and {Berdyugina}, Svetlana V. and {Jermak}, Helen and {Jonker}, Peter G. and {Kajava}, Jari J.~E. and {Kosenkov}, Ilia A. and {Kravtsov}, Vadim and {Piirola}, Vilppu and {Shrestha}, Manisha and {Perez Torres}, Manuel A. and {Tsygankov}, Sergey S.},
        title = "{Black hole spin─orbit misalignment in the x-ray binary MAXI J1820+070}",
      journal = {Science},
         year = 2022,
        month = feb,
       volume = {375},
       number = {6583},
        pages = {874-876},
          doi = {10.1126/science.abl4679},
archivePrefix = {arXiv},
       eprint = {2109.07511},
 primaryClass = {astro-ph.HE},
       adsurl = {https://ui.adsabs.harvard.edu/abs/2022Sci...375..874P}
}

@ARTICLE{Vigna-Gomez25,
       author = {{Vigna-G{\'o}mez}, Alejandro},
        title = "{The impact of natal kicks on black hole binaries}",
      journal = {\aap},
         year = 2025,
        month = sep,
       volume = {701},
          eid = {L3},
        pages = {L3},
          doi = {10.1051/0004-6361/202556051},
archivePrefix = {arXiv},
       eprint = {2507.07573},
 primaryClass = {astro-ph.HE},
       adsurl = {https://ui.adsabs.harvard.edu/abs/2025A&A...701L...3V}
}

@ARTICLE{h-amr,
       author = {{Liska}, M.~T.~P. and {Chatterjee}, K. and {Issa}, D. and {Yoon}, D. and {Kaaz}, N. and {Tchekhovskoy}, A. and {van Eijnatten}, D. and {Musoke}, G. and {Hesp}, C. and {Rohoza}, V. and {Markoff}, S. and {Ingram}, A. and {van der Klis}, M.},
        title = "{H-AMR: A New GPU-accelerated GRMHD Code for Exascale Computing with 3D Adaptive Mesh Refinement and Local Adaptive Time Stepping}",
      journal = {\apjs},
         year = 2022,
        month = dec,
       volume = {263},
       number = {2},
          eid = {26},
        pages = {26},
          doi = {10.3847/1538-4365/ac9966},
archivePrefix = {arXiv},
       eprint = {1912.10192},
 primaryClass = {astro-ph.HE},
       adsurl = {https://ui.adsabs.harvard.edu/abs/2022ApJS..263...26L}
}

@article{scheck_multidimensional_2006,
	title = {Multidimensional supernova simulations with approximative neutrino transport: {I}. {Neutron} star kicks and the anisotropy of neutrino-driven explosions in two spatial dimensions},
	volume = {457},
	issn = {0004-6361, 1432-0746},
	shorttitle = {Multidimensional supernova simulations with approximative neutrino transport},
	url = {http://www.aanda.org/10.1051/0004-6361:20064855},
	doi = {10.1051/0004-6361:20064855},
	language = {en},
	number = {3},
	urldate = {2024-05-15},
	journal = {A\&A},
	author = {Scheck, L. and Kifonidis, K. and Janka, H.-Th. and Müller, E.},
	month = oct,
	year = {2006},
	pages = {963--986},
}

@inbook{doi:10.1142/9789811282676_0003,
	author = {Alexander Heger and Bernhard M{\"u}ller and Ilya Mandel},
	booktitle = {The Encyclopedia of Cosmology},
	chapter = {Chapter 3},
	doi = {10.1142/9789811282676_0003},
	eprint = {https://www.worldscientific.com/doi/pdf/10.1142/9789811282676_0003},
	pages = {61-111},
	title = {Black Holes as the End State of Stellar Evolution: Theory and Simulations},
	url = {https://www.worldscientific.com/doi/abs/10.1142/9789811282676_0003},
    publisher = {World Scientific Series in Astrophysics},
    year = {2023}}

@ARTICLE{Gottlieb2019,
       author = {{Gottlieb}, Ore and {Levinson}, Amir and {Nakar}, Ehud},
        title = "{High efficiency photospheric emission entailed by formation of a collimation shock in gamma-ray bursts}",
      journal = {\mnras},
         year = 2019,
        month = sep,
       volume = {488},
       number = {1},
        pages = {1416-1426},
          doi = {10.1093/mnras/stz1828},
archivePrefix = {arXiv},
       eprint = {1904.07244},
 primaryClass = {astro-ph.HE},
       adsurl = {https://ui.adsabs.harvard.edu/abs/2019MNRAS.488.1416G}
}

@ARTICLE{Gottlieb2022,
       author = {{Gottlieb}, Ore and {Lalakos}, Aretaios and {Bromberg}, Omer and {Liska}, Matthew and {Tchekhovskoy}, Alexander},
        title = "{Black hole to breakout: 3D GRMHD simulations of collapsar jets reveal a wide range of transients}",
      journal = {\mnras},
         year = 2022,
        month = mar,
       volume = {510},
       number = {4},
        pages = {4962-4975},
          doi = {10.1093/mnras/stab3784},
archivePrefix = {arXiv},
       eprint = {2109.14619},
 primaryClass = {astro-ph.HE},
       adsurl = {https://ui.adsabs.harvard.edu/abs/2022MNRAS.510.4962G}
}

@article{wongwathanarat_three-dimensional_2013,
	title = {Three-dimensional neutrino-driven supernovae: {Neutron} star kicks, spins, and asymmetric ejection of nucleosynthesis products},
	volume = {552},
	issn = {0004-6361, 1432-0746},
	shorttitle = {Three-dimensional neutrino-driven supernovae},
	url = {http://www.aanda.org/10.1051/0004-6361/201220636},
	doi = {10.1051/0004-6361/201220636},
	language = {en},
	urldate = {2024-05-15},
	journal = {A\&A},
	author = {Wongwathanarat, A. and Janka, H.-Th. and Müller, E.},
	month = apr,
	year = {2013},
	pages = {A126},
}

@ARTICLE{belczynski_binaries,
       author = {{Belczynski}, Krzysztof and {Sadowski}, Aleksander and {Rasio}, Frederic A. and {Bulik}, Tomasz},
        title = "{Initial Populations of Black Holes in Star Clusters}",
      journal = {\apj},
         year = 2006,
        month = oct,
       volume = {650},
       number = {1},
        pages = {303-325},
          doi = {10.1086/506186},
archivePrefix = {arXiv},
       eprint = {astro-ph/0508005},
 primaryClass = {astro-ph},
       adsurl = {https://ui.adsabs.harvard.edu/abs/2006ApJ...650..303B}
}

@article{blandford_electromagnetic_1977,
	title = {Electromagnetic extraction of energy from {Kerr} black holes.},
	volume = {179},
	issn = {0035-8711},
	url = {https://ui.adsabs.harvard.edu/abs/1977MNRAS.179..433B},
	doi = {10.1093/mnras/179.3.433},
	urldate = {2024-06-05},
	journal = {Monthly Notices of the Royal Astronomical Society},
	author = {Blandford, R. D. and Znajek, R. L.},
	month = may,
	year = {1977},
	note = {Publisher: OUP
ADS Bibcode: 1977MNRAS.179..433B},
	pages = {433--456},
}

@article{wysocki_explaining_2018,
	title = {Explaining {LIGO}’s observations via isolated binary evolution with natal kicks},
	volume = {97},
	issn = {2470-0010, 2470-0029},
	url = {https://link.aps.org/doi/10.1103/PhysRevD.97.043014},
	doi = {10.1103/PhysRevD.97.043014},
	language = {en},
	number = {4},
	urldate = {2025-06-10},
	journal = {Phys. Rev. D},
	author = {Wysocki, Daniel and Gerosa, Davide and O’Shaughnessy, Richard and Belczynski, Krzysztof and Gladysz, Wojciech and Berti, Emanuele and Kesden, Michael and Holz, Daniel E.},
	month = feb,
	year = {2018},
	pages = {043014},
}

@ARTICLE{Boco2026,
       author = {{Boco}, Lumen and {Bosi}, Michele and {Sgalletta}, Cecilia and {Romagnolo}, Amedeo and {Mapelli}, Michela},
        title = "{A binary black hole merger rate comparison within the same metallicity - star formation rate framework}",
      journal = {arXiv e-prints},
         year = 2026,
        month = aug,
          eid = {arXiv:2608.13648},
        pages = {arXiv:2608.13648},
          doi = {10.48550/arXiv.2608.13648},
archivePrefix = {arXiv},
       eprint = {2608.13648},
 primaryClass = {astro-ph.HE},
       adsurl = {https://ui.adsabs.harvard.edu/abs/2026arXiv260813648B}
}

@ARTICLE{Fruchter2006,
       author = {{Fruchter}, A.~S. and {Levan}, A.~J. and {Strolger}, L. and {Vreeswijk}, P.~M. and {Thorsett}, S.~E. and {Bersier}, D. and {Burud}, I. and {Castro Cer{\'o}n}, J.~M. and {Castro-Tirado}, A.~J. and {Conselice}, C. and {Dahlen}, T. and {Ferguson}, H.~C. and {Fynbo}, J.~P.~U. and {Garnavich}, P.~M. and {Gibbons}, R.~A. and {Gorosabel}, J. and {Gull}, T.~R. and {Hjorth}, J. and {Holland}, S.~T. and {Kouveliotou}, C. and {Levay}, Z. and {Livio}, M. and {Metzger}, M.~R. and {Nugent}, P.~E. and {Petro}, L. and {Pian}, E. and {Rhoads}, J.~E. and {Riess}, A.~G. and {Sahu}, K.~C. and {Smette}, A. and {Tanvir}, N.~R. and {Wijers}, R.~A.~M.~J. and {Woosley}, S.~E.},
        title = "{Long {\ensuremath{\gamma}}-ray bursts and core-collapse supernovae have different environments}",
      journal = {\nat},
         year = 2006,
        month = may,
       volume = {441},
       number = {7092},
        pages = {463-468},
          doi = {10.1038/nature04787},
archivePrefix = {arXiv},
       eprint = {astro-ph/0603537},
 primaryClass = {astro-ph},
       adsurl = {https://ui.adsabs.harvard.edu/abs/2006Natur.441..463F}
}

@ARTICLE{Broekgaarden2026,
       author = {{Broekgaarden}, Floor S.},
        title = "{Lower Your Rates: On Claims of a Binary Black Hole Merger-Rate Crisis}",
      journal = {arXiv e-prints},
         year = 2026,
        month = jun,
          eid = {arXiv:2606.28515},
        pages = {arXiv:2606.28515},
          doi = {10.48550/arXiv.2606.28515},
archivePrefix = {arXiv},
       eprint = {2606.28515},
 primaryClass = {astro-ph.HE},
       adsurl = {https://ui.adsabs.harvard.edu/abs/2026arXiv260628515B}
}

@article{farr_distinguishing_2017,
	title = {Distinguishing spin-aligned and isotropic black hole populations with gravitational waves},
	volume = {548},
	issn = {0028-0836, 1476-4687},
	url = {https://www.nature.com/articles/nature23453},
	doi = {10.1038/nature23453},
	language = {en},
	number = {7668},
	urldate = {2024-09-04},
	journal = {Nature},
	author = {Farr, Will M. and Stevenson, Simon and Miller, M. Coleman and Mandel, Ilya and Farr, Ben and Vecchio, Alberto},
	month = aug,
	year = {2017},
	pages = {426--429},
}

@article{callister_state_2021,
	title = {State of the {Field}: {Binary} {Black} {Hole} {Natal} {Kicks} and {Prospects} for {Isolated} {Field} {Formation} after {GWTC}-2},
	volume = {920},
	issn = {0004-637X, 1538-4357},
	shorttitle = {State of the {Field}},
	url = {https://iopscience.iop.org/article/10.3847/1538-4357/ac1347},
	doi = {10.3847/1538-4357/ac1347},
	language = {en},
	number = {2},
	urldate = {2025-06-10},
	journal = {ApJ},
	author = {Callister, Thomas A. and Farr, Will M. and Renzo, Mathieu},
	month = oct,
	year = {2021},
	pages = {157},
}

@ARTICLE{Renzo2026,
       author = {{Renzo}, M. and {Gottlieb}, O. and {Chan}, H.~S. and {Goldberg}, J.~A. and {Grichener}, A. and {Sen}, K. and {Shah}, N. and {Farag}, E. and {Cantiello}, Matteo},
        title = "{A grid of fast-rotating, chemically-homogeneous, supernova and/or long-GRB progenitors}",
      journal = {arXiv e-prints},
         year = 2026,
        month = jun,
          eid = {arXiv:2606.21824},
        pages = {arXiv:2606.21824},
          doi = {10.48550/arXiv.2606.21824},
archivePrefix = {arXiv},
       eprint = {2606.21824},
 primaryClass = {astro-ph.HE},
       adsurl = {https://ui.adsabs.harvard.edu/abs/2026arXiv260621824R}
}

@misc{renzo24,
      title={Pair-instability evolution and explosions in massive stars}, 
      author={M. Renzo and N. Smith},
      year={2024},
      eprint={2407.16113},
      archivePrefix={arXiv},
      primaryClass={astro-ph.HE},
      url={https://arxiv.org/abs/2407.16113}, 
}

@article{fryer_explaining_2025,
	title = {Explaining {Nonmerger} {Gamma}-{Ray} {Bursts} and {Broad}-lined {Supernovae} with {Close} {Binary} {Progenitors} with {Black} {Hole} {Central} {Engines}},
	volume = {986},
	issn = {0004-637X},
	url = {https://doi.org/10.3847/1538-4357/add474},
	doi = {10.3847/1538-4357/add474},
	language = {en},
	number = {2},
	urldate = {2025-11-12},
	journal = {The Astrophysical Journal},
	author = {Fryer, Christopher L. and Burns, Eric and Ho, Anna Y. Q. and Corsi, Alessandra and Lien, Amy Y. and Perley, Daniel A. and Vail, Jada L. and Villar, V. Ashley},
	month = jun,
	year = {2025},
	note = {Publisher: The American Astronomical Society},
	pages = {185},
}

@ARTICLE{Du+2018,
       author = {{Du}, Shuang and {Li}, Xiao-Dong and {Hu}, Yi-Ming and {Peng}, Fang-Kun and {Li}, Miao},
        title = "{Gravitational waves induced by the asymmetric jets of gamma-ray bursts}",
      journal = {\mnras},
         year = 2018,
        month = oct,
       volume = {480},
       number = {1},
        pages = {402-406},
          doi = {10.1093/mnras/sty1800},
archivePrefix = {arXiv},
       eprint = {1807.00437},
 primaryClass = {astro-ph.HE},
       adsurl = {https://ui.adsabs.harvard.edu/abs/2018MNRAS.480..402D}
}

@article{Sedda2021,
    author = "Sedda, Manuel Arca and Mapelli, Michela and Benacquista, Matthew and Spera, Mario",
    title = "{Isolated and dynamical black hole mergers with B-POP: the role of star formation and dynamics, star cluster evolution, natal kicks, mass and spins, and hierarchical mergers}",
    eprint = "2109.12119",
    archivePrefix = "arXiv",
    primaryClass = "astro-ph.GA",
    doi = "10.1093/mnras/stad331",
    journal = "Mon. Not. Roy. Astron. Soc.",
    volume = "520",
    number = "4",
    pages = "5259--5282",
    year = "2023"
}

@ARTICLE{GW231123,
       author = {{Abac}, A.~G. and {Abouelfettouh}, I. and {Acernese}, F. and {Ackley}, K. and {Adamcewicz}, C. and {Adhicary}, S. and {Adhikari}, D. and {Adhikari}, N. and {Adhikari}, R.~X. and {Adkins}, V.~K. and {Afroz}, S. and {Agapito}, A. and {Agarwal}, D. and {Agathos}, M. and {Aggarwal}, N. and {Aggarwal}, S. and {Aguiar}, O.~D. and {Ahrend}, I.-L. and {Aiello}, L. and {Ain}, A. and {Ajith}, P. and {Akutsu}, T. and {Albanesi}, S. and {Ali}, W. and {Al-Kershi}, S. and {All{\'e}n{\'e}}, C. and {Allocca}, A. and {Al-Shammari}, S. and {Altin}, P.~A. and {Alvarez-Lopez}, S. and {Amar}, W. and {Amarasinghe}, O. and {Amato}, A. and {Amicucci}, F. and {Amra}, C. and {Ananyeva}, A. and {Anderson}, S.~B. and {Anderson}, W.~G. and {Andia}, M. and {Ando}, M. and {Andr{\'e}s-Carcasona}, M. and {Andri{\'c}}, T. and {Anglin}, J. and {Ansoldi}, S. and {Antelis}, J.~M. and {Antier}, S. and {Aoumi}, M. and {Appavuravther}, E.~Z. and {Appert}, S. and {Apple}, S.~K. and {Arai}, K. and {Alvarez}, C. Araujo and {Araya}, A. and {Araya}, M.~C. and {Sedda}, M. Arca and {Areeda}, J.~S. and {Aritomi}, N. and {Armato}, F. and {Armstrong}, S. and {Arnaud}, N. and {Arogeti}, M. and {Aronson}, S.~M. and {Arun}, K.~G. and {Ashton}, G. and {Aso}, Y. and {Asprea}, L. and {Assiduo}, M. and {Assis de Souza Melo}, S. and {Aston}, S.~M. and {Astone}, P. and {Attadio}, F. and {Aubin}, F. and {AultONeal}, K. and {Avallone}, G. and {Avila}, E.~A. and {Babak}, S. and {Badger}, C. and {Bae}, S. and {Bagnasco}, S. and {Baiotti}, L. and {Bajpai}, R. and {Baka}, T. and {Baker}, A.~M. and {Baker}, K.~A. and {Baker}, T. and {Baldi}, G. and {Baldicchi}, N. and {Ball}, M. and {Ballardin}, G. and {Ballmer}, S.~W. and {Banagiri}, S. and {Banerjee}, B. and {Bankar}, D. and {Baptiste}, T.~M. and {Baral}, P. and {Baratti}, M. and {Barayoga}, J.~C. and {Barish}, B.~C. and {Barker}, D. and {Barman}, N. and {Barneo}, P. and {Barone}, F. and {Barr}, B. and {Barsotti}, L. and {Barsuglia}, M. and {Barta}, D. and {Bartoletti}, A.~M. and {Barton}, M.~A. and {Bartos}, I. and {Basalaev}, A. and {Bassiri}, R. and {Basti}, A. and {Bawaj}, M. and {Baxi}, P. and {Bayley}, J.~C. and {Baylor}, A.~C. and {Baynard}, II, P.~A. and {Bazzan}, M. and {Bedakihale}, V.~M. and {Beirnaert}, F. and {Bejger}, M. and {Belardinelli}, D. and {Bell}, A.~S. and {Bellie}, D.~S. and {Bellizzi}, L. and {Benoit}, W. and {Bentara}, I. and {Bentley}, J.~D. and {Ben Yaala}, M. and {Bera}, S. and {Bergamin}, F. and {Berger}, B.~K. and {Bernuzzi}, S. and {Beroiz}, M. and {Berry}, C.~P.~L. and {Bersanetti}, D. and {Bertheas}, T. and {Bertolini}, A. and {Betzwieser}, J. and {Beveridge}, D. and {Bevilacqua}, G. and {Bevins}, N. and {Bhandare}, R. and {Bhatt}, R. and {Bhattacharjee}, D. and {Bhattacharyya}, S. and {Bhaumik}, S. and {Bhagwat}, S. and {Biancalana}, V. and {Bianchi}, A. and {Bilenko}, I.~A. and {Billingsley}, G. and {Binetti}, A. and {Bini}, S. and {Binu}, C. and {Biot}, S. and {Birnholtz}, O. and {Biscoveanu}, S. and {Bisht}, A. and {Bitossi}, M. and {Bizouard}, M.-A. and {Blaber}, S. and {Blackburn}, J.~K. and {Blagg}, L.~A. and {Blair}, C.~D. and {Blair}, D.~G. and {Bode}, N. and {Boettner}, N. and {Boileau}, G. and {Boldrini}, M. and {Bolingbroke}, G.~N. and {Bolliand}, A. and {Bonavena}, L.~D. and {Bondarescu}, R. and {Bondu}, F. and {Bonilla}, E. and {Bonilla}, M.~S. and {Bonino}, A. and {Bonnand}, R. and {Borchers}, A. and {Borhanian}, S. and {Boschi}, V. and {Bose}, S. and {Bossilkov}, V. and {Bothra}, Y. and {Boudon}, A. and {Bourg}, L. and {Boyle}, M. and {Bozzi}, A. and {Bradaschia}, C. and {Brady}, P.~R. and {Branch}, A. and {Branchesi}, M. and {Braun}, I. and {Briant}, T. and {Brillet}, A. and {Brinkmann}, M. and {Brockill}, P. and {Brockmueller}, E. and {Brooks}, A.~F.},
        title = "{GW231123: A Binary Black Hole Merger with Total Mass 190─265 M$_{{\ensuremath{\odot}}}$}",
      journal = {\apjl},
         year = 2025,
        month = nov,
       volume = {993},
       number = {1},
          eid = {L25},
        pages = {L25},
          doi = {10.3847/2041-8213/ae0c9c},
archivePrefix = {arXiv},
       eprint = {2507.08219},
 primaryClass = {astro-ph.HE},
       adsurl = {https://ui.adsabs.harvard.edu/abs/2025ApJ...993L..25A}
}

@article{Chatterjee_magneto-spin,
  title = {Misaligned magnetized accretion flows onto spinning black holes: Magneto-spin alignment, outflow power, and intermittent jets},
  author = {Chatterjee, Koushik and Kaaz, Nicholas and Liska, Matthew and Tchekhovskoy, Alexander and Markoff, Sera},
  journal = {Phys. Rev. D},
  volume = {112},
  issue = {6},
  pages = {063013},
  numpages = {19},
  year = {2025},
  month = {Sep},
  publisher = {American Physical Society},
  doi = {10.1103/hgj9-v4fk},
  url = {https://link.aps.org/doi/10.1103/hgj9-v4fk}
}

@ARTICLE{Skoutnev+24,
       author = {{Skoutnev}, Valentin A. and {Beloborodov}, Andrei M.},
        title = "{Zones of Tayler Instability in Stars}",
      journal = {\apj},
         year = 2025,
        month = aug,
       volume = {988},
       number = {2},
          eid = {195},
        pages = {195},
          doi = {10.3847/1538-4357/ade547},
archivePrefix = {arXiv},
       eprint = {2411.08492},
 primaryClass = {astro-ph.SR},
       adsurl = {https://ui.adsabs.harvard.edu/abs/2025ApJ...988..195S}
}

@article{popov_natal_2025,
	title = {Natal kicks of compact objects},
	volume = {101},
	issn = {1387-6473},
	url = {https://www.sciencedirect.com/science/article/pii/S1387647325000132},
	doi = {10.1016/j.newar.2025.101734},
	urldate = {2025-11-11},
	journal = {New Astronomy Reviews},
	author = {Popov, Sergei and Müller, Bernhard and Mandel, Ilya},
	month = dec,
	year = {2025},
	pages = {101734},
}

@ARTICLE{1918PhyZ...19..156L,
       author = {{Lense}, Josef and {Thirring}, Hans},
        title = "{{\"U}ber den Einflu{\ss} der Eigenrotation der Zentralk{\"o}rper auf die Bewegung der Planeten und Monde nach der Einsteinschen Gravitationstheorie}",
      journal = {Physikalische Zeitschrift},
         year = 1918,
        month = jan,
       volume = {19},
        pages = {156},
       adsurl = {https://ui.adsabs.harvard.edu/abs/1918PhyZ...19..156L}
}

@book{MTW,
    author = "Misner, Charles W. and Thorne, K. S. and Wheeler, J. A.",
    title = "{Gravitation}",
    isbn = "978-0-7167-0344-0, 978-0-691-17779-3",
    publisher = "W. H. Freeman",
    address = "San Francisco",
    year = "1973"
}

@ARTICLE{Pavlik2018,
       author = {{Pavl{\'\i}k}, V{\'a}clav and {Je{\v{r}}{\'a}bkov{\'a}}, Tereza and {Kroupa}, Pavel and {Baumgardt}, Holger},
        title = "{The black hole retention fraction in star clusters}",
      journal = {\aap},
         year = 2018,
        month = sep,
       volume = {617},
          eid = {A69},
        pages = {A69},
          doi = {10.1051/0004-6361/201832919},
archivePrefix = {arXiv},
       eprint = {1806.05192},
 primaryClass = {astro-ph.GA},
       adsurl = {https://ui.adsabs.harvard.edu/abs/2018A&A...617A..69P}
}

@ARTICLE{jacquemin_ide_2025,
       author = {{Jacquemin-Ide}, Jonatan and {Begelman}, Mitchell C. and {Lowell}, Beverly and {Liska}, Matthew and {Dexter}, Jason and {Tchekhovskoy}, Alexander},
        title = "{Demystifying flux eruptions: Magnetic flux transport in magnetically arrested disks}",
      journal = {arXiv e-prints},
         year = 2025,
        month = oct,
          eid = {arXiv:2510.25842},
        pages = {arXiv:2510.25842},
          doi = {10.48550/arXiv.2510.25842},
archivePrefix = {arXiv},
       eprint = {2510.25842},
 primaryClass = {astro-ph.HE},
       adsurl = {https://ui.adsabs.harvard.edu/abs/2025arXiv251025842J}
}

@ARTICLE{OConnor2010,
       author = {{O'Connor}, Evan and {Ott}, Christian D.},
        title = "{A new open-source code for spherically symmetric stellar collapse to neutron stars and black holes}",
      journal = {Classical and Quantum Gravity},
         year = 2010,
        month = jun,
       volume = {27},
       number = {11},
          eid = {114103},
        pages = {114103},
          doi = {10.1088/0264-9381/27/11/114103},
archivePrefix = {arXiv},
       eprint = {0912.2393},
 primaryClass = {astro-ph.HE},
       adsurl = {https://ui.adsabs.harvard.edu/abs/2010CQGra..27k4103O}
}

@article{magneto-spin,
	author = {Jonathan C. McKinney and Alexander Tchekhovskoy and Roger D. Blandford},
	doi = {10.1126/science.1230811},
	eprint = {https://www.science.org/doi/pdf/10.1126/science.1230811},
	journal = {Science},
	number = {6115},
	pages = {49-52},
	title = {Alignment of Magnetized Accretion Disks and Relativistic Jets with Spinning Black Holes},
	url = {https://www.science.org/doi/abs/10.1126/science.1230811},
	volume = {339},
	year = {2013}}

@ARTICLE{blaauw,
       author = {{Blaauw}, A.},
        title = "{On the origin of the O- and B-type stars with high velocities (the ``run-away'' stars), and some related problems}",
      journal = {\bain},
         year = 1961,
        month = may,
       volume = {15},
        pages = {265},
       adsurl = {https://ui.adsabs.harvard.edu/abs/1961BAN....15..265B}
}

@ARTICLE{banagiri2023,
       author = {{Banagiri}, Sharan and {Doctor}, Zoheyr and {Kalogera}, Vicky and {Kimball}, Chase and {Andrews}, Jeff J.},
        title = "{Direct Statistical Constraints on the Natal Kick Velocity of a Black Hole in an X-Ray Quiet Binary}",
      journal = {\apj},
         year = 2023,
        month = dec,
       volume = {959},
       number = {2},
          eid = {106},
        pages = {106},
          doi = {10.3847/1538-4357/ad0557},
archivePrefix = {arXiv},
       eprint = {2211.16361},
 primaryClass = {astro-ph.HE},
       adsurl = {https://ui.adsabs.harvard.edu/abs/2023ApJ...959..106B}
}

@ARTICLE{Marchant2016,
       author = {{Marchant}, Pablo and {Langer}, Norbert and {Podsiadlowski}, Philipp and {Tauris}, Thomas M. and {Moriya}, Takashi J.},
        title = "{A new route towards merging massive black holes}",
      journal = {\aap},
         year = 2016,
        month = apr,
       volume = {588},
          eid = {A50},
        pages = {A50},
          doi = {10.1051/0004-6361/201628133},
archivePrefix = {arXiv},
       eprint = {1601.03718},
 primaryClass = {astro-ph.SR},
       adsurl = {https://ui.adsabs.harvard.edu/abs/2016A&A...588A..50M}
}

@ARTICLE{deMink2009,
       author = {{de Mink}, S.~E. and {Cantiello}, M. and {Langer}, N. and {Pols}, O.~R. and {Brott}, I. and {Yoon}, S.-Ch.},
        title = "{Rotational mixing in massive binaries. Detached short-period systems}",
      journal = {\aap},
         year = 2009,
        month = apr,
       volume = {497},
       number = {1},
        pages = {243-253},
          doi = {10.1051/0004-6361/200811439},
archivePrefix = {arXiv},
       eprint = {0902.1751},
 primaryClass = {astro-ph.SR},
       adsurl = {https://ui.adsabs.harvard.edu/abs/2009A&A...497..243D}
}

@ARTICLE{Polko2017,
       author = {{Polko}, Peter and {McKinney}, Jonathan C.},
        title = "{Electromagnetic versus Lense-Thirring alignment of black hole accretion discs}",
      journal = {\mnras},
         year = 2017,
        month = jan,
       volume = {464},
       number = {3},
        pages = {2660-2671},
          doi = {10.1093/mnras/stw1875},
archivePrefix = {arXiv},
       eprint = {1512.07969},
 primaryClass = {astro-ph.HE},
       adsurl = {https://ui.adsabs.harvard.edu/abs/2017MNRAS.464.2660P}
}

@ARTICLE{Song2016,
       author = {{Song}, H.~F. and {Meynet}, G. and {Maeder}, A. and {Ekstr{\"o}m}, S. and {Eggenberger}, P.},
        title = "{Massive star evolution in close binaries. Conditions for homogeneous chemical evolution}",
      journal = {\aap},
         year = 2016,
        month = jan,
       volume = {585},
          eid = {A120},
        pages = {A120},
          doi = {10.1051/0004-6361/201526074},
archivePrefix = {arXiv},
       eprint = {1508.06094},
 primaryClass = {astro-ph.SR},
       adsurl = {https://ui.adsabs.harvard.edu/abs/2016A&A...585A.120S}
}

@ARTICLE{BP_effect,
       author = {{Bardeen}, James M. and {Petterson}, Jacobus A.},
        title = "{The Lense-Thirring Effect and Accretion Disks around Kerr Black Holes}",
      journal = {\apjl},
         year = 1975,
        month = jan,
       volume = {195},
        pages = {L65},
          doi = {10.1086/181711},
       adsurl = {https://ui.adsabs.harvard.edu/abs/1975ApJ...195L..65B}
}

@ARTICLE{Jermyn2023,
       author = {{Jermyn}, Adam S. and {Bauer}, Evan B. and {Schwab}, Josiah and {Farmer}, R. and {Ball}, Warrick H. and {Bellinger}, Earl P. and {Dotter}, Aaron and {Joyce}, Meridith and {Marchant}, Pablo and {Mombarg}, Joey S.~G. and {Wolf}, William M. and {Sunny Wong}, Tin Long and {Cinquegrana}, Giulia C. and {Farrell}, Eoin and {Smolec}, R. and {Thoul}, Anne and {Cantiello}, Matteo and {Herwig}, Falk and {Toloza}, Odette and {Bildsten}, Lars and {Townsend}, Richard H.~D. and {Timmes}, F.~X.},
        title = "{Modules for Experiments in Stellar Astrophysics (MESA): Time-dependent Convection, Energy Conservation, Automatic Differentiation, and Infrastructure}",
      journal = {\apjs},
         year = 2023,
        month = mar,
       volume = {265},
       number = {1},
          eid = {15},
        pages = {15},
          doi = {10.3847/1538-4365/acae8d},
archivePrefix = {arXiv},
       eprint = {2208.03651},
 primaryClass = {astro-ph.SR},
       adsurl = {https://ui.adsabs.harvard.edu/abs/2023ApJS..265...15J}
}

@ARTICLE{Oded2011,
       author = {{Papish}, Oded and {Soker}, Noam},
        title = "{Exploding core collapse supernovae with jittering jets}",
      journal = {\mnras},
         year = 2011,
        month = sep,
       volume = {416},
       number = {3},
        pages = {1697-1702},
          doi = {10.1111/j.1365-2966.2011.18671.x},
archivePrefix = {arXiv},
       eprint = {1103.1554},
 primaryClass = {astro-ph.HE},
       adsurl = {https://ui.adsabs.harvard.edu/abs/2011MNRAS.416.1697P}
}

@ARTICLE{Paxton2019,
       author = {{Paxton}, Bill and {Smolec}, R. and {Schwab}, Josiah and {Gautschy}, A. and {Bildsten}, Lars and {Cantiello}, Matteo and {Dotter}, Aaron and {Farmer}, R. and {Goldberg}, Jared A. and {Jermyn}, Adam S. and {Kanbur}, S.~M. and {Marchant}, Pablo and {Thoul}, Anne and {Townsend}, Richard H.~D. and {Wolf}, William M. and {Zhang}, Michael and {Timmes}, F.~X.},
        title = "{Modules for Experiments in Stellar Astrophysics (MESA): Pulsating Variable Stars, Rotation, Convective Boundaries, and Energy Conservation}",
      journal = {\apjs},
         year = 2019,
        month = jul,
       volume = {243},
       number = {1},
          eid = {10},
        pages = {10},
          doi = {10.3847/1538-4365/ab2241},
archivePrefix = {arXiv},
       eprint = {1903.01426},
 primaryClass = {astro-ph.SR},
       adsurl = {https://ui.adsabs.harvard.edu/abs/2019ApJS..243...10P}
}

@ARTICLE{Paxton2018,
       author = {{Paxton}, Bill and {Schwab}, Josiah and {Bauer}, Evan B. and {Bildsten}, Lars and {Blinnikov}, Sergei and {Duffell}, Paul and {Farmer}, R. and {Goldberg}, Jared A. and {Marchant}, Pablo and {Sorokina}, Elena and {Thoul}, Anne and {Townsend}, Richard H.~D. and {Timmes}, F.~X.},
        title = "{Modules for Experiments in Stellar Astrophysics (MESA): Convective Boundaries, Element Diffusion, and Massive Star Explosions}",
      journal = {\apjs},
         year = 2018,
        month = feb,
       volume = {234},
       number = {2},
          eid = {34},
        pages = {34},
          doi = {10.3847/1538-4365/aaa5a8},
archivePrefix = {arXiv},
       eprint = {1710.08424},
 primaryClass = {astro-ph.SR},
       adsurl = {https://ui.adsabs.harvard.edu/abs/2018ApJS..234...34P}
}

@ARTICLE{Paxton2015,
       author = {{Paxton}, Bill and {Marchant}, Pablo and {Schwab}, Josiah and {Bauer}, Evan B. and {Bildsten}, Lars and {Cantiello}, Matteo and {Dessart}, Luc and {Farmer}, R. and {Hu}, H. and {Langer}, N. and {Townsend}, R.~H.~D. and {Townsley}, Dean M. and {Timmes}, F.~X.},
        title = "{Modules for Experiments in Stellar Astrophysics (MESA): Binaries, Pulsations, and Explosions}",
      journal = {\apjs},
         year = 2015,
        month = sep,
       volume = {220},
       number = {1},
          eid = {15},
        pages = {15},
          doi = {10.1088/0067-0049/220/1/15},
archivePrefix = {arXiv},
       eprint = {1506.03146},
 primaryClass = {astro-ph.SR},
       adsurl = {https://ui.adsabs.harvard.edu/abs/2015ApJS..220...15P}
}

@ARTICLE{Paxton2013,
       author = {{Paxton}, Bill and {Cantiello}, Matteo and {Arras}, Phil and {Bildsten}, Lars and {Brown}, Edward F. and {Dotter}, Aaron and {Mankovich}, Christopher and {Montgomery}, M.~H. and {Stello}, Dennis and {Timmes}, F.~X. and {Townsend}, Richard},
        title = "{Modules for Experiments in Stellar Astrophysics (MESA): Planets, Oscillations, Rotation, and Massive Stars}",
      journal = {\apjs},
         year = 2013,
        month = sep,
       volume = {208},
       number = {1},
          eid = {4},
        pages = {4},
          doi = {10.1088/0067-0049/208/1/4},
archivePrefix = {arXiv},
       eprint = {1301.0319},
 primaryClass = {astro-ph.SR},
       adsurl = {https://ui.adsabs.harvard.edu/abs/2013ApJS..208....4P}
}

@ARTICLE{Paxton2011,
       author = {{Paxton}, Bill and {Bildsten}, Lars and {Dotter}, Aaron and {Herwig}, Falk and {Lesaffre}, Pierre and {Timmes}, Frank},
        title = "{Modules for Experiments in Stellar Astrophysics (MESA)}",
      journal = {\apjs},
         year = 2011,
        month = jan,
       volume = {192},
       number = {1},
          eid = {3},
        pages = {3},
          doi = {10.1088/0067-0049/192/1/3},
archivePrefix = {arXiv},
       eprint = {1009.1622},
 primaryClass = {astro-ph.SR},
       adsurl = {https://ui.adsabs.harvard.edu/abs/2011ApJS..192....3P}
}

@ARTICLE{nakamura_1998bw_2001,
       author = {{Nakamura}, Takayoshi and {Mazzali}, Paolo A. and {Nomoto}, Ken'ichi and {Iwamoto}, Koichi},
        title = "{Light Curve and Spectral Models for the Hypernova SN 1998BW Associated with GRB 980425}",
      journal = {\apj},
         year = 2001,
        month = apr,
       volume = {550},
       number = {2},
        pages = {991-999},
          doi = {10.1086/319784},
archivePrefix = {arXiv},
       eprint = {astro-ph/0007010},
 primaryClass = {astro-ph},
       adsurl = {https://ui.adsabs.harvard.edu/abs/2001ApJ...550..991N}
}

@ARTICLE{1998bw,
       author = {{Galama}, T.~J. and {Vreeswijk}, P.~M. and {van Paradijs}, J. and {Kouveliotou}, C. and {Augusteijn}, T. and {B{\"o}hnhardt}, H. and {Brewer}, J.~P. and {Doublier}, V. and {Gonzalez}, J.-F. and {Leibundgut}, B. and {Lidman}, C. and {Hainaut}, O.~R. and {Patat}, F. and {Heise}, J. and {in't Zand}, J. and {Hurley}, K. and {Groot}, P.~J. and {Strom}, R.~G. and {Mazzali}, P.~A. and {Iwamoto}, K. and {Nomoto}, K. and {Umeda}, H. and {Nakamura}, T. and {Young}, T.~R. and {Suzuki}, T. and {Shigeyama}, T. and {Koshut}, T. and {Kippen}, M. and {Robinson}, C. and {de Wildt}, P. and {Wijers}, R.~A.~M.~J. and {Tanvir}, N. and {Greiner}, J. and {Pian}, E. and {Palazzi}, E. and {Frontera}, F. and {Masetti}, N. and {Nicastro}, L. and {Feroci}, M. and {Costa}, E. and {Piro}, L. and {Peterson}, B.~A. and {Tinney}, C. and {Boyle}, B. and {Cannon}, R. and {Stathakis}, R. and {Sadler}, E. and {Begam}, M.~C. and {Ianna}, P.},
        title = "{An unusual supernova in the error box of the {\ensuremath{\gamma}}-ray burst of 25 April 1998}",
      journal = {\nat},
         year = 1998,
        month = oct,
       volume = {395},
       number = {6703},
        pages = {670-672},
          doi = {10.1038/27150},
archivePrefix = {arXiv},
       eprint = {astro-ph/9806175},
 primaryClass = {astro-ph},
       adsurl = {https://ui.adsabs.harvard.edu/abs/1998Natur.395..670G}
}

@ARTICLE{Palmerio2019,
       author = {{Palmerio}, J.~T. and {Vergani}, S.~D. and {Salvaterra}, R. and {Sanders}, R.~L. and {Japelj}, J. and {Vidal-Garc{\'\i}a}, A. and {D'Avanzo}, P. and {Corre}, D. and {Perley}, D.~A. and {Shapley}, A.~E. and {Boissier}, S. and {Greiner}, J. and {Le Floc'h}, E. and {Wiseman}, P.},
        title = "{Are long gamma-ray bursts biased tracers of star formation? Clues from the host galaxies of the Swift/BAT6 complete sample of bright LGRBs. III. Stellar masses, star formation rates, and metallicities at z > 1}",
      journal = {\aap},
         year = 2019,
        month = mar,
       volume = {623},
          eid = {A26},
        pages = {A26},
          doi = {10.1051/0004-6361/201834179},
archivePrefix = {arXiv},
       eprint = {1901.02457},
 primaryClass = {astro-ph.HE},
       adsurl = {https://ui.adsabs.harvard.edu/abs/2019A&A...623A..26P}
}

@ARTICLE{henisey_variability_2012,
       author = {{Henisey}, Ken B. and {Blaes}, Omer M. and {Fragile}, P. Chris},
        title = "{Variability from Non-axisymmetric Fluctuations Interacting with Standing Shocks in Tilted Black Hole Accretion Disks}",
      journal = {\apj},
         year = 2012,
        month = dec,
       volume = {761},
       number = {1},
          eid = {18},
        pages = {18},
          doi = {10.1088/0004-637X/761/1/18},
archivePrefix = {arXiv},
       eprint = {1211.2273},
 primaryClass = {astro-ph.HE},
       adsurl = {https://ui.adsabs.harvard.edu/abs/2012ApJ...761...18H}
}

@article{fragile_hydrodynamic_2005,
	title = {Hydrodynamic {Simulations} of {Tilted} {Thick}-{Disk} {Accretion} onto a {Kerr} {Black} {Hole}},
	volume = {623},
	issn = {0004-637X},
	url = {https://iopscience.iop.org/article/10.1086/428433/meta},
	doi = {10.1086/428433},
	language = {en},
	number = {1},
	urldate = {2025-09-16},
	journal = {The Astrophysical Journal},
	author = {Fragile, P. Chris and Anninos, Peter},
	month = apr,
	year = {2005},
	note = {Publisher: IOP Publishing},
	pages = {347},
}

@article{sykes_long-time_2025,
	title = {Long-time {3D} supernova simulations of nonrotating progenitors with magnetic fields},
	volume = {111},
	url = {https://link.aps.org/doi/10.1103/PhysRevD.111.063042},
	doi = {10.1103/PhysRevD.111.063042},
	number = {6},
	urldate = {2025-11-11},
	journal = {Physical Review D},
	author = {Sykes, Bailey and Müller, Bernhard},
	month = mar,
	year = {2025},
	note = {Publisher: American Physical Society},
	pages = {063042},
}

@ARTICLE{Burrows+2024b,
       author = {{Burrows}, Adam and {Wang}, Tianshu and {Vartanyan}, David and {Coleman}, Matthew S.~B.},
        title = "{A Theory for Neutron Star and Black Hole Kicks and Induced Spins}",
      journal = {\apj},
         year = 2024,
        month = mar,
       volume = {963},
       number = {1},
          eid = {63},
        pages = {63},
          doi = {10.3847/1538-4357/ad2353},
       adsurl = {https://ui.adsabs.harvard.edu/abs/2024ApJ...963...63B}
}

@ARTICLE{Fuller+19,
       author = {{Fuller}, Jim and {Piro}, Anthony L. and {Jermyn}, Adam S.},
        title = "{Slowing the spins of stellar cores}",
      journal = {\mnras},
         year = 2019,
        month = may,
       volume = {485},
       number = {3},
        pages = {3661-3680},
          doi = {10.1093/mnras/stz514},
archivePrefix = {arXiv},
       eprint = {1902.08227},
 primaryClass = {astro-ph.SR},
       adsurl = {https://ui.adsabs.harvard.edu/abs/2019MNRAS.485.3661F}
}

@article{jacquemin-ide_collapsar_2024,
	title = {Collapsar {Gamma}-{Ray} {Bursts} {Grind} {Their} {Black} {Hole} {Spins} to a {Halt}},
	volume = {961},
	issn = {0004-637X, 1538-4357},
	url = {https://iopscience.iop.org/article/10.3847/1538-4357/ad02f0},
	doi = {10.3847/1538-4357/ad02f0},
	language = {en},
	number = {2},
	urldate = {2024-05-15},
	journal = {ApJ},
	author = {Jacquemin-Ide, Jonatan and Gottlieb, Ore and Lowell, Beverly and Tchekhovskoy, Alexander},
	month = feb,
	year = {2024},
	pages = {212},
}

@article{repetto_galactic_2017,
	title = {The {Galactic} distribution of {X}-ray binaries and its implications for compact object formation and natal kicks},
	volume = {467},
	issn = {0035-8711},
	url = {https://doi.org/10.1093/mnras/stx027},
	doi = {10.1093/mnras/stx027},
	number = {1},
	urldate = {2026-08-26},
	journal = {Monthly Notices of the Royal Astronomical Society},
	author = {Repetto, Serena and Igoshev, Andrei P. and Nelemans, Gijs},
	month = may,
	year = {2017},
	pages = {298--310},
}

@article{gottlieb_collapsar_2023,
	title = {Collapsar {Black} {Holes} {Are} {Likely} {Born} {Slowly} {Spinning}},
	volume = {952},
	issn = {2041-8205, 2041-8213},
	url = {https://iopscience.iop.org/article/10.3847/2041-8213/ace779},
	doi = {10.3847/2041-8213/ace779},
	language = {en},
	number = {2},
	urldate = {2025-06-17},
	journal = {ApJL},
	author = {Gottlieb, Ore and Jacquemin-Ide, Jonatan and Lowell, Beverly and Tchekhovskoy, Alexander and Ramirez-Ruiz, Enrico},
	month = aug,
	year = {2023},
	pages = {L32},
}

@ARTICLE{Kissin&Thompson18,
       author = {{Kissin}, Yevgeni and {Thompson}, Christopher},
        title = "{Rotation and Magnetism of Massive Stellar Cores}",
      journal = {\apj},
         year = 2018,
        month = aug,
       volume = {862},
       number = {2},
          eid = {111},
        pages = {111},
          doi = {10.3847/1538-4357/aab1fb},
archivePrefix = {arXiv},
       eprint = {1705.07906},
 primaryClass = {astro-ph.SR},
       adsurl = {https://ui.adsabs.harvard.edu/abs/2018ApJ...862..111K}
}

@ARTICLE{Cantiello+07,
       author = {{Cantiello}, M. and {Yoon}, S. -C. and {Langer}, N. and {Livio}, M.},
        title = "{Binary star progenitors of long gamma-ray bursts}",
      journal = {\aap},
         year = 2007,
        month = apr,
       volume = {465},
       number = {2},
        pages = {L29-L33},
          doi = {10.1051/0004-6361:20077115},
archivePrefix = {arXiv},
       eprint = {astro-ph/0702540},
 primaryClass = {astro-ph},
       adsurl = {https://ui.adsabs.harvard.edu/abs/2007A&A...465L..29C}
}

@ARTICLE{Langer92,
       author = {{Langer}, N.},
        title = "{Helium enrichment in massive early type stars.}",
      journal = {\aap},
         year = 1992,
        month = nov,
       volume = {265},
        pages = {L17-L20},
       adsurl = {https://ui.adsabs.harvard.edu/abs/1992A&A...265L..17L}
}

@ARTICLE{Maeder87,
       author = {{Maeder}, A.},
        title = "{Evidences for a bifurcation in massive star evolution. The ON-blue stragglers.}",
      journal = {\aap},
         year = 1987,
        month = may,
       volume = {178},
        pages = {159-169},
       adsurl = {https://ui.adsabs.harvard.edu/abs/1987A&A...178..159M}
}

@article{chatterjee_flux_2022,
	title = {Flux {Eruption} {Events} {Drive} {Angular} {Momentum} {Transport} in {Magnetically} {Arrested} {Accretion} {Flows}},
	volume = {941},
	issn = {0004-637X, 1538-4357},
	url = {https://iopscience.iop.org/article/10.3847/1538-4357/ac9d97},
	doi = {10.3847/1538-4357/ac9d97},
	language = {en},
	number = {1},
	urldate = {2024-09-06},
	journal = {ApJ},
	author = {Chatterjee, K. and Narayan, R.},
	month = dec,
	year = {2022},
	pages = {30},
}

@article{ripperda_black_2022,
	title = {Black {Hole} {Flares}: {Ejection} of {Accreted} {Magnetic} {Flux} through {3D} {Plasmoid}-mediated {Reconnection}},
	volume = {924},
	issn = {2041-8205, 2041-8213},
	shorttitle = {Black {Hole} {Flares}},
	url = {https://iopscience.iop.org/article/10.3847/2041-8213/ac46a1},
	doi = {10.3847/2041-8213/ac46a1},
	language = {en},
	number = {2},
	urldate = {2025-01-13},
	journal = {ApJL},
	author = {Ripperda, B. and Liska, M. and Chatterjee, K. and Musoke, G. and Philippov, A. A. and Markoff, S. B. and Tchekhovskoy, A. and Younsi, Z.},
	month = jan,
	year = {2022},
	pages = {L32},
}

@article{narayan_jets_2022,
	title = {Jets in magnetically arrested hot accretion flows: geometry, power, and black hole spin-down},
	volume = {511},
	copyright = {https://academic.oup.com/journals/pages/open\_access/funder\_policies/chorus/standard\_publication\_model},
	issn = {0035-8711, 1365-2966},
	shorttitle = {Jets in magnetically arrested hot accretion flows},
	url = {https://academic.oup.com/mnras/article/511/3/3795/6524200},
	doi = {10.1093/mnras/stac285},
	language = {en},
	number = {3},
	urldate = {2025-07-04},
	journal = {Monthly Notices of the Royal Astronomical Society},
	author = {Narayan, Ramesh and Chael, Andrew and Chatterjee, Koushik and Ricarte, Angelo and Curd, Brandon},
	month = feb,
	year = {2022},
	pages = {3795--3813},
}
\bibliographystyle{aasjournalv7}

\appendix
\begin{figure}
    \centering
    \includegraphics[width=\linewidth]{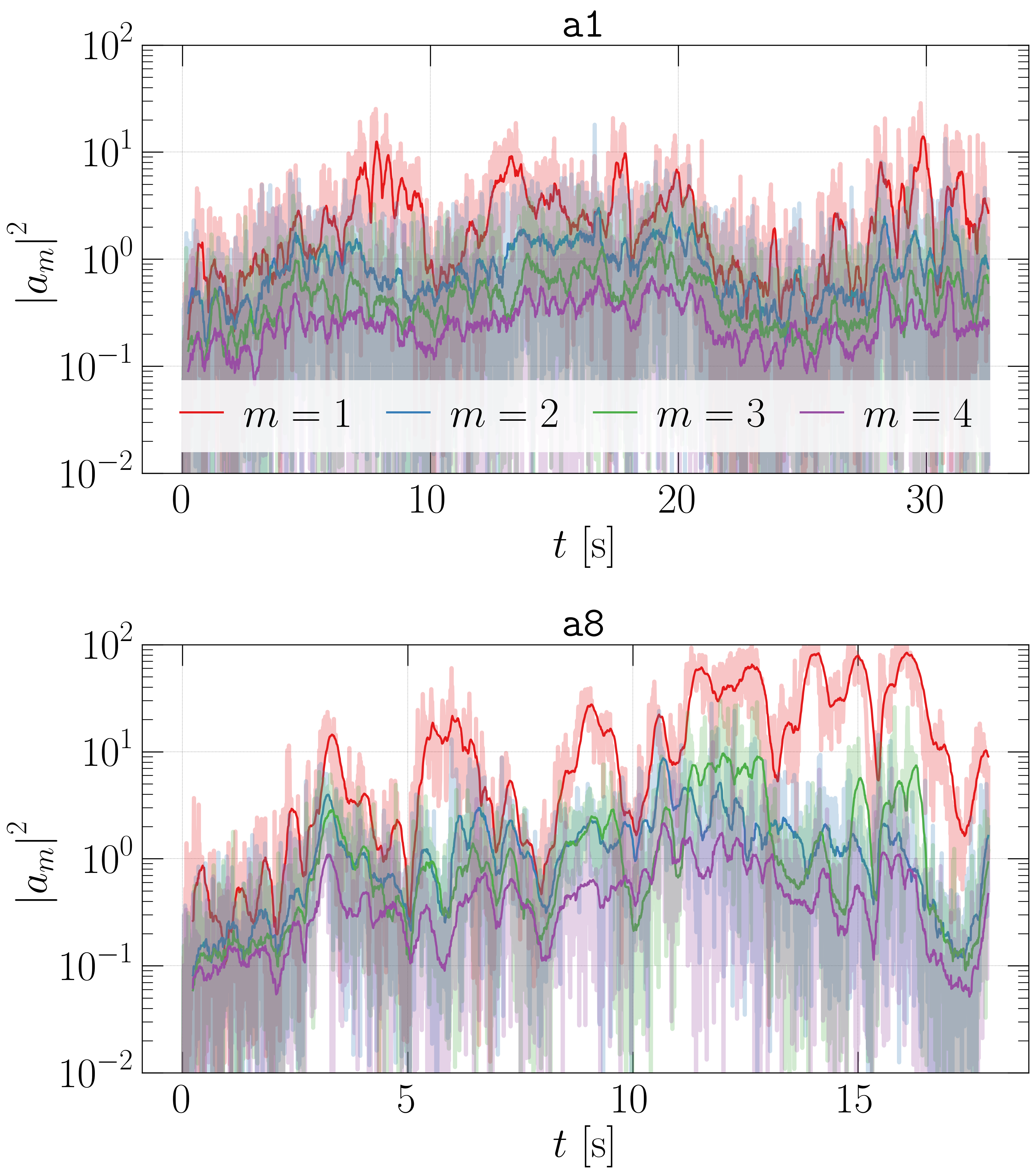}
    \caption{Normalized power spectrum of azimuthal modes $m \in \{1,2,3,4\}$ of the vertical displacement of the accretion disk from its midplane for fiducial models \texttt{a1} (top) and \texttt{a8} (bottom).}
    \label{fig:modes}
\end{figure}
\begin{figure*}
    \centering \includegraphics[width=\linewidth]{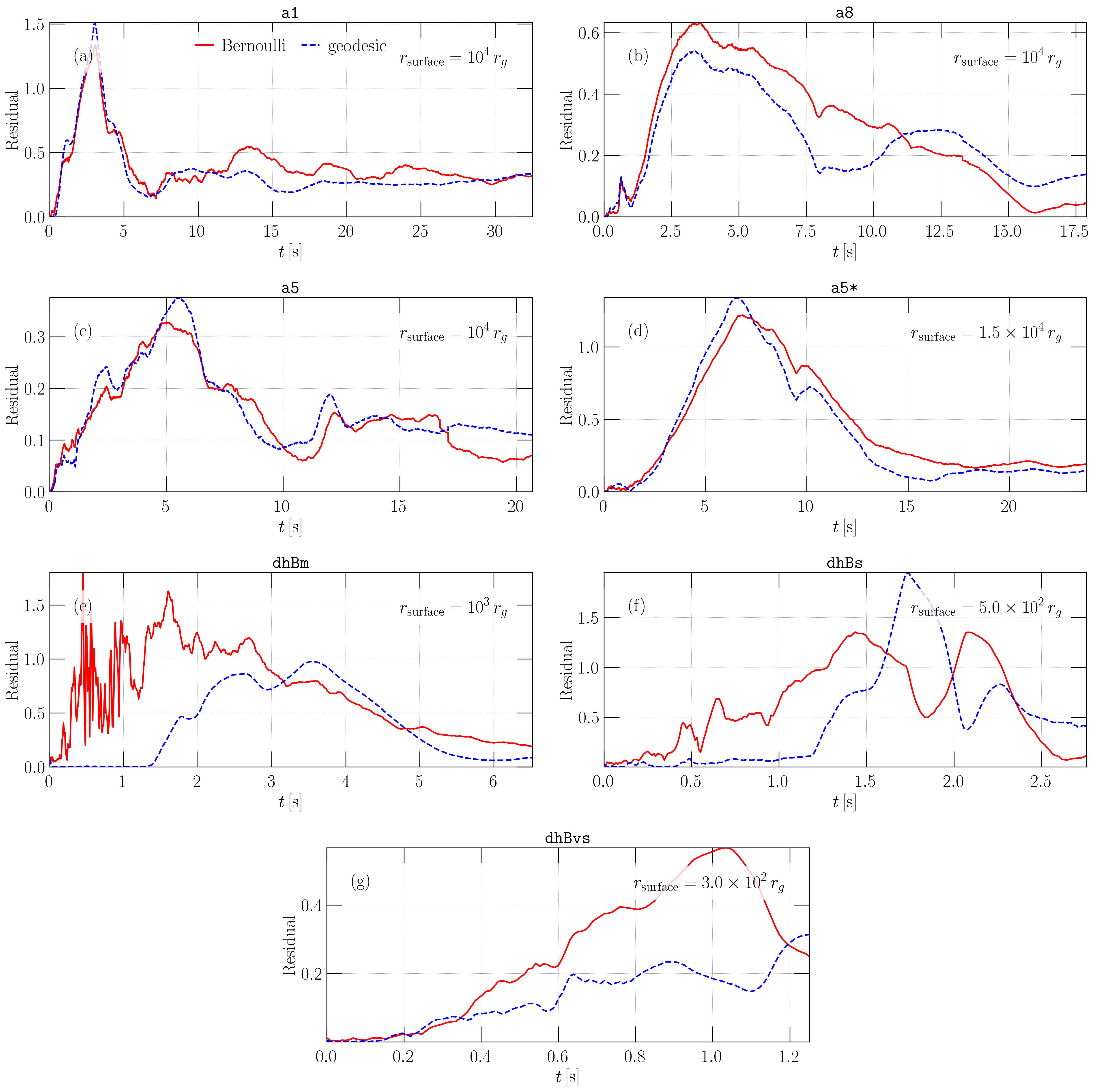} 
    \caption{Approximate ``Gauss's law" tests for our simulations. For both the Bernoulli and geodesic bound matter criteria, we compute the vector norm of the residual between the volume-integrated ejecta momentum (corrected for losses through the outer boundary) and the momentum that crosses a surface of radius $r_{\rm surface}$, given in the top-right corner of each panel. We normalize the residual to the magnitude of the BH momentum inferred from \eq{p-bh} at the end of the simulation. The surface radii are chosen for being large enough to enclose most of the bound matter but small enough to capture most of the outflows before the simulations end. In general, the discrepancies are $\lesssim 40\%$ of the total BH momentum and can be attributed to momentum exchange via gravity and outflows launched at late times that have not yet crossed the chosen surface.}
    \label{fig:gauss}
\end{figure*}
\begin{figure}
    \centering
    \includegraphics[width=\linewidth]{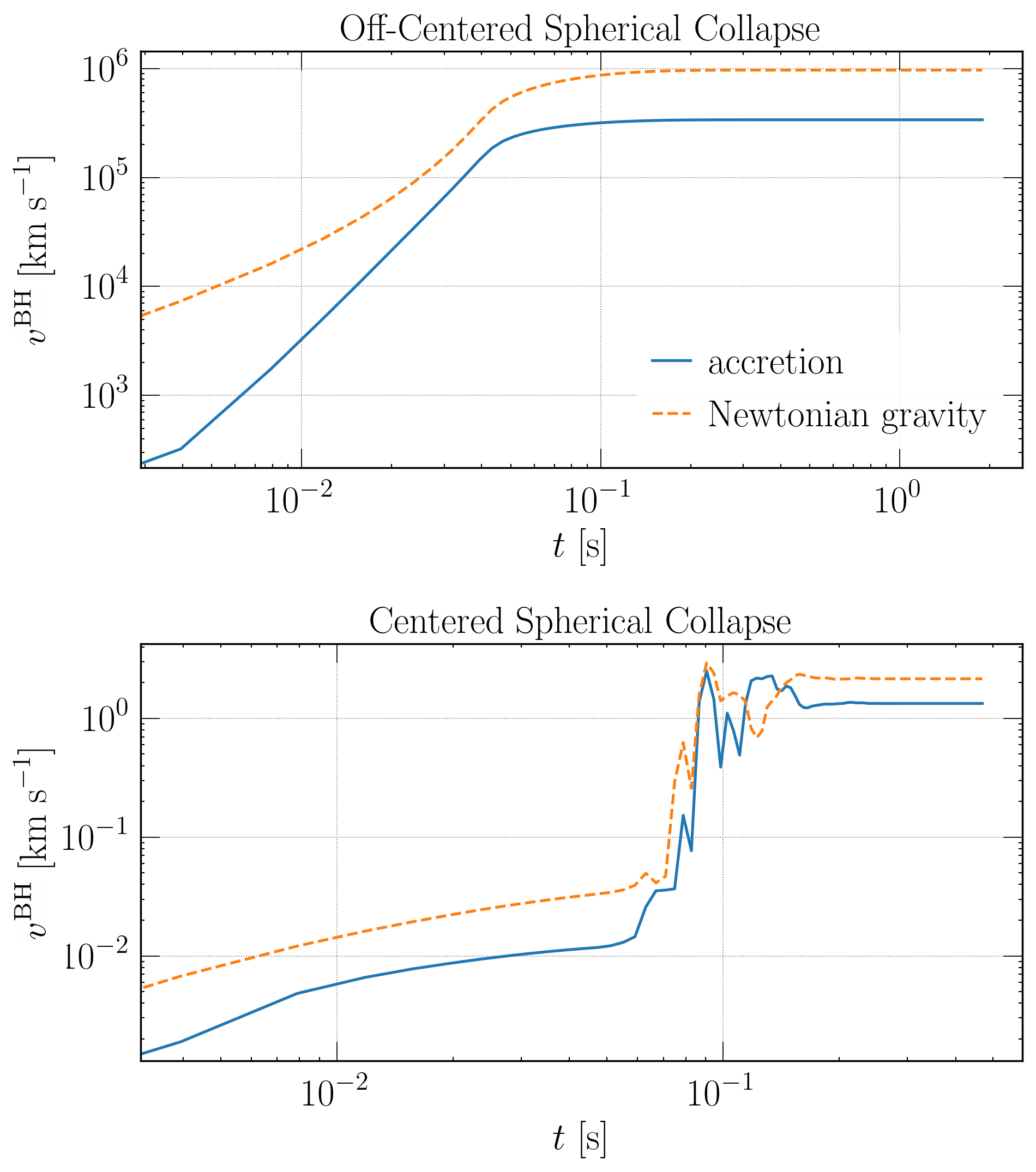}
    \caption{Magnitudes of the kick due to momentum accretion and Newtonian gravity versus time for test simulations in which a small non-rotating, unmagnetized star accretes onto a Schwarzschild BH. In the top panel the center of the star is initially displaced along the $z$-axis and in the bottom panel the star is centered around the BH. In the off-centered simulation, Newtonian gravity dominates over horizon flux, so that they combine to give a superluminal BH kick. In the centered simulation, the net kick is $\mathcal{O}(1\,\rm km\,s^{-1})$ due to numerical error alone.} 
    \label{fig:grav-tests}
\end{figure}
\begin{figure*}
    \centering
    \includegraphics[width=\linewidth]{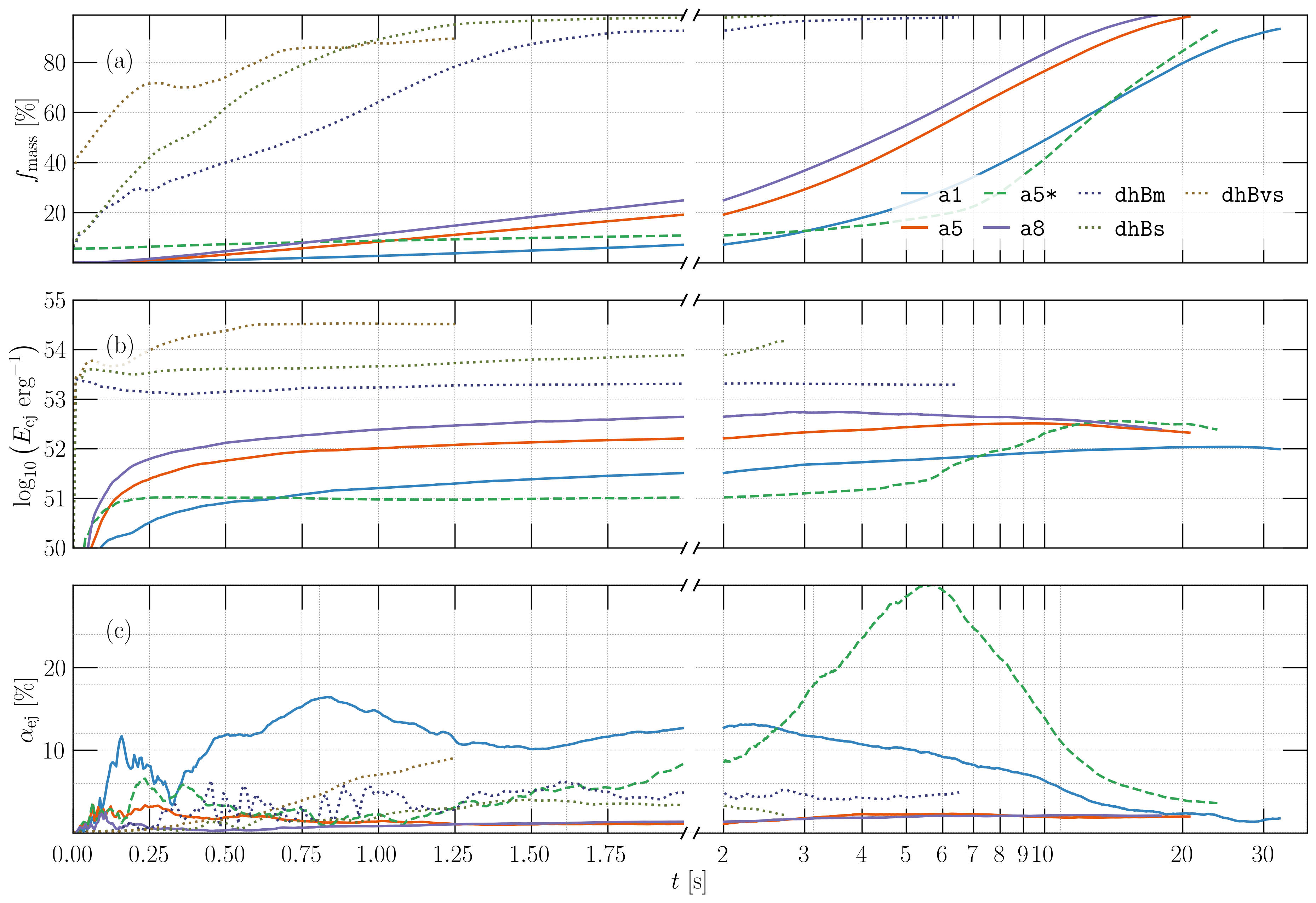}
    \caption{Time series of the processed (accreted or unbound) mass fraction $f_{\rm mass}$, diagnostic explosion energy $E_{\rm ej}$, and momentum asymmetry parameter $\alpha_{\rm ej}$. (Here $f_{\rm mass}$ and $E_{\rm ej}$ computed using the Bernoulli criterion for bound matter.) For visibility we plot on a linear time scale until $t = 2\,\rm{s}$ and a logarithmic scale thereafter.}
    \label{fig:diag_evolution}
\end{figure*}

\section{Diagnostic Quantities}\label{app:diag}
 
We further characterize the outflows in our simulations with the mass-averaged asymptotic speed of the ejecta, 
\begin{align}
    \bar v_{\rm ej,\inf} \equiv M_{\rm ej}^{-1}\int D \sqrt{1-\Gamma_\inf^{-2}}\sqrt{-g}\dd{r}\dd{\theta}\dd{\phi},
\end{align}
where the asymptotic Lorentz factor $\Gamma_\inf = \mathcal{B} +1$ and the integration is restricted to $\mathcal{B} > 0$ and $u^r > 0$. We find that these diagnostics do not differ significantly between the geodesic and Bernoulli criteria, so we only report them for the latter.

To measure the tilt and precession of the accretion disk, we calculate the angular momentum vector of the disk in the innermost $100\, r_g$ following \citet{fragile_hydrodynamic_2005}, i.e.
\begin{align}
    J^{\rm disk}_\mu = \frac{\epsilon_{\alpha\beta\gamma\mu} L^{\alpha\beta} S^\gamma}{2\sqrt{-S^\nu S_\nu}},
\end{align}
where 
\begin{align}
    L^{\mu\nu} = \int_{r_+<r<100r_g} \qty(x^\mu T^{\nu t} - x^\nu T^{\mu t})\sqrt{-g}\dd{r}\dd{\theta}\dd{\phi},
\end{align}
and 
\begin{align}
    S^\mu = \int_{r_+<r<100r_g} T^{\mu t}\sqrt{-g}\dd{r}\dd{\theta}\dd{\phi}.
\end{align}
The tilt (nutation) angle relative to the $z$-axis is 
\begin{align}
    \theta_{\rm disk} = \arccos(J^{\rm disk}_z/J^{\rm disk}), \label{eq:disk-tilt}
\end{align}
where 
\begin{align}
    J^2_{\rm disk} \equiv \sum_{i\in\{x,y,z\}} (J^{\rm disk}_i)^2
\end{align}
and the precession angle is given by
\begin{align}
    \phi_{\rm disk} = \arctan(-J^{\rm disk}_x/J^{\rm disk}_y). \label{eq:disk-prec}
\end{align}

If launched, the two jets may become arbitrarily tilted as the disk orientation changes, so we refer to them as ``$+$" and ``$-$" rather than ``northern" and ``southern", except in cases where they are persistently oriented in the $\pm z$-direction. We define jet ``$+$" (``$-$") as the jet whose axis makes a positive (negative) inner product with the disk angular momentum, where we define the jet axes as originating from the origin and intercepting the center of magnetic energy on either side of the disk plane, 
\begin{align}
    x^i_{b,\pm} = \frac{\int_{\pm} x^i b^2 \sqrt{-g}\dd{r}\dd{\theta}\dd{\phi}}{\int_{\pm} b^2 \sqrt{-g}\dd{r}\dd{\theta}\dd{\phi}}.
\end{align}
We compute the tilt and precession of the jet axes via \eq{disk-tilt} and \eq{disk-prec}, replacing $(J_{\rm disk})_i$ with $x^i_{b,\pm}$. Rather than plotting the precession angles as a function of time directly, we show the cumulative azimuthal phase of the disk and jets (equal to the instantaneous precession angle up to integer multiples of $2\pi$) to more clearly visualize the time evolution. We track the misalignment between the jet axes by defining a ``separation" angle
\begin{align}
    \delta = \arccos(-\frac{\sum_i x^i_{b,+} x^i_{b,-}}{\sqrt{\sum_i(x^i_{b,+})^2\sum_i(x^i_{b,-})^2}}),
\end{align}
such that $\delta = 0$ when the jets are diametrically opposed.

We compute the power for jet $\pm$ by integrating positive contributions to the radial Poynting flux at the BH horizon over the hemisphere $\mathcal{H}_\pm$,
\begin{align}
    P_\pm = \oint_{r_+,\,\mathcal{H}_\pm} \max(0,-(\tensor{T}{^r_{t}})_{\rm EM})\sqrt{-g}\dd{\theta}\dd{\phi}, \label{eq:p-jet}
\end{align}
where $\mathcal{H}_+$ and $\mathcal{H}_-$ are separated by the accretion disk plane. The total jet power is $P_{\rm jet} = P_+ + P_-$.

\section{Momentum Conservation Equation}\label{app:cons}

The MHD evolution is subject to the conservative equation,
\begin{align}
    0 &= \nabla_{\mu}\tensor{T}{^\mu_\nu}.
\end{align}
Expanding the covariant divergence, we have 
\begin{align}
    0 &= \partial_\mu \tensor{T}{^\mu_\nu} + \Gamma^\mu_{\mu\lambda}\tensor{T}{^\lambda_\nu} - \Gamma^\lambda_{\mu\nu}\tensor{T}{^\mu_\lambda}.
\end{align}
Using the metric determinant to group the first two terms, we have 
\begin{align}
    0 &= \frac{1}{\sqrt{-g}}\partial_\mu(\tensor{T}{^\mu_\nu}\sqrt{-g}) - \Gamma^\lambda_{\mu\nu}\tensor{T}{^\mu_\lambda}.
\end{align}
Multiplying both sides by $\sqrt{-g}$, splitting the time and space components, and rearranging gives
\begin{align}
    \partial_t(\tensor{T}{^t_\nu}\sqrt{-g}) = -\partial_i(\tensor{T}{^i_\nu}\sqrt{-g}) + \Gamma^\lambda_{\mu\nu}\tensor{T}{^\mu_\lambda}\sqrt{-g}. \label{eq:cons_intermediate}
\end{align}
In terms of the metric, the Christoffel symbol is given by
\begin{align}
    \Gamma^\lambda_{\mu\nu} = \frac{1}{2}g^{\lambda\delta}(g_{\mu\delta,\nu} + g_{\nu\delta,\mu} - g_{\mu\nu,\delta}),
\end{align}
where commas denote partial derivatives. Raising the index $\lambda$ on the stress-energy tensor, the second term on the RHS of \eq{cons_intermediate} becomes
\begin{align}
    \Gamma^\lambda_{\mu\nu}\tensor{T}{^\mu_\lambda}\sqrt{-g} &= \frac{1}{2}\tensor{T}{^{\mu\delta}}\sqrt{-g}(g_{\mu\delta,\nu} + g_{\nu\delta,\mu} - g_{\mu\nu,\delta}).
\end{align}
Together, the last two terms in parentheses are antisymmetric under $\mu\leftrightarrow\delta$, so they vanish when contracted with $T^{\mu\delta}$. Thus \eq{cons_intermediate} now reads,
\begin{align}
    \partial_t(\tensor{T}{^t_\nu}\sqrt{-g}) = -\partial_i(\tensor{T}{^i_\nu}\sqrt{-g}) + \frac{1}{2}g_{\mu\delta,\nu}\tensor{T}{^{\mu\delta}}\sqrt{-g}.
\end{align}
Integrating this equation over all space and applying the divergence theorem on the RHS, we obtain
\begin{align}
    \begin{split}
        \partial_t\int\tensor{T}{^t_\nu}\sqrt{-g}\dd[3]{x} &= -\oint \tensor{T}{^i_\nu}\sqrt{-g}\dd{\Sigma}_i\\ &\phantom{=~} + \int \frac{1}{2}g_{\mu\delta,\nu}\tensor{T}{^{\mu\delta}}\sqrt{-g}\dd[3]{x},
    \end{split}
\end{align}
where $\Sigma_i$ is the boundary of the spatial volume. For $\nu = j$ this equation expresses spatial momentum conservation. For our calculation, the first term on the RHS corresponds to the momentum flux through the horizon and the outer boundary of the grid, and the second term is the change in the gas momentum due to the gravity of the BH.

\section{Kick Consistency Check}\label{app:consistency}

As a sanity check, we perform a rough ``Gauss's law" test by comparing the momentum flux through a coordinate sphere of moderate radius with the total momentum of the ejecta (unbound material), computed via \eq{p-ej}, for both bound criteria. In Fig.~\ref{fig:gauss}, we show the time evolution of the residuals between the ejecta momentum and the momentum that crosses a sphere of chosen radius $r_{\rm surface}$ normalized to the final BH momentum given by \eq{p-bh}. We choose $r_{\rm surface}$ to contain most of the bound matter but also capture most of the outflows by the end of the simulations. For most models, the two curves for the Bernoulli and geodesic criteria are in reasonable agreement, and the final residuals are of order $\rm{few}\times10\%$ of the BH kick momentum. Discrepancies are expected due to momentum exchange via gravity and outflows launched at late times that have not yet crossed $r_{\rm surface}$ by the end of the simulation.

\section{Newtonian Gravity}\label{app:gravity}

The effect of the gas gravity on the BH can be estimated using a Newtonian calculation,
\begin{align}
    \dot p_i^{\rm grav,\,Newton} = \int D\frac{x_i}{r^3}\sqrt{-g}\dd{r}\dd{\theta}\dd{\phi}, \label{eq:Newton}
\end{align}
where $x^i \equiv (x, y, z)$. Because \eq{Newton} is Newtonian while the gas responds relativistically to the BH spacetime and the BH is held fixed at the origin, we do not expect exact cancellation between $\dot p^{\rm acc}_i$ and $\dot p^{\rm grav,\, Newton}_i$ for matter with zero initial radial momentum that falls into the BH. This discrepancy makes \eq{Newton} inaccurate for our simulations, especially because the density near the BH is orders of magnitude higher in collapsars than in ordinary CCSN explosions due to the presence of a disk. Moreover, gravitational forces from aspherical bound matter that has yet to be accreted by the end of the simulation can exaggerate the apparent kick.

Indeed, we find that the apparent BH kick due to the Newtonian integral given by \eq{Newton} differs by a factor $\sim 2$--$4$ from the implicit gravitational kick obtained by integrating \eq{f-grav}, and in some cases has one or more components with opposite sign. The greatest discrepancy arises in our DH models, which feature much greater mass accretion rates. Were we to calculate the total kick by adding this Newtonian gravity contribution to the kick from horizon fluxes, we would, for some models, obtain kicks of order $0.1c$, which is clearly unphysical for natal kicks. We attribute this behavior to GR corrections for matter in the dense collapsar disk near the BH as well as our inability to self-consistently track the BH motion relative to the progenitor, which would introduce additional drag and dynamical friction and potentially enhance the accretion rate. To illustrate the limitations of the Newtonian integral, we simulate the accretion of a small, non-rotating star with zero magnetic fields onto a Schwarzschild BH displaced from its center and compare the measured kicks from \eq{Newton} and \eq{f-mom}. As the top panel of \fig{grav-tests} shows, gravity strongly dominates, yielding an impossible apparent net kick of $v > c$. To show that this error is not mainly associated with numerical error such as imperfections in the grid, we also simulate the collapse of the same spherical progenitor centered around the BH and obtain near-zero accretion and Newtonian gravity kick contributions of $\mathcal{O}(1\,\rm km\,s^{-1})$, as the bottom panel of \fig{grav-tests} shows.

\section{Disk Warping}\label{app:warp}

Our collapsar disks are warped on the BH horizon, that is, the height of the disk above the midplane (whose normal is given by the mean angular momentum in the innermost $100\,r_g$ of the disk) varies non-axisymmetrically. To quantify the warp, at each timestep we compute the $\rho$-weighted mean disk height as a function of azimuth in coordinates aligned with the instantaneous disk angular momentum. We then extract the azimuthal modes as well as the RMS disk height integrated over the disk. We show the evolution of the power for modes $m \in \{1,2,3,4\}$ for models \texttt{a1} and \texttt{a8} in \fig{modes}. The $m = 1$ mode dominates, but higher harmonics also carry significant sustained power, especially for the high-spin BH. We find no significant correlation between the power spectrum and the linear momentum transfer to the BH; it is possible that their relationship is evident only at frequencies higher than are practical to resolve in these long-duration simulations.

\section{Evolution of Explosion Properties}\label{app:explosion}

In \fig{diag_evolution}, we plot the time evolution of the accreted or ejected mass fraction $f_{\rm mass}$, explosion energy $\rm E_{\rm ej}$, and momentum asymmetry parameter $\alpha_{\rm ej}$ for all our simulations. Due to greater jet power at early times, fiducial models \texttt{a5} and \texttt{a8} converge much faster than \texttt{a1} and \texttt{a5*}; see \fig{diag_evolution}(a). The DH models converge very quickly compared to the low-mass models due to enormous explosion energy. The values of $E_{\rm ej}$ (\fig{diag_evolution}[b]) saturate within the first few seconds for the low-mass collapsars, with the exception of $\texttt{a5*}$, in which the delayed revival of the jets at around $t\approx 6\,\rm{s}$ drives $E_{\rm ej}$ up by around an order of magnitude. For DH model \texttt{dhBs}, $E_{\rm ej}$ appears to still be rising by the end of the simulation. Note that losses through the outer boundary of the grid are responsible for the shallow decline in $E_{\rm ej}$ apparent in models \texttt{a5*}, \texttt{a5}, and \texttt{a8}, but this does not affect our overall conclusions. For models \texttt{a1} and \texttt{a5*}, the substantial decline in $\alpha_{\rm ej}$ (\fig{diag_evolution}[c]) is a result of highly asymmetric outflows at early times becoming diluted by more symmetric outflows at later times, increasing the denominator in \eq{alpha-ej} while keeping the numerator relatively constant, or asymmetries aligned opposite their earlier counterparts.

\end{document}